\documentclass[twocolumn]{aastex63}
\usepackage{natbib}
\usepackage{color} 
\usepackage{aas_macros} 
\usepackage{ulem}
\usepackage{hyperref} 
\usepackage{amsmath}
\accepted{}
\submitjournal{ApJ}

\shorttitle{proportion of halo stars belonging to \textit{Gaia}-Sausage}
\shortauthors{Wenbo et al.}
\begin{document}
{
\title{Contribution of Gaia Sausage to the Galactic Stellar Halo Revealed by K Giants and Blue
	Horizontal Branch Stars from the Large Sky Area Multi-Object Fiber Spectroscopic
	Telescope, Sloan Digital Sky Survey, and Gaia}

\correspondingauthor{Gang Zhao}
\email{gzhao@nao.cas.cn}
\author{Wenbo Wu}
\affil{CAS Key Laboratory of Optical Astronomy, National Astronomical Observatories, Chinese Academy of Sciences, Beijing 100101, People's Republic of China; gzhao@nao.cas.cn}
\affil{School of Astronomy and Space Science, University of Chinese Academy of Sciences, Beijing 100049, People's Republic of China}
\author{Gang Zhao}
\affil{CAS Key Laboratory of Optical Astronomy, National Astronomical Observatories, Chinese Academy of Sciences, Beijing 100101, People's Republic of China; gzhao@nao.cas.cn}
\affil{School of Astronomy and Space Science, University of Chinese Academy of Sciences, Beijing 100049, People's Republic of China}
\author{Xiang-Xiang Xue} \affil{CAS Key Laboratory of Optical Astronomy, National Astronomical
  Observatories, Chinese Academy of Sciences, Beijing 100101, People's Republic of China; gzhao@nao.cas.cn} \affil{School of Astronomy and Space Science, University of Chinese Academy of Sciences, Beijing 100049, People's Republic of China}
\author{Sarah A. Bird} \affil{CAS Key Laboratory of Optical Astronomy, National Astronomical
	Observatories, Chinese Academy of Sciences, Beijing 100101, People's Republic of China; gzhao@nao.cas.cn} \affil{China Three Gorges University, Yichang 443002, People's Republic of China} \affil{Center for Astronomy and Space Sciences, China Three Gorges University, Yichang 443002, People's Republic of China} 
\author{Chengqun Yang}
\affiliation{Shanghai Astronomical Observatory, Chinese Academy of Sciences, 80 Nandan Road, Shanghai 200030, People’s Republic of China}



\begin{abstract} \label{abstract}

We explore the contribution of the \textit{Gaia}-Sausage to the stellar halo of the Milky Way by making use of a Gaussian Mixture model (GMM) and applying it to halo star samples of LAMOST K giants, SEGUE K giants, and SDSS blue horizontal branch stars. The GMM divides the stellar halo into two parts, of which one represents a more metal-rich and highly radially biased component associated with an ancient, head-on collision referred to as the \textit{Gaia}-Sausage, and the other one is a more metal-poor and isotropic halo. A symmetric bimodal Gaussian is used to describe the distribution of spherical velocity of the \textit{Gaia}-Sausage, and we find that the mean absolute radial velocity of the two lobes decreases with Galactocentric radius. We find that the \textit{Gaia}-Sausage contributes about $41\%-74\%$ of the inner (Galactocentric radius $r_\mathrm{gc} < 30$ kpc) stellar halo. The fraction of stars of the \textit{Gaia}-Sausage starts to decline beyond $r_\mathrm{gc} \sim$ $25-30$ kpc, and the outer halo is found to be significantly less influenced by the \textit{Gaia}-Sausage than the inner halo. After the removal of halo substructures found by integrals of motion, the contribution of the \textit{Gaia}-Sausage falls slightly within $r_\mathrm{gc} \sim$ 25 kpc, but is still as high as $30\%-63\%$. Finally, we select several possible Sausage-related substructures consisting of stars on highly eccentric orbits. The GMM/Sausage component agrees well with the selected substructure stars in their chemodynamical properties, which increases our confidence in the reliability of the GMM fits.

\end{abstract}

\keywords{\textit{Unified Astronomy Thesaurus concepts}: Milky Way stellar halo (1060), Galaxy mergers (608), Milky Way formation (1053), Milky Way Galaxy physics (1056)}


\section{Introduction} \label{sec:intro}

In the $\Lambda$ CDM cosmological paradigm, Milky Way sized halos are built from the mergers of smaller satellite galaxies \citep{1978MNRAS.183..341W}. According to the stellar mass ratio of the satellite to the host galaxy, mergers can be categorized into major or minor types. These mergers usually happen in the early epoch of galaxy formation, and contribute much to the stellar halo population. Understanding the motions and chemical properties of halo stars can help us disentangle the Galaxy's detailed accretion history.

The combination of data from the \textit{Gaia} satellite and large spectroscopic surveys has provided new insights into the assembly history of the Milky Way. One of the most impressive findings is that the local stellar halo is dominated by the stars left from a major merger event named as the \textit{Gaia}-Sausage \citep{2018MNRAS.478..611B}, or the \textit{Gaia}-Enceladus \citep{2018Natur.563...85H}. From the study of the chemodynamic properties of its possible members, the \textit{Gaia}-Sausage is thought to be a large dwarf galaxy with a highly eccentric orbit and a stellar mass on the order of $10^9 - 10^{10}\,\text{M}_\odot$ \citep{2019MNRAS.487L..47V,2019MNRAS.490.3426D,2019MNRAS.484.4471F,2019MNRAS.482.3426M}, and the metallicity distribution function of the \textit{Gaia}-Sausage stars has a peak [Fe/H] ranging from $-1.4$ to $-1.2$ dex \citep{2019ApJ...881L..10S,2019NatAs...3..932G,2020MNRAS.493.5195D,2020MNRAS.497..109F,2020ApJ...901...48N}. Recent studies found that the expected head-on collision of the ancient $Gaia$-Sausage and the Galaxy happened around $8-10$ Gyr ago \citep{2019ApJ...881L..10S,2019ApJ...883L...5B,2020ApJ...897L..18B}.

The Galactic halo is expected to be heavily influenced by the tidal debris from the \textit{Gaia}-Sausage considering the large size of this satellite. \citet{2018MNRAS.478..611B} found a component of stars with highly radial orbits and a metallicity of [Fe/H] $\geq -1.7$ dex in the local stellar halo (within $\sim$ 10 kpc from the Sun). This component is further confirmed to extend out to a Galactocentric radius $r_\mathrm{gc} \sim 25$ kpc \citep{2019MNRAS.486..378L,2019AJ....157..104B,2021ApJ...919...66B}. \citet{2018ApJ...862L...1D} showed that highly eccentric (eccentricity $\mathrm{ec} \sim$ 0.9) stars share an apocenter ($r_\mathrm{apo}$) of about $20-25$ kpc, which is also the ``break radius'' of the density profile of the stellar halo \citep{2009MNRAS.398.1757W,2011ApJ...731....4S,2015A&A...579A..38P,2015ApJ...809..144X}. This anisotropic component is thought to be the tidal debris of the \textit{Gaia}-Sausage which is mixed together with a more metal-poor and isotropic component of the Galactic halo \citep{2018ApJ...856L..26M,2019ApJ...874....3N,2019MNRAS.486..378L,2021MNRAS.502.5686I}.

The highly radially-biased portion, which can be interpreted as the debris of the ancient massive merger know as the \textit{Gaia}-Sausage, is thought to contribute a large number stars to the Galactic stellar halo. \citet{2018ApJ...862L...1D} studied the fraction of main sequence stars with $\mathrm{ec} \geq 0.9$ and $10 \leq r_\mathrm{apo}/\text{kpc} \leq 25$ as a function of metallicity. Such stars peak at a fraction of 20\% in their SDSS DR9 main sequence star sample and have a mean metallicity of [Fe/H] $= -1.4$ dex. \citet{2019ApJ...874....3N} showed that the \textit{Gaia}-Sausage stars peak at [Fe/H] = $-1.4$ dex, and contribute $60\%-80\%$ of main sequence stars in the SDSS-\textit{Gaia} DR2 sample within $7.5 \leq r_\mathrm{gc}/\text{kpc} \leq 10$ and $|z| > 2.5$ kpc after the exclusion of disc stars. In the work of \citet{2019MNRAS.486..378L}, stars originating from the \textit{Gaia}-Sausage account for at least 50\% of their halo blue horizontal branch (BHB) star sample within the inner stellar halo ($r_\mathrm{gc} \leq 25$ kpc), while the fraction sharply diminishes in the outer halo ($r_\mathrm{gc} \geq 30$ kpc). The mean metallicity of these \textit{Gaia}-Sausage BHB stars is $-1.62$ dex. \citet{2021MNRAS.502.5686I} studied the chemo-kinematics of the RR Lyrae stars in the \textit{Gaia} DR2 catalog, and they found that the \textit{Gaia}-Sausage contributes $50\%-80\%$ of the halo RR Lyrae at $5 < r_\mathrm{gc}/\text{kpc} < 25$. The metallicity of RR Lyrae belonging to the radially-biased halo ranges from $-1.7$ to $-1.2$ dex. $N$-body simulations run by \citet{2021ApJ...923...92N} implied that the \textit{Gaia}-Sausage merger delivered about half of the Milky Way's stellar halo. Many studies of cosmological hydrodynamic simulations also support the view that the inner stellar halo of the Milky Way is dominated by the last significant accretion event \citep{2019MNRAS.484.4471F,2019MNRAS.485.2589M,2020MNRAS.495...29E}.

The contribution of the \textit{Gaia}-Sausage can also be found as halo substructures. \citet{2019ApJ...886...76D} and \citet{2019MNRAS.482..921S} suggested that the Virgo Overdensity of stars is caused by a radial dwarf galaxy merger. The study of low-$\alpha$ metal-rich stars on eccentric orbits provides further indication of the connection between the \textit{Gaia}-Sausage and the Virgo Overdensity, also called the Virgo Radial Merger \citep{2021SCPMA..6439562Z}. Besides the Virgo Overdensity, many halo substructures are likely associated with the \textit{Gaia}-Sausage as indicated by similarities in the halo star phase-space distribution \citep{2019A&A...631L...9K,2020ApJ...898L..37Y,2020ApJ...891...39Y,2021ApJ...908..191C}. \citet{2021ApJ...919...66B} selected several substructures characterized by stars on highly eccentric orbits based on the work of Xue et al. (2021, in preparation). They found that stars belonging to the highly radial substructures share similar kinematic properties with the radial Sausage component of \citet{2019MNRAS.486..378L}. These selected stars contribute around 60\% of all substructures found in Xue et al. (2021, in preparation). 

The removal of halo substructures in \citet{2021ApJ...919...66B} results in a slightly less radial stellar halo. However, the decline of the anisotropy is limited, which suggests that considerable numbers of the \textit{Gaia}-Sausage stars are not obviously selected as substructure stars. \citet{2019MNRAS.486..378L} used a Gaussian mixture model to study the contribution of the \textit{Gaia}-Sausage to the stellar halo after the removal of the Sagittarius stream members (Sgr-removed halo) by halo BHB stars. The question still remains as to how many stars deposited by the \textit{Gaia}-Sausage contribute the smooth stellar halo (defined as the remaining halo stars after removing substructures). Inspired by these works, we will study the contribution of the \textit{Gaia}-Sausage both in the Sgr-removed and the smooth stellar halo of K giants and BHB stars. BHB stars are located at the lower end of the metallicity distribution of Galactic halo stars, which could lead to an underestimation of the contribution of the \textit{Gaia}-Sausage. Compared to BHB stars, K giants better represent of the overall metallicity distribution of Galactic halo stars. Therefore, the inclusion of K giants in this study is expected to bring a less biased result.

In Section~\ref{sec:data_model} we introduce the Sgr-removed and the smooth stellar halo data used in this study. Then, we describe two halo models in Section~\ref{sec:model}. Our analysis of the contribution of the \textit{Gaia}-Sausage to the Sgr-removed and the smooth halo is presented in Section~\ref{sub:diffuse} and ~\ref{sub:smooth}.  In Section~\ref{sub:substructures}, we select several possible Sausage-related substructures by their dynamical state. The shape of velocity ellipsoid and the metallicity distribution of the selected substructure stars are compared to the GMM/Sausage component. The influence of the selection effect of spectroscopic surveys is discussed in Section~\ref{Sec:selection}. Finally, conclusions are made in Section~\ref{dis_con}.

\section{Data} \label{sec:data_model}
\subsection{Halo Samples} \label{sec:data}

Three stellar samples consisting of K giants and BHB stars are included in this study. K giants are obtained from SEGUE \citep{2009AJ....137.4377Y} and LAMOST DR5 \citep{2006ChJAA...6..265Z,2012RAA....12.1197C,2012RAA....12.1243L,2012RAA....12..723Z}. Distances of these K giants are estimated by a Bayesian method presented in \citet{2014ApJ...784..170X}. The SEGUE K giants are obtained from the catalog of \citet{2014ApJ...784..170X} which contains stars from SEGUE-\uppercase\expandafter{\romannumeral1} and -\uppercase\expandafter{\romannumeral2}. The LAMOST K giants are selected using the method of \citet{2014ApJ...790..110L}. The specific selection criteria are 4000 $\leq$ $T_{\textup{eff}}/\textup{K}$ $\leq$ 4600 with log $g$ $\leq$ 3.5 dex and 4600 $\leq$ $T_{\textup{eff}}/\textup{K}$ $\leq$ 5600 with log $g$ $\leq$ 4 dex. As shown in \citet{2014ApJ...790..110L}, the contamination by stars other than K giants is smaller than 2.5\%. The third sample consists of 4985 BHB stars with distances estimated \citep{2008ApJ...684.1143X,2011ApJ...738...79X}. The photometric and spectroscopic information used for choosing BHB stars are publicly released in SDSS DR8 \citep{2011ApJS..193...29A}. Line-of-sight velocity and stellar metallicity are provided by the published spectroscopic surveys, and proper motions are taken from \textit{Gaia} EDR3 \citep{2021A&A...649A...1G} by cross-matching with a radius of $1^{\prime \prime}$.

The three stellar samples provide 7D information for the halo stars, including equatorial coordinate information ($\alpha$, $\delta$), heliocentric distance d, line-of-sight velocities $v_\text{los}$, proper motions ($\mu_{\alpha}$, $\mu_{\delta}$), and stellar metallicity [Fe/H]. For measurement uncertainties, the SEGUE K giants have a mean distance precision ($\delta d/d$) of 16\%, mean $v_\text{los}$ uncertainty of 2 km $\textup{s}^{-1}$, and mean metallicity error of 0.12 dex. The LAMOST K giants have a mean distance precision of 13\%, mean $v_\text{los}$ uncertainty of 7 km $\textup{s}^{-1}$, and mean metallicity error of 0.14 dex. The SDSS BHB stars have a mean distance precision of 5\%, a mean $v_\text{los}$ uncertainty of 4 km $\textup{s}^{-1}$, and mean metallicity error of 0.21 dex. The proper motion uncertainties of 95\% of these halo stars range from 0.01 to 0.10 mas $\textup{yr}^{-1}$. 
	
The 7D information is transformed to the spherical Galactocentric coordinate system using the conventions in \texttt{astropy} \citep{2018AJ....156..123A}. In Galactocentric Cartesian coordinates, we use the values of the Solar Galactocentric distance $r_{\mathrm{gc},\odot}$ = 8.122 kpc \citep{2018A&A...615L..15G}, and height $Z_{\odot}$ = 20.8 pc \citep{2019MNRAS.482.1417B}. We adopt a Solar motion of (+12.9, +245.6, +7.78) km $\textup{s}^{-1}$ \citep{2004ApJ...616..872R,2018A&A...615L..15G,2018RNAAS...2..210D}. To transform to the Galactocentric spherical coordinate system (\textit{r, $\theta, \phi$}) and ($v_{r}, v_{\theta}, v_{\phi}$), we follow Equation.~\ref{eq:1} $-$ \ref{eq:spherical},
\begin{align}
& r = \sqrt{X^2 + Y^2 + Z^2} \label{eq:1} \\
& \theta = \pi/2 - \mathrm{atan}(Z/\sqrt{X^2 + Y^2}) \\
& \phi = \mathrm{atan2}(Y, X) \\
& v_r = (U \mathrm{cos}\,\phi + V \mathrm{sin}\,\phi)\mathrm{sin}\,\theta + W \mathrm{cos}\,\theta \\
& v_\theta = -(U \mathrm{cos}\,\phi + V \mathrm{sin}\,\phi)\mathrm{cos}\,\theta + W \mathrm{sin}\,\theta \\
& v_\phi = -U \mathrm{sin}\,\phi + V \mathrm{cos}\,\phi, 
\label{eq:spherical}
\end{align}
where (\textit{X, Y, Z}) and (\textit{U, V, W}) are within the Galactocentric Cartesian coordinate system. The uncertainties in distance, line-of-sight radial velocity, and proper motion (including the measurement error and covariance of proper motions) are propagated using Monte Carlo sampling in order to estimate the median, standard error, and covariance of the velocities ($v_{r}, v_{\theta}, v_{\phi}$) in Galactocentric spherical polar coordinates. 

We first select stars with a relative distance precision better than 30\% ($\delta{d}/d < 0.3$). To exclude possible disc stars, we retain stars with $|Z|$ $>$ 5 kpc and [Fe/H] $< -0.5$ dex. To eliminate egregious outliers, we require the absolute spherical velocities to be $<$ 400 km $\text{s}^{-1}$ and the uncertainties $<$ 150 km $\text{s}^{-1}$. \cite{2019MNRAS.486..378L} showed that the existence of stars with high velocities in their BHB star sample are possibly caused by the contamination of blue straggler stars. Therefore, we remove all stars with SDSS colors satisfying $u - g$ $<$ 1.15 and $g - r$ $> -0.07$ as well as stars satisfying $u - g <$ 1.15 and c($\gamma$) $<$ 0.925 for our SDSS BHB sample. Here c($\gamma$) is a description of the shape of H($\gamma$) line. 

\subsection{Sgr-Removed and Smooth Halo Samples} \label{subsec:sgrre}

An integral of motion (IoM) in a given force field is any function of the phase space coordinates that remains constant along an orbit. Stars with similar integrals of motion also have similar orbits. Stars clumping together in IoM phase space coordinates are likely to share a common origin. Therefore, integrals of motion can be used to search for substructures within the stellar halo.

The substructures are identified in IoM space following the method of Xue et al. (2021, in preparation). The substructure is composed of stars on similar orbits characterized by their IoM. Xue et al. (2021, in preparation) assumed that the Galactic halo can be described by a spherical potential. In such case the orbit of a star can be characterized by four integrals of motion, i.e., energy \textit{E} and the angular momentum vector $\vec{\textit{L}}$. The exact coordinates adopted by Xue et al. (2021, in preparation) are eccentricity ec, semi-major axis $a$, angular direction of the orbital pole relative to the defined Galactic coordinate plane ($l_\text{orb}$, $b_\text{orb}$), and angular direction of apocenter $l_\text{apo}$ (angle between apocenter and the projection of the $X$-axis onto the orbital plane), which are calculated through \textit{E} and $\vec{\textit{L}}$, and the position of the star. As a note, the value of $l_\text{apo}$ changes with each period but it remains constant within one certain period, thus $l_\text{apo}$ can be used to separate stars in the same stream but belonging to the different epochs (e.g., Sgr-leading and trailing arms). Considering the errors of the stellar observations, the orbital distribution of a star is defined as the normalized 5D histogram of {ec, $a$, $l_\text{orb}, b_\text{orb}, l_\text{apo}$}. Stars on similar orbits should have more overlap of their orbital distributions. By defining a statistic to describe the similarity of the orbital distributions of two stars and using the friends-of-friends group finding algorithm to identify groups of stars in IoM space, we define groups with more than six members as belonging to a common substructure.

We name the halo star samples with Sagittarius stream members removed as the Sgr-removed stellar halo. The Sagittarius stream \citep{1994Natur.370..194I}, the most prominent substructure in the Milky Way, is still unrelaxed, which may lead to some bias when studying the \textit{Gaia}-Sausage. To attain an unbiased measurement of the shape of the velocity ellipsoid and the metallicity distribution, we remove the Sgr stream members published by \citet{2019ApJ...886..154Y}, who analyzed the chemodynamical properties of about 3000 Sgr stream members selected from LAMOST DR5 K and M giants, SEGUE K giants, and SDSS BHB stars by Xue et al. (2021, in preparation). Applying certain selection criteria, 10419 LAMOST K giants, 4571 SEGUE K giants, and 3517 SDSS BHB stars remain. \citet{2019MNRAS.486..378L} removed the Sgr stream members by cutting the positions and distances of stars, and their Sgr-removed halo sample contains 3112 SDSS BHB stars. Results obtained from different SDSS BHB Sgr-removed halo samples might be slightly different.

We name the halo star samples with all obvious substructures (including the Sgr stream members) removed as the smooth stellar halo. After the removal of substructures, 6644 LAMOST K giants, 3375 SEGUE K giants, and 2281 SDSS BHB stars remain (Xue et al. 2021, in preparation). The spatial distributions ($\sqrt{X^2+Y^2}$-$Z$ plane) of the Sgr-removed and the smooth halo samples are shown in Figure~\ref{fig:spatial}. Histograms of the spherical velocities and metallicity of the six halo samples are shown in Figure~\ref{fig:sgr_smooth}. After the removal of halo substructures, the distribution of metallicity remains similar but the tangential velocity distribution becomes weaker in the central region ($v_\theta = 0, v_\phi = 0$). BHB stars have already passed the main-sequence and giant-branch stages of evolution. Therefore, halo BHB stars are generally older than halo K giants, which is exemplified by the lower metallicity of our SDSS BHB star samples compared to our K giant samples. The LAMOST K giant samples have a truncation of metallicity at [Fe/H] = $-2.5$ dex because of the pipeline used \citep{2011RAA....11..924W,2015RAA....15.1095L}. The SEGUE K giants are mainly selected through the $l$-color method targeting stars with $l >$ 0.07 for SEGUE-\uppercase\expandafter{\romannumeral1} and $l > 0.09$ for SEGUE-\uppercase\expandafter{\romannumeral2}\footnote[1]{$l=-0.436u+1.129g-0.119r-0.574i+0.1984$. The specific selection criteria can be found at \url{https://www.sdss3.org/dr9/algorithms/segue_target_selection.php/\#SEGUEts1.}}.} As \citet{2009AJ....137.4377Y} and \citet{2014ApJ...784..170X} pointed out, the $l$-color selected K giants tend to cluster at bluer $(u - g)$ color where larger values of $l$-color correspond to more metal poor stars. Since the \textit{l}-color method prefers metal poor K giants, our SEGUE K giant sample is more metal poor than the LAMOST sample.

 \begin{figure*}
 	\centering
 	\includegraphics[width = 0.8\textwidth]{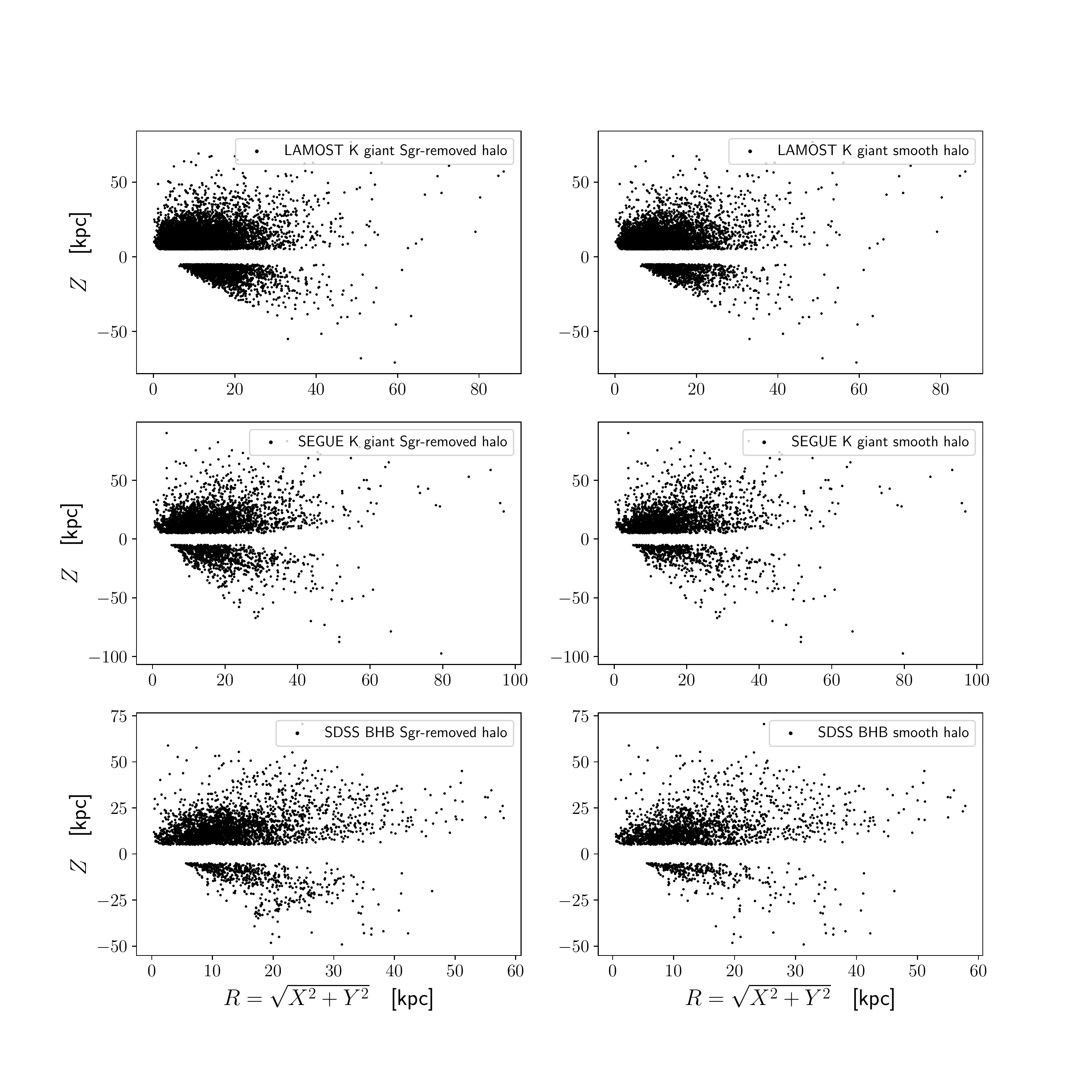}
 	\caption{Spatial distributions of the six halo samples.}
 	\label{fig:spatial}
 \end{figure*}

\begin{figure*}
	\centering
	\includegraphics[width = 1.0\textwidth]{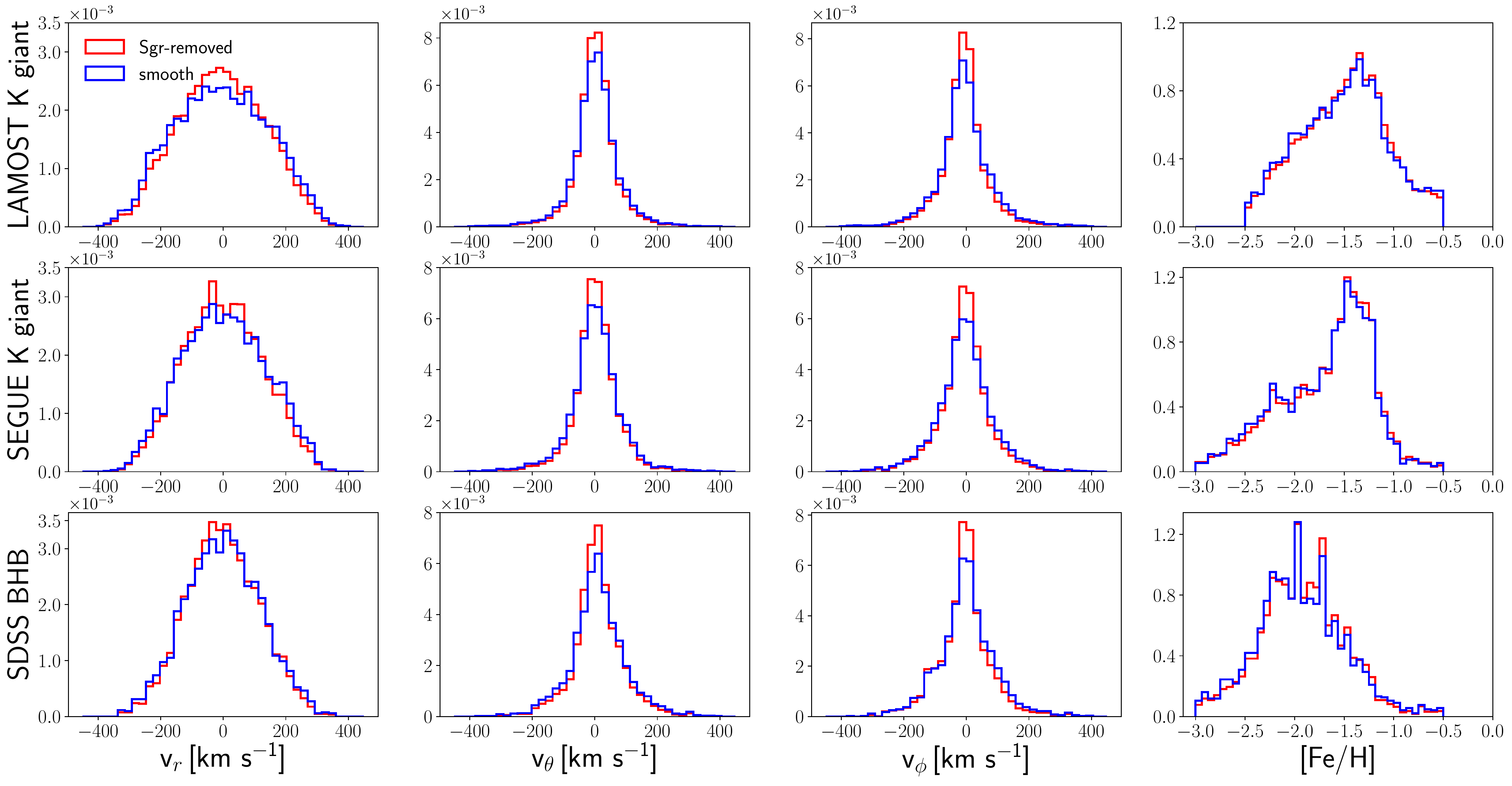}
	\caption{Normalized histograms of the spherical velocity components and metallicity of the Sgr-removed (red) and smooth halo (blue) samples for LAMOST K giants (top row), SEGUE K giants (middle row), and SDSS BHB stars (bottom row). Compared to the Sgr-removed halo samples, the smooth halo samples have a weaker distribution of tangential velocity in the central region ($v_\theta = 0, v_\phi = 0$).}
	\label{fig:sgr_smooth}
\end{figure*}

\section{Models} \label{sec:model}

To understand how the shape of velocity ellipsoid evolves with Galactocentric radius, two different models are used in this study. The first one is a single Gaussian model, and the second one is a Gaussian mixture model (GMM). The single Gaussian model treats the stellar halo as one component, and the Gaussian mixture model divides the stellar halo into two components of which one represents the \textit{Gaia}-Sausage. \cite{2019MNRAS.486..378L} and \cite{2019ApJ...874....3N} showed that the GMM provides a better fit to the shape of velocity ellipsoid when compared to the single Gaussian model. We next introduce our two models in detail and apply them to our stellar halo samples.

\subsection{Single Gaussian Model} \label{subsec:iso}

\begin{figure*}
	\centering
	\includegraphics[width = 1\textwidth]{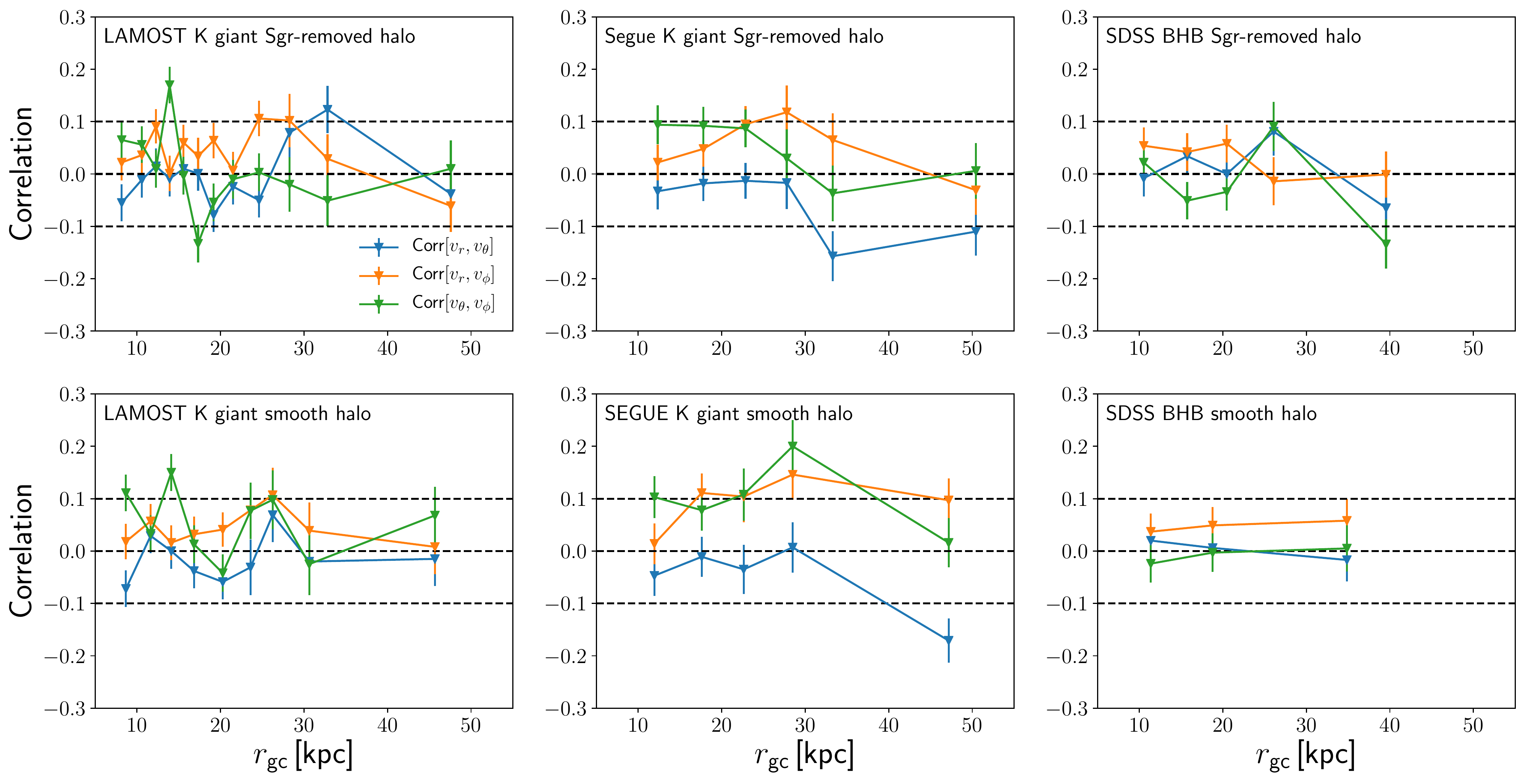}
	\caption{Correlation coefficient between the spherical velocity components in different distance bins of the Sgr-removed and smooth halo samples. The correlation coefficient is defined as Corr$[v_i, v_j] = \frac{\mathrm{Cov}(v_i, v_j)}{\sigma_{v_i}\sigma_{v_j}}$, where the covariance and velocity dispersion are obtained from the posterior distribution of the parameters of single Gaussian model.}
	\label{fig:correlation}
\end{figure*}

In the single Gaussian model (SGM), the shape of velocity ellipsoid is fit with a three-dimensional normal distribution $\mathcal{N}(\boldsymbol{v_i}|\boldsymbol{V}, \Sigma_i^\mathrm{sgm})$, and the metallicity distribution is fit with a one-dimensional normal distribution $\mathcal{N}(\text{[Fe/H]}_i|\mu_\text{[Fe/H]}^\mathrm{sgm}, {\sigma_{\text{[Fe/H]},i}^\mathrm{sgm}}^2)$. The likelihood of observing one star $D_i (v_{r,i}, v_{\theta,i}, v_{\phi,i}, \mathrm{[Fe/H]}_i)$ is defined as
\begin{equation}
\begin{aligned}
	\mathcal{L}_\mathrm{sgm}(D_i|\theta) = \mathcal{N}(\boldsymbol{v_i}|\boldsymbol{V}, \Sigma_i^\mathrm{sgm})\mathcal{N}(\text{[Fe/H]}_i|\mu_\text{[Fe/H]}^\mathrm{sgm}, {\sigma_{{\text{[Fe/H]},i}}^\mathrm{sgm}}^2),
	\end{aligned}
	\label{eq:single}
\end{equation}
  where $\boldsymbol{v_i}$ is the Galactocentric spherical velocity $(v_{r,i}, v_{\theta,i}, v_{\phi,i})$ of one star $i$, and $\boldsymbol{V}$ represents the mean velocities (${\langle v_r \rangle}$, ${\langle v_\theta \rangle}$, ${\langle v_\phi \rangle}$). The covariance matrix in velocity space $\Sigma_i^\mathrm{sgm}$ is a sum of the covariance matrix of the model being fit $\Sigma^\mathrm{sgm}$ and the covariance matrix of measurement error $\Sigma_i$. Six free parameters are included in $\Sigma^\mathrm{sgm}$, which are the intrinsic velocity dispersions $\sigma_{v_r}$, $\sigma_{v_\theta}$, $\sigma_{v_\phi}$, and the covariances of spherical velocity Cov($v_r, v_\theta$), Cov($v_\theta, v_\phi$), Cov($v_r, v_\phi$). In the studies of \citet{2009ApJ...698.1110S} and \citet{2016MNRAS.456.4506E}, the covariances of spherical velocity components are believed to be small. This view has been further supported from the study of the dynamical state of RR Lyrae stars, BHB stars, and K giants of the stellar halo after the data release of \textit{Gaia} DR2 \citep{2019MNRAS.485.3296W, 2019MNRAS.486..378L,2021ApJ...919...66B}. The comparison of the correlation coefficient between the components of spherical velocity with Galactocentric radius is shown in Figure~\ref{fig:correlation}. The magnitude of the correlation is smaller than 0.1 for most distances.

We define $\textup{[Fe/H]}_i$ as the stellar metallicity of one star $i$, and $\mu_\text{[Fe/H]}^\mathrm{sgm}$ as the mean metallicity. The full metallicity dispersion $\sigma_{\text{[Fe/H]},i}^\mathrm{sgm}$ for a single star is a combination (in quadrature) of the measurement error $\sigma_{\text{[Fe/H]},i}$ for the star and the intrinsic metallicity dispersion $\sigma_{\text{[Fe/H]}}^\mathrm{sgm}$ of stars in the corresponding distance bin($\sigma_{\text{[Fe/H]},i}^\mathrm{sgm} = \sqrt{{\sigma_{\text{[Fe/H]},i}}^2 + {\sigma_{\text{[Fe/H]}}^\mathrm{sgm}}^2}$). 
 
 In the single Gaussian model, the halo of the Milky Way is treated as one simple component, and there are a total of 11 free parameters. The likelihood of the single Gaussian model for a halo sample containing $N$ stars is defined as
 
\begin{equation}
	\mathcal{L}({D}|\theta) = \prod_{i=1}^{N}\mathcal{L}_\mathrm{sgm}(D_i|\theta).
\end{equation}

 \subsection{Gaussian Mixture Model}\label{sub:GMM}
 
 The Gaussian mixture model divides the stellar halo into two components: one is a more metal-poor, isotropic stellar halo referred to as the GMM/isotropic component, and the other represents a more metal-rich, highly radially anisotropic stellar halo referred to as the GMM/anisotropic or GMM/Sausage component. The likelihoods of the isotropic and the anisotropic stellar halos are shown in Equation.~\ref{eq:iso} and Equation.~\ref{eq:ani},
 
 \begin{equation}
 \mathcal{L}_\mathrm{iso}(D_i|\theta) = \mathcal{N}(\boldsymbol{v_i}|\mathbf{0}, \Sigma_{i}^\mathrm{iso})\mathcal{N}(\text{[Fe/H]}_{i}|\mu_\mathrm{[Fe/H]}^\mathrm{iso}, {\sigma_{\mathrm{[Fe/H]},i}^\mathrm{iso}}^2)
 \label{eq:iso}
 \end{equation}
 
 \begin{equation}
 \begin{aligned}
 	\mathcal{L}_\mathrm{an}(D_i|\theta) = \frac{1}{2}[\mathcal{N}(\boldsymbol{v_i}|\boldsymbol{V}^{\widetilde{\mathrm{an}}}, \Sigma_{i}^\mathrm{an}) + \mathcal{N}(\boldsymbol{v_i}|\boldsymbol{V}^\text{an}, \Sigma_{i}^\text{an})]\\
 	\times\mathcal{N}(\text{[Fe/H]}_{i}|\mu_\mathrm{[Fe/H]}^\mathrm{an}, {\sigma_{\text{[Fe/H]},i}^\mathrm{an}}^2).
 	\end{aligned}
 	\label{eq:ani}
 \end{equation} 
  
 The isotropic halo is similar to the single Gaussian model except for several points. In Equation.~\ref{eq:iso}, the mean spherical velocities are set to be zero. We force the velocity tensor in the tangential direction to be isotropic by setting $\sigma_{v_\theta;\mathrm{iso}} = \sigma_{v_\phi;\mathrm{iso}}$, and $\sigma_{t;\mathrm{iso}}$ is used as a unified representation of the velocity dispersion in the tangential direction. The velocity dispersion of the GMM/isotropic component $\Sigma_{i}^\mathrm{iso}$ is a combination of the covariance matrix of measurement error $\Sigma_i$ and the intrinsic dispersion diag$(\sigma_{v_r;\mathrm{iso}}^2, \sigma_{t;\mathrm{iso}}^2, \sigma_{t;\mathrm{iso}}^2)$. The mean metallicity $\mu_\mathrm{[Fe/H]}^\mathrm{iso}$ and the intrinsic metallicity dispersion $\sigma_{\mathrm{[Fe/H]}}^{\mathrm{iso}}$ are considered as two free parameters in the isotropic halo model. The full dispersion $\sigma_{\text{[Fe/H]},i}^\text{iso}$ is a combination (in quadrature) of the individual measurement error $\sigma_{\text{[Fe/H]},i}$ and the intrinsic metallicity dispersion $\sigma_{\text{[Fe/H]}}^\text{iso}$.
 
 The anisotropic halo model is motivated by previous studies on the chemodynamical properties of the \textit{Gaia}-Sausage. To track the behavior of the dynamical state of the stellar halo, \cite{2018MNRAS.478..611B} fitted the shape of velocity ellipsoid of SDSS-\textit{Gaia} main sequence stars with a zero-mean multi-variate Gaussian model. They found that the residuals show clear over-densities of stars with high positive and negative radial velocity, especially in the most metal-rich stellar halo. A massive, highly radial merger is a possible explanation of the two distinct high $v_r$ lobes. We assume that the GMM/Sausage component is built from two Gaussians with an equal mixing fraction. To match the two $v_r$ lobes, the mean radial velocities of the two Gaussians are set to be $+\langle v_{r}^\mathrm{an} \rangle$ and $-\langle v_{r}^\mathrm{an} \rangle$. The mean rotation velocity $\langle v_\phi^\mathrm{an} \rangle$ of the \textit{Gaia}-Sausage is not forced to be zero based on the analyses of \citet{2018MNRAS.478..611B}, \citet{2018Natur.563...85H}, and \citet{2019MNRAS.486..378L}. In the two Gaussians, the mean velocity $\boldsymbol{V}^{\text{an}}$ is set to be $(\langle v_{r}^\mathrm{an} \rangle, 0, \langle v_\phi^\mathrm{an} \rangle)$, and $\boldsymbol{V}^{\widetilde{\text{an}}}$ is $(-\langle v_r^\mathrm{an} \rangle, 0, \langle v_\phi^\mathrm{an} \rangle)$. The velocity dispersion of the GMM/Sausage component $\Sigma_{i}^\text{an}$ is a combination of the covariance matrix of measurement error  and the intrinsic dispersion diag$(\sigma_{v_r;\text{an}}^2, \sigma_{t;\text{an}}^2, \sigma_{t;\text{an}}^2)$. We define $\mu_\mathrm{[Fe/H]}^\mathrm{an}$ as the mean metallicity of the anisotropic stellar halo, and $\sigma_{\text{[Fe/H]}}^\text{an}$ as the intrinsic dispersion. The full metallicity dispersion $\sigma_{\text{[Fe/H]},i}^\text{an}$ is a combination (in quadrature) of the measurement error $\sigma_{\text{[Fe/H]},i}$ and the intrinsic metallicity dispersion $\sigma_{\text{[Fe/H]}}^\text{an}$.

 The GMM has a total of 11 free parameters. The likelihood of the GMM for a halo sample containing $N$ stars is defined as
 \begin{equation}
 \begin{aligned}
 	\mathcal{L}(D|\theta) = \prod_{i=1}^{N}(f_\mathrm{iso}\mathcal{L}_\mathrm{iso}(D_i|\theta) + f_\mathrm{an}\mathcal{L}_\mathrm{an}(D_i|\theta)),
 	 \end{aligned}
 	 \label{eq:GMM}
 \end{equation}
 where $f_\mathrm{iso}$ is the contribution of the isotropic component, and $f_\mathrm{an}$ is the contribution of the anisotropic component, such that $f_\mathrm{iso} + f_\mathrm{an} = 1$. 
 
 \section{Results}\label{sec:ratio}
 \subsection{Contribution of the \textit{Gaia}-Sausage to the Sgr-removed Halo}\label{sub:diffuse}
 
 Our main aim is to explore the proportion of stars originating from the major merger event as a function of Galactocentric radius. To achieve this, halo stars are distributed to different distance bins. For the LAMOST K giant Sgr-removed halo, 12 bins are defined, the edges of which are $r_\mathrm{gc}$ = 5.40, 9.60, 11.46, 13.12, 14.72, 16.46, 18.22, 20.24, 22.87, 26.79, 30.00, 36.95, and 102.93 kpc. Each of the first nine bins contain 1000 stars, and the last three bins have 455, 500, and 464 stars, respectively. For the SEGUE K giant Sgr-removed halo, the edges of the distance bins are $r_\mathrm{gc}$ = 6.09, 15.43, 20.18, 25.85, 30.00, 37.31, and 125.02 kpc. Each of the first three bins contain 1000 stars, and the last three bins have 480, 500 and 591 stars. For the SDSS BHB Sgr-removed halo, we define five bins with edges of $r_\mathrm{gc}$ = 5.45, 13.45, 17.91, 23.34, 30.00, and 60.00 kpc. Each of the first three bins contain 800 stars, and the last two bins have 505 and 612 stars. In the work of \citet{2018ApJ...862L...1D}, $r_\mathrm{gc}\approx$ 30 kpc is believed to be the outermost apocenter of the Sausage debris. \citet{2019MNRAS.486..378L} found that the fraction of the \textit{Gaia}-Sausage stars drops sharply beyond 30 kpc for the SDSS halo BHB stars. Motivated by these works, the stellar halo within $r_\mathrm{gc}$ of 30 kpc is defined as the inner halo, and further than $r_\mathrm{gc}$ of 30 kpc is defined as the outer halo.
 
 \texttt{Emcee}, which is an implementation of Goodman \& Weare’s Affine Invariant Markov chain Monte Carlo (MCMC) Ensemble sampler, is adopted in the fitting process \citep{2013PASP..125..306F}. We use 200 walkers and 1000 steps as the burn-in, then followed by 3000 steps to get the posterior distributions for each bin. We use the 50th percentile of the marginalized posterior distributions as the best estimated value and the 16th and 84th as the uncertainties.
 
\begin{figure*}
 	\centering
 	\includegraphics[width = 1\textwidth]{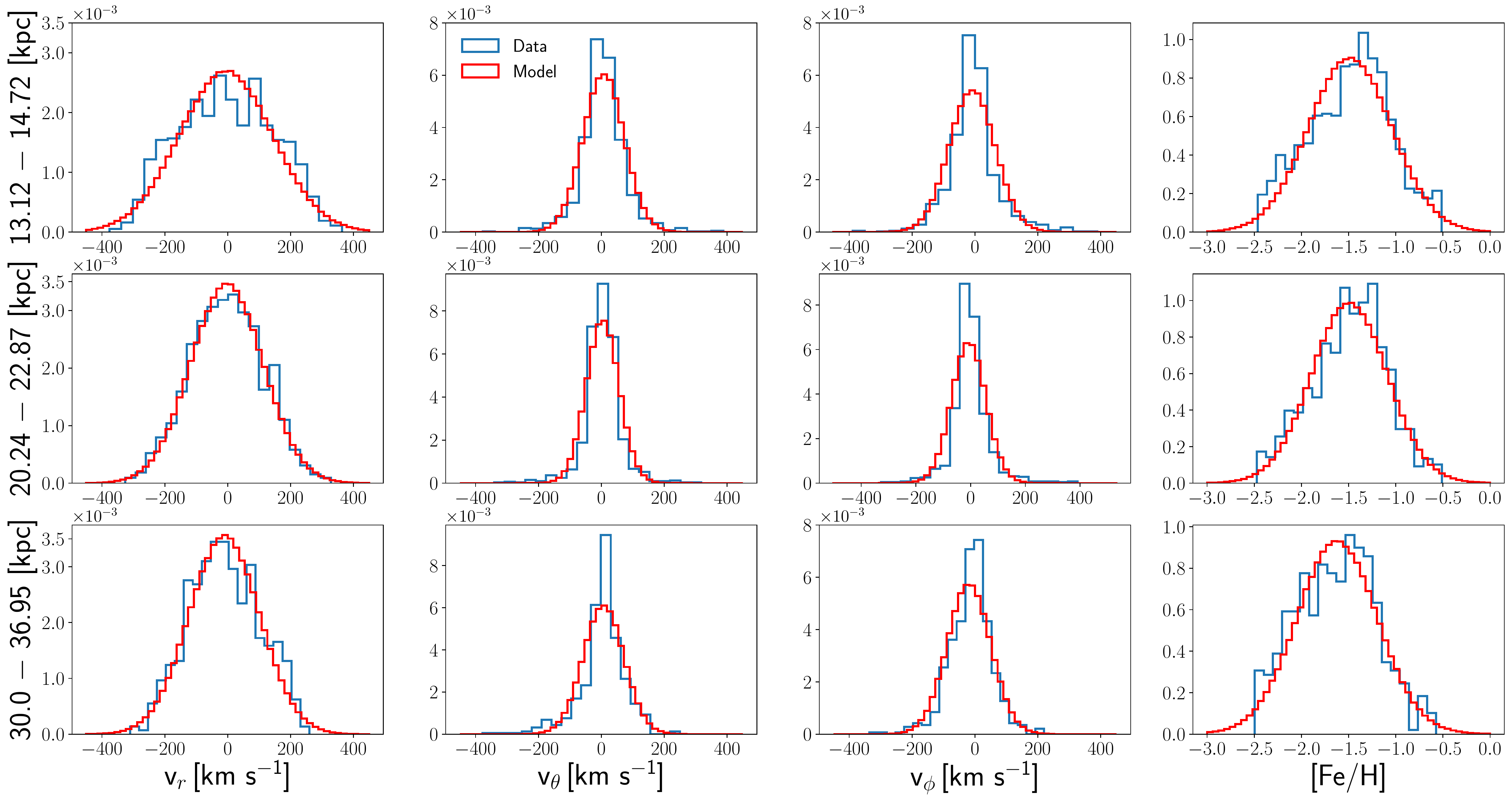}
 	\caption{SGM fitting result of the LAMOST Sgr-removed halo. The red normalized histograms represent the mock data sampled from the best estimate parameters of SGM, and blue histograms represent the observational data. The likelihoods ln $\mathcal{L}_\mathrm{MAP}$ of the two distance bins (top and middle rows) in the inner halo are $-18369$ and $-17665$, respectively. For the distance bin (bottom row) in the outer halo the ln $\mathcal{L}_\mathrm{MAP} = -9140$.}
 	\label{fig:single_se}
 	\includegraphics[width = 1\textwidth]{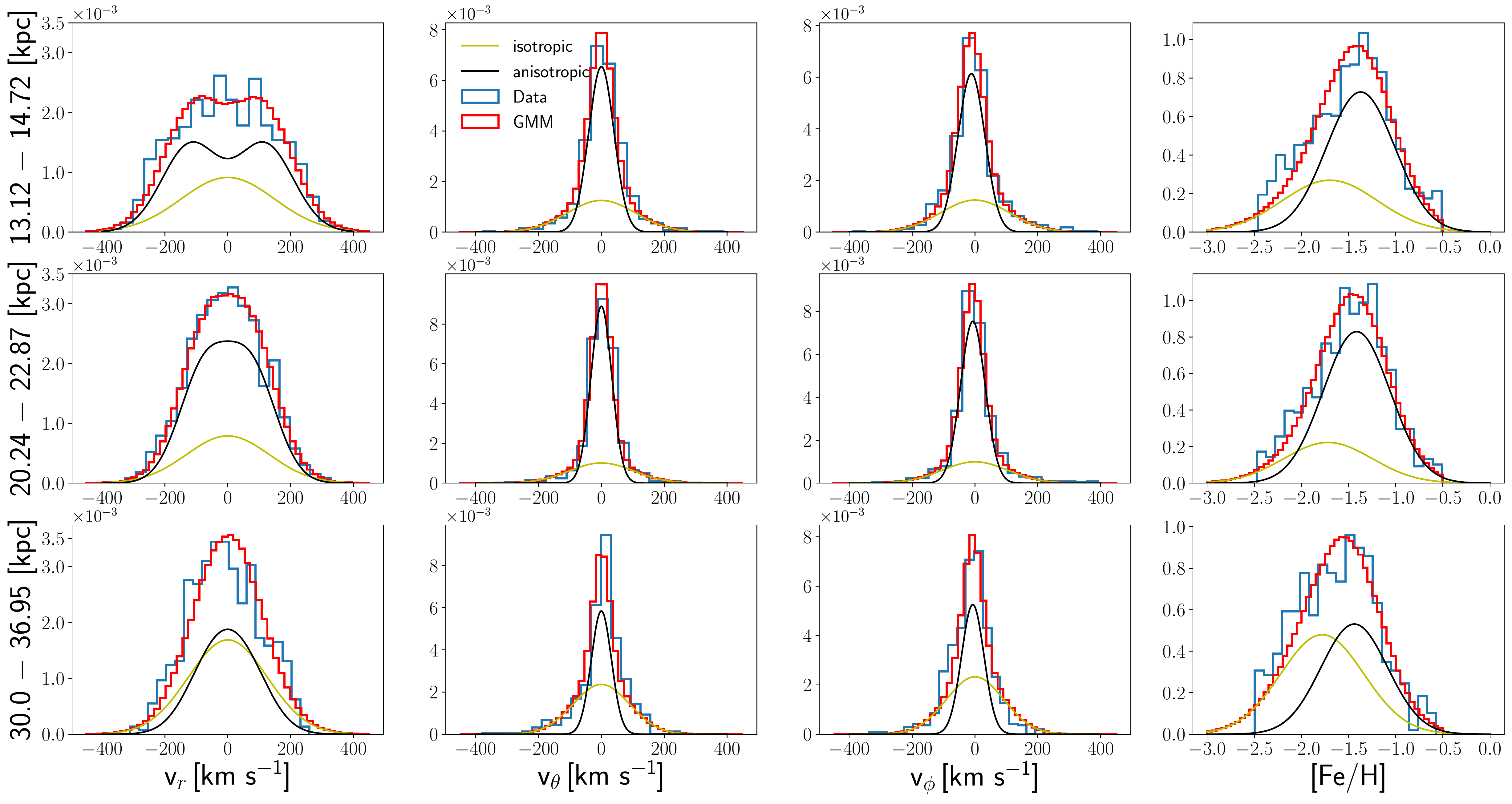}
 	\caption{GMM fitting result of the LAMOST Sgr-removed halo. The red normalized histograms represent the mock data sampled from the best estimate parameters of GMM, and blue histograms represent the observational data. The yellow and black lines represent the GMM/isotropic and GMM/anisotropic components of the mock data respectively. The likelihoods ln $\mathcal{L}_\mathrm{MAP}$ of the top, middle, and bottom rows are $-18111, -17332$, and $-9051$, respectively. The GMM has a lower BIC value than SGM, and the $\Delta$BIC values of the two models for the three distance bins are 258, 332, and 89, which indicates that the GMM is a better fit to the data than the single Gaussian model.}
 	\label{fig:gmm_la2}
 \end{figure*}

 We first apply the single Gaussian model to the LAMOST K giant Sgr-removed halo, and get the best estimated parameters. We use the model parameters to specify the velocity and metallicity distributions, and then draw random samples from the specified distributions to make mock data with two million fake stars by the random sampling function \texttt{Numpy.random.multivariate\_normal()} in the \texttt{Numpy} package \citep{harris2020array}. Each sampled point is convolved with a covariance matrix of measurement error selected randomly from the observational data of that distance bin. The comparison between the observational data and the mock data is shown in Figure~\ref{fig:single_se} for three respective radial bins of the inner and outer halo. It is clear that the single Gaussian model fails in fitting the distributions, especially for the distributions of the tangential velocities. We can see that the distributions of the tangential velocities are much stronger in the central region ($v_{\theta} = 0, v_{\phi} = 0$) for the observational data. The existence of some stars with high eccentricity probably results in the failure of the fit.
 
\begin{figure*}
 \centering
 \includegraphics[width = 1\textwidth]{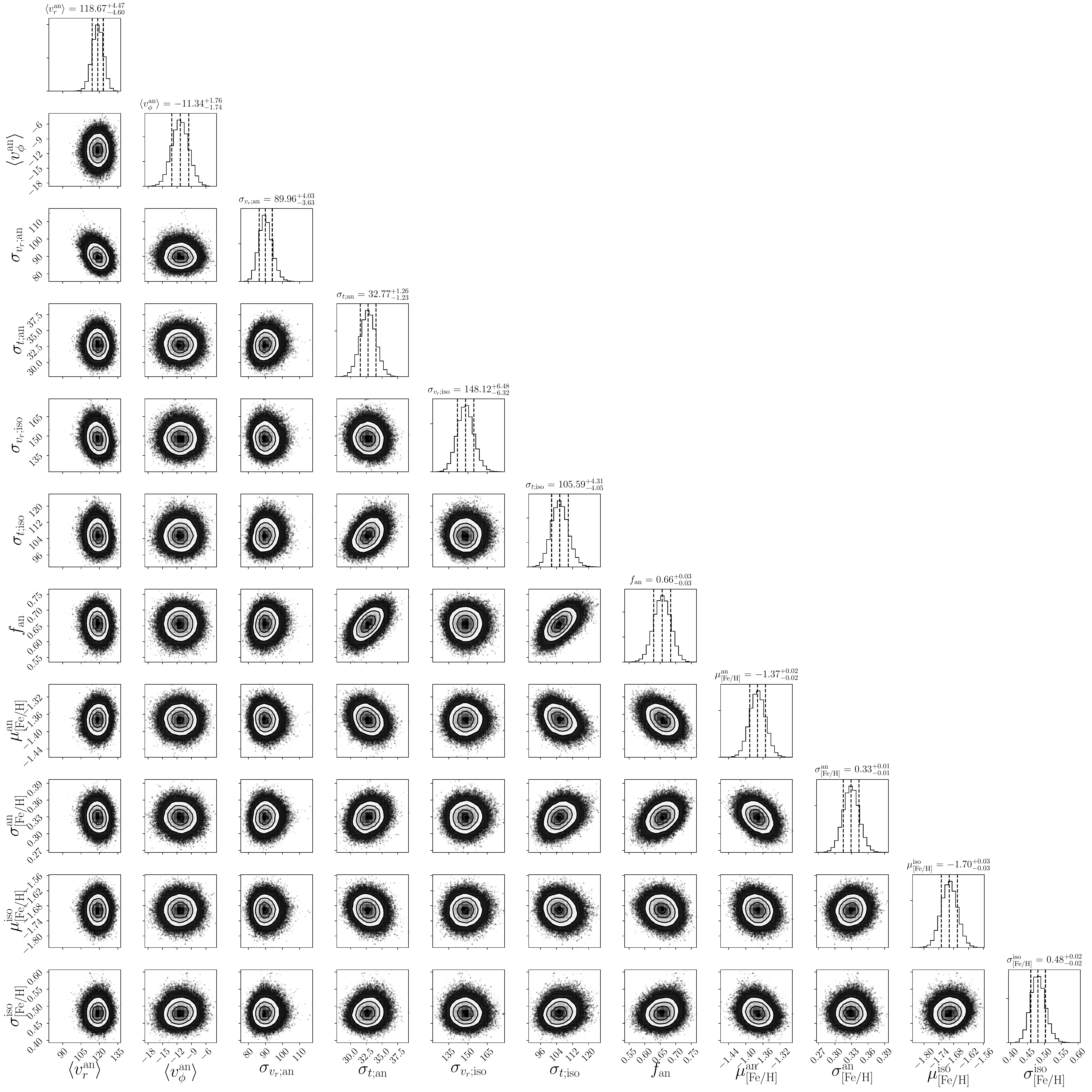}
\caption{Corner plot for the GMM parameters in the region $r_\mathrm{gc} \in [13.12, 14.72]$ kpc for the LAMOST K giant Sgr-removed stellar halo. The dashed lines show the 16th,
50th, 84th percentiles of the marginalized distribution of each parameter.} 
\label{fig:corner_plot}
\end{figure*}

\begin{figure*}
 	\centering
 	\includegraphics[width = 1.0\textwidth]{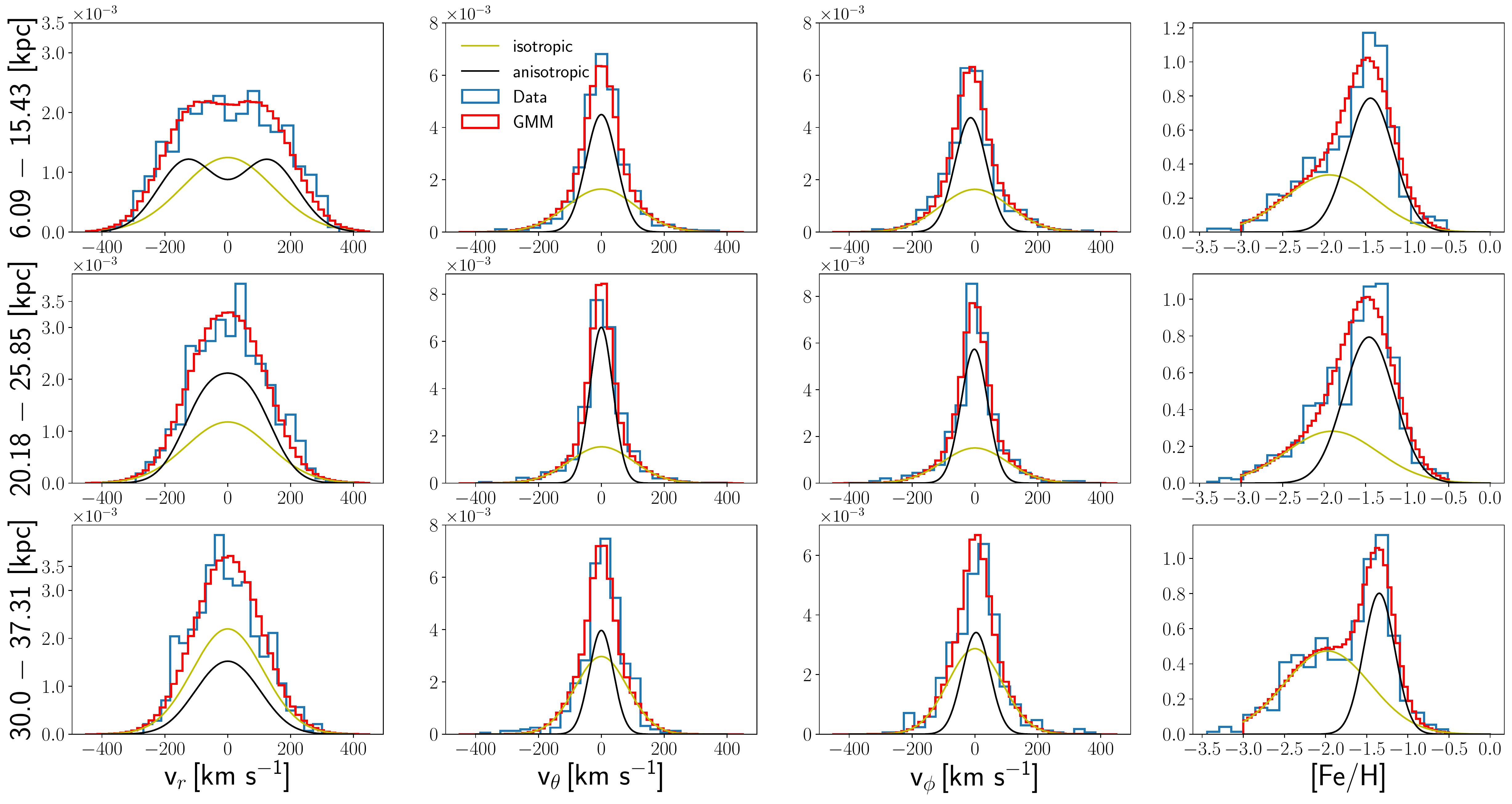}
 	\caption{GMM fitting result of the SEGUE K giant Sgr-removed stellar halo.}
 	\label{fig:gmm_se}
 	\includegraphics[width = 1.0\textwidth]{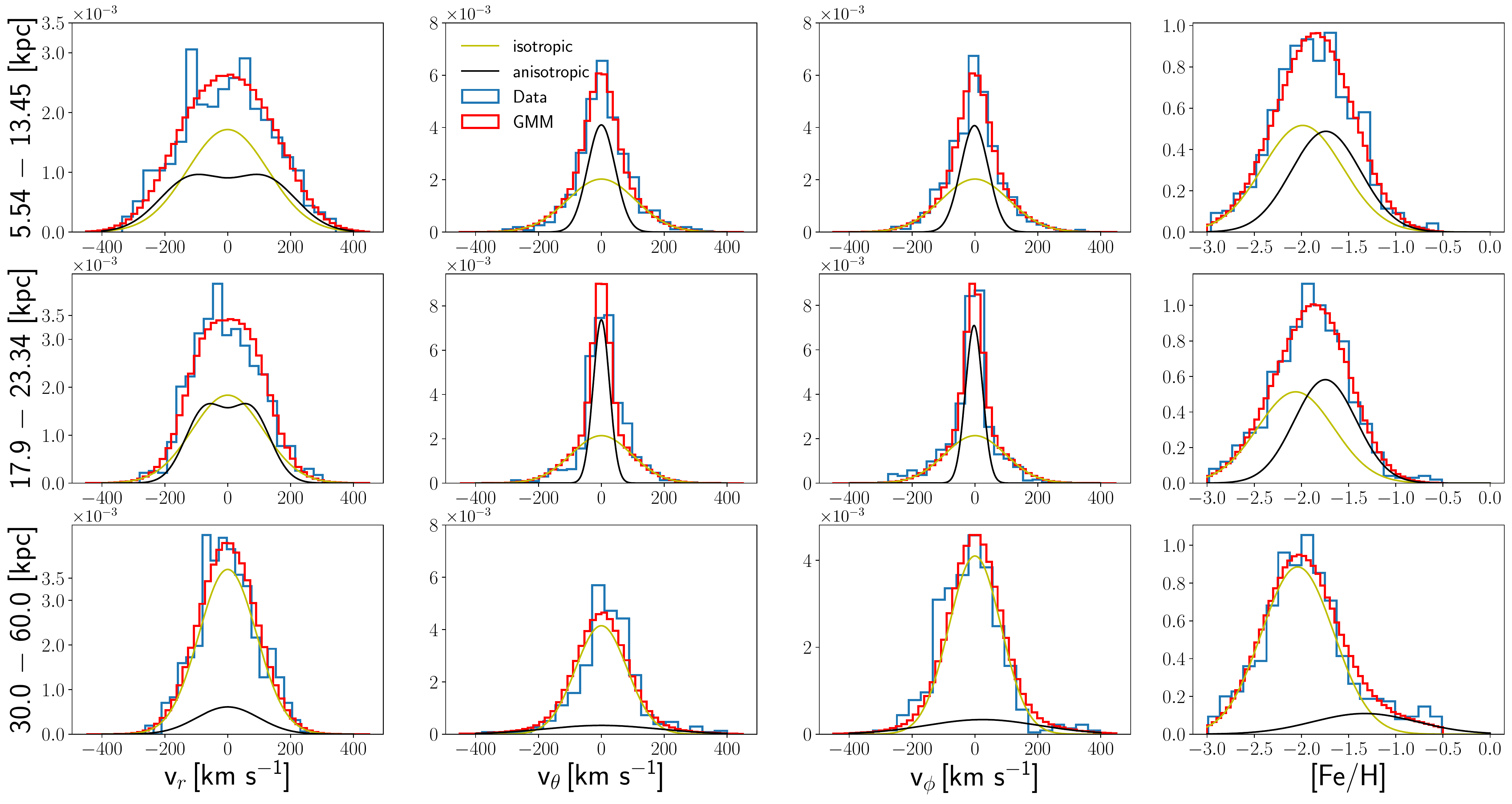}
 	\caption{GMM fitting result of the SDSS BHB Sgr-removed stellar halo.}
 	\label{fig:gmm_bhb}
 \end{figure*}

 
We then apply the GMM to the LAMOST K giant Sgr-removed halo sample, and get the best estimated GMM parameters. Mock data based on the best estimated GMM parameters are obtained using the same method as the SGM with two million fake stars. Since the GMM is composed of multiple Gaussian functions, we draw random samples from each Gaussian function, and then mix them according to their fractions. The comparison between the observational data and the mock data is shown in Figure~\ref{fig:gmm_la2}. The corner plot for the GMM parameters in the region $r_\mathrm{gc} \in [13.12, 14.72]$ kpc is shown in Figure~\ref{fig:corner_plot}. The significance of the two models is analyzed with the Bayesian Information Criterion (BIC)
\begin{equation}
	\text{BIC} = k\,\mathrm{ln} N - 2\,\mathrm{ln} \mathcal{L}_\mathrm{MAP},
\end{equation}
where $k$ is the number of free parameters, $N$ is the number of data points, and $\mathcal{L}_\mathrm{MAP}$ is the maximum a posteriori (MAP) of the likelihood for the model. A lower BIC value implies a better fit, and we consider the model with a lower \text{BIC} value to be significantly better than the model with a higher BIC value when $\Delta$BIC (BIC(higher) $-$ BIC(lower)) is larger than 10. Our two models have the same number of free parameters and data points but different MAP values of the likelihood. The GMM has a smaller BIC, and $\Delta$BIC values for the three distance bins in Figure~\ref{fig:gmm_la2} are 258, 332, and 89. We show the $\Delta$BIC of all distance bins of the Sgr-removed halo in Table~\ref{tab:Sgr_bic} and the smooth halo in Table~\ref{tab:smooth_bic}. According to $\Delta$BIC, the two-component Gaussian mixture model provides a significantly better fit to the chemodynamical properties of the halo K giants and BHB stars. The predicted distributions of the mock data are compared to the observational data for the other two Sgr-removed halo samples in Figure~\ref{fig:gmm_se} and ~\ref{fig:gmm_bhb}. The best estimated parameters of the GMM are show in Table~\ref{tab:Sgr-removed}. The mean metallicity $\mu_\text{[Fe/H]}^\mathrm{an}$ of the GMM/Sausage component of the two K giant samples ranges from $-1.50$ to $-1.30$, which is in good agreement with the previous estimates \citep{2018ApJ...862L...1D,2019ApJ...881L..10S,2019ApJ...874....3N,2019NatAs...3..932G,2020MNRAS.493.5195D} using stars such as main sequence stars and red giants. The SDSS BHB star sample is much more metal-poor than the K giant and main sequence star samples, and the mean metallicity $\mu_\text{[Fe/H]}^\mathrm{an}$ is $-1.75$. The GMM/Sausage component has a smaller tangential velocity dispersion and is more metal-rich when compared to the GMM/isotropic component and the observational stellar halo.

 The kinematic properties of the stellar halo can be well summarized by the anisotropy parameter $\beta$ defined as
 \begin{equation}
     \beta = 1 - \frac{\sigma_\theta^2 + \sigma_\phi^2}{2\sigma_r^2},
     \label{eq:beta}
 \end{equation}
 where $\beta$ quantifies the degree of velocity anisotropy of a system of  stellar orbits: radially biased ($\beta > 0$, perfectly radial $\beta = 1$), perfectly isotropic ($\beta = 0$), tangentially biased ($\beta < 0$, perfectly circular $\beta = -\infty$).
 
\begin{figure*}
\centering
	\includegraphics[width = 1.0\textwidth]{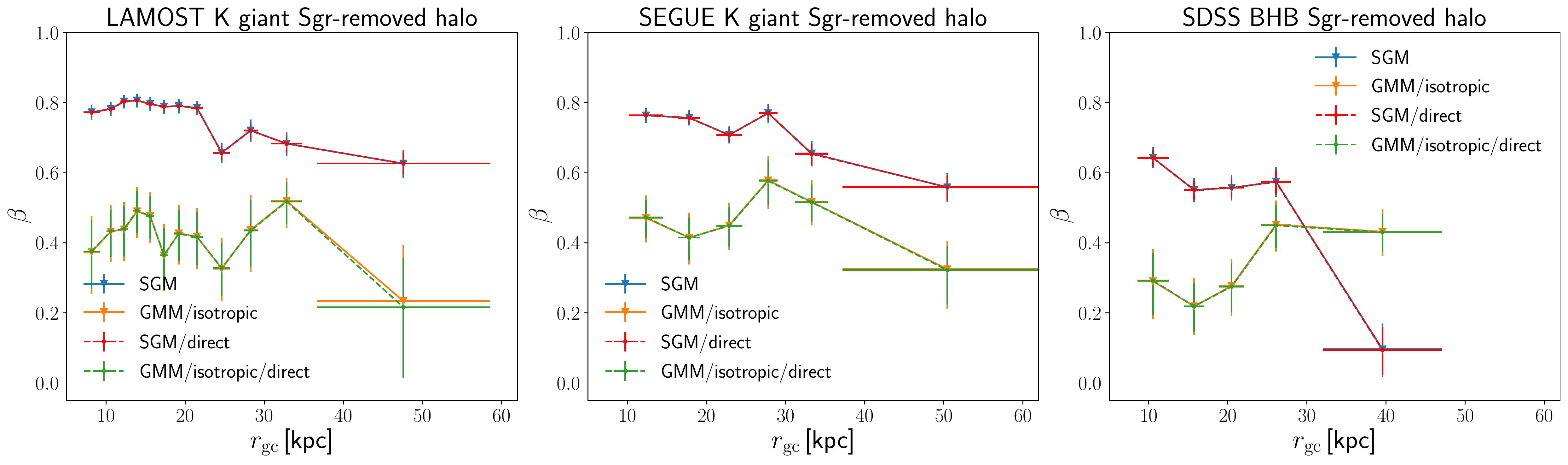}
	\caption{Anisotropy of the SGM and GMM/isotropic component obtained from the mock datasets  (solid lines) and directly from the MCMC chains (dashed lines). We find that these two methods lead to almost the same resulting anisotropy.}	
	\label{fig:direct_indirect}
\end{figure*}

\begin{figure}
  	\centering
  	\includegraphics[width = 0.5\textwidth]{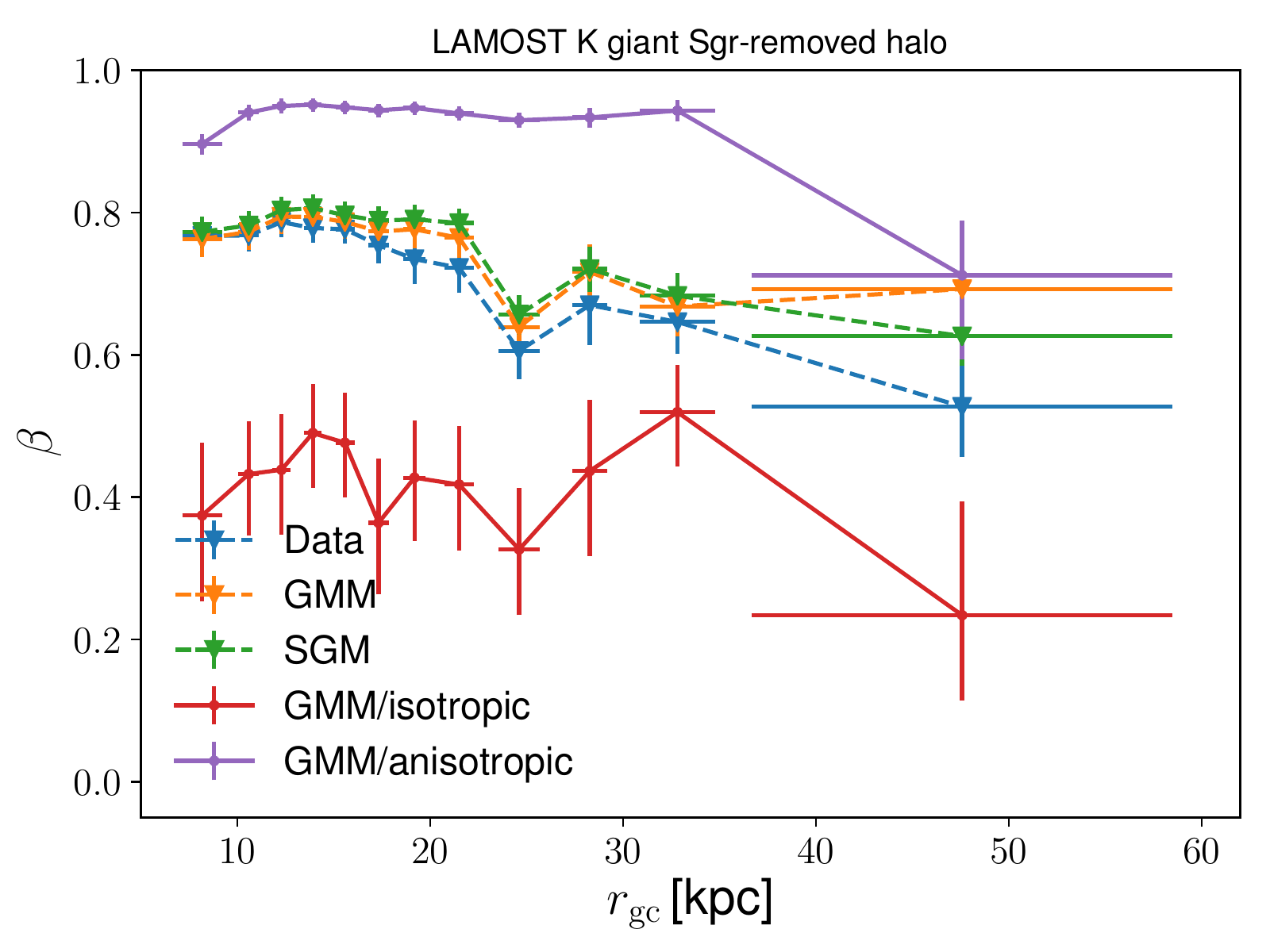}
  	  	\includegraphics[width = 0.5\textwidth]{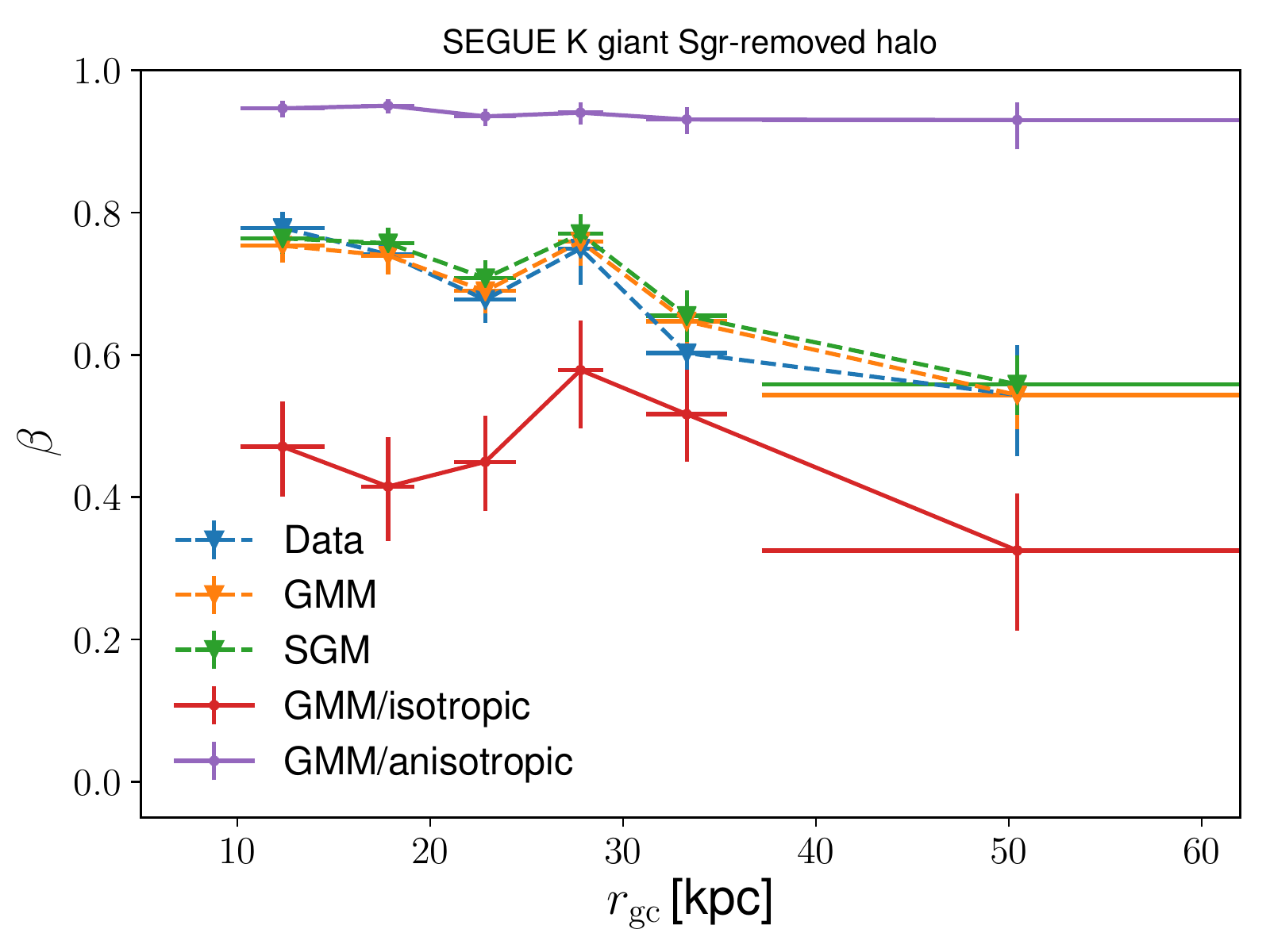}
  	  	  	\includegraphics[width = 0.5\textwidth]{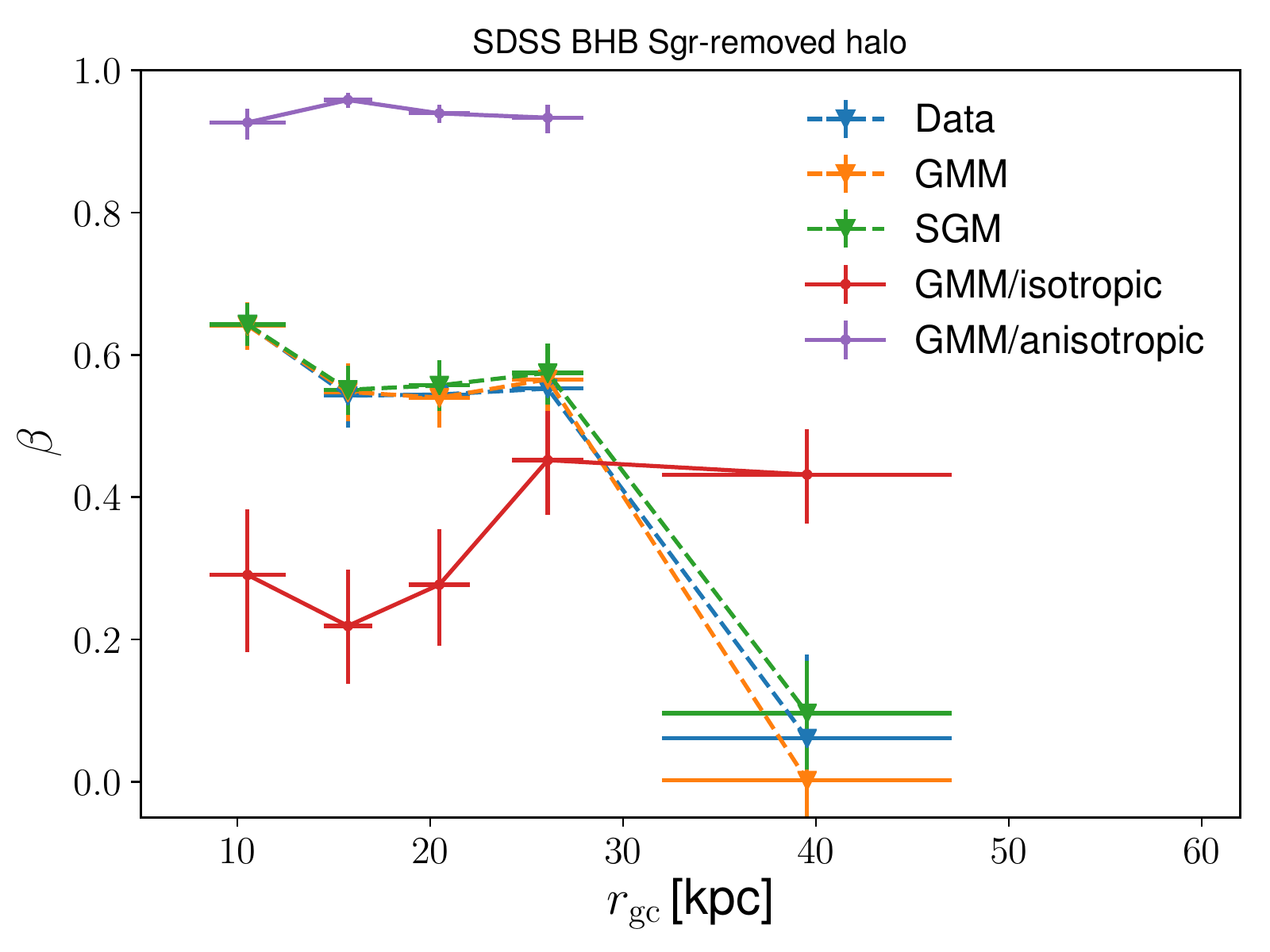}
  	\caption{Anisotropy parameter $\beta$ versus $r_\mathrm{gc}$ for the Sgr-removed halo samples and the mock data. Errors of ${\beta}$ of the observational data are obtained through a bootstrap method. Errors of the other four ${\beta}$ are obtained through the construction of mock datasets specified by model parameters drawn randomly from the posterior distributions. The lower and upper errors of anisotropy are obtained from the 16th, 50th, 84th percentiles of the marginalized distribution. The standard deviation of ${r_{gc}}$ is adopted as the error bar of ${r_{gc}}$ of that distance bin.}
  	\label{fig:beta_diffuse}
  \end{figure}

 The SGM and GMM/isotropic component are described by standard Gaussian functions, and the velocity dispersions used for calculating the anisotropy can be directly obtained from the MCMC chains. However, for the GMM and GMM/anisotropic component we can not directly obtain the velocity dispersions from the MCMC chains. The GMM/anisotropic component is composed of two lobes with positive and negative mean radial velocity as seen in Equation~\ref{eq:ani}. The radial dispersion $\sigma_{v_r;\mathrm{an}}$ in the MCMC chains only describes the radial velocity dispersion of one $v_r$ lobe, which is not suitable for the whole GMM/anisotropic component since the whole is a combination of the two lobes. Similarity for the anisotropy of the total GMM, the MCMC chains provide the velocity dispersions of the GMM/isotropic and GMM/anisotropic (only one lobe) components, but there are no direct velocity dispersions for the GMM as a combination of the two components. Therefore, we do not measure the anisotropy of the full GMM (all components combined) and GMM/anisotropic directly from the MCMC chains, but instead use an indirect approach and derive the anisotropy from the mock datasets. We generate 8000 mock datasets using \texttt{Numpy} random sampling function where the model parameters used for specifying the distributions are selected randomly from the MCMC chains. Each mock datasets has 100000 fake stars, and every sampled point is convolved with a covariance matrix of measurement error selected randomly from the observational data in the corresponding radial bins. We define the anisotropy of the mock datasets as the anisotropy of the models, and then obtain the median and errors of anisotropy of these mock datasets. Anisotropy of the observational and mock data is obtained with mean velocity measurement error subtracted in quadrature. Although we can calculate the anisotropy directly from the MCMC chains for the SGM and GMM/isotropic component, we still use the indirect method to remain consistent with the GMM and GMM/anisotropic component. Figure~\ref{fig:direct_indirect} shows that the difference of anisotropy between these two methods is almost negligible. Errors of the anisotropy of the observational data are obtained using a bootstrap method. We define $[{D}_i]_{i=1,N}$ as the catalog of stars in a distance bin where \textit{N} is the number of stars. An integer catalog $[{N_j}]_{j=1,N}$ is generated from a random sample of a uniform distribution U(1, N). A new synthetic catalog of stars is defined as $[{D_{N_j}}]_{j=1,N}$. We generate 200 synthetic catalogs in such manner. We propagate all sources of error as in Section~\ref{sec:data} to obtain the distributions of the velocity dispersion and anisotropy in each synthetic catalog, and then obtain errors from the distributions of anisotropy. Figure~\ref{fig:beta_diffuse} shows the anisotropy profiles of the observational and the mock datasets. In general, the stellar halo inferred from the GMM and SGM agrees well with the observational halo data. In the GMM, the velocity tensor of the two components in the tangential direction is set to be equal, while in the observational data slight differences between $\sigma_\theta$ and $\sigma_\phi$ exist. This could causes slight differences in anisotropy between the mock and observation halo data. The anisotropy of the GMM/isotropic and the GMM/Sausage components for all distance bins are recorded in Table~\ref{tab:Sgr_bic}. Anisotropy of the GMM/Sausage component is usually larger than 0.9, which agrees well with the previous studies that the \textit{Gaia}-Sausage debris is a highly radially biased system. On the other hand, anisotropy of the GMM/isotropic component is smaller than 0.6. We also notice that $\beta$ of the GMM/Sausage component is nearly always constant, while $\beta$ of the GMM/isotropic component changes more sporadically with $r_\mathrm{gc}$. We interpret the GMM/anisotropic component as originating from one single major merger event, and the GMM/isotropic component as originating from in-situ processes, gas accretion, multiple minor mergers, etc. The complex formation processes of the GMM/isotropic component may cause the large variation of the $\beta$ profile, which could further introduce bumps and wiggles in the total profile of the Galactic stellar halo. 

\begin{figure}
     \centering
     \includegraphics[width=0.5\textwidth]{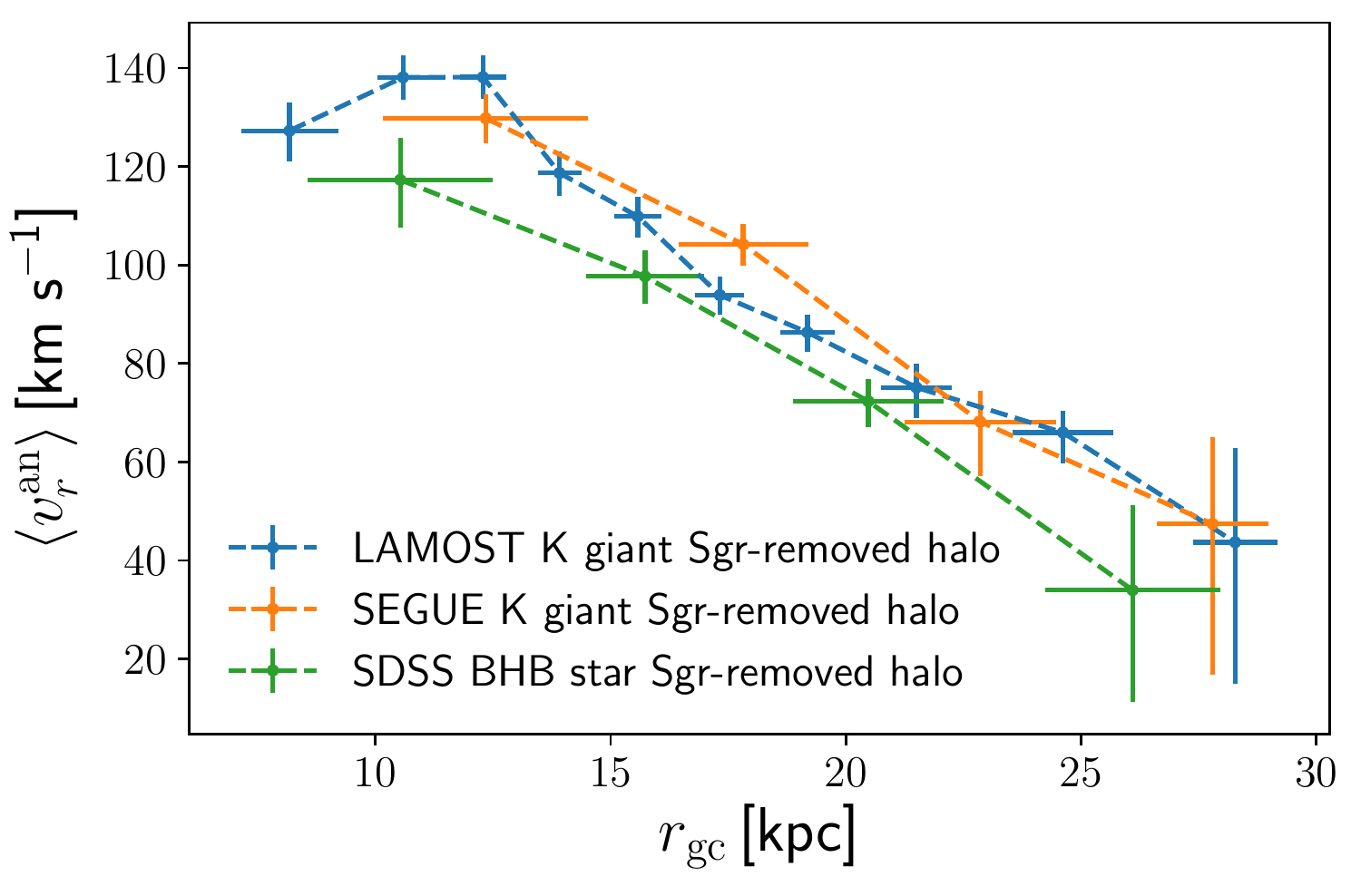}
     \caption{Absolute mean radial velocity $\langle v_{r}^\mathrm{an} \rangle$ of the two lobes as a function of $r_\mathrm{gc}$ in the inner halo.}
     \label{fig:caustics_all}
 \end{figure}

 \citet{2013MNRAS.435..378S} analyzed the tidal features known as ``shells'' caused by high-mass-ratio mergers with satellite galaxies on nearly radial orbits. They found that the absolute radial velocity $v_r$ of the shell structures declines with increasing $r_\mathrm{gc}$. Figure~\ref{fig:caustics_all} shows the absolute mean radial velocity $\langle v_r^\mathrm{an}\rangle$ as a function of $r_\mathrm{gc}$ in the inner halo. We find that the absolute $\langle v_r^\mathrm{an}\rangle$ of the two $v_r$ lobes also decreases with increasing $r_\mathrm{gc}$. The two lobes can be physically interpreted as the infalling and outgoing parts of the highly radial merger event \citep{2019MNRAS.486..378L}. According to $\beta$ of the GMM/Sausage component, most of the stars of the debris of the \textit{Gaia}-Sausage are on highly radial orbits. These highly eccentric stars move fast with large positive or negative $v_r$ near the Galactic Center, which is revealed by the large absolute $\langle v_{r}^\mathrm{an}\rangle$ and small $r_\mathrm{gc}$ in Figure~\ref{fig:caustics_all}.  
 
\begin{figure}
 	\centering
 	\includegraphics[width = 0.5\textwidth]{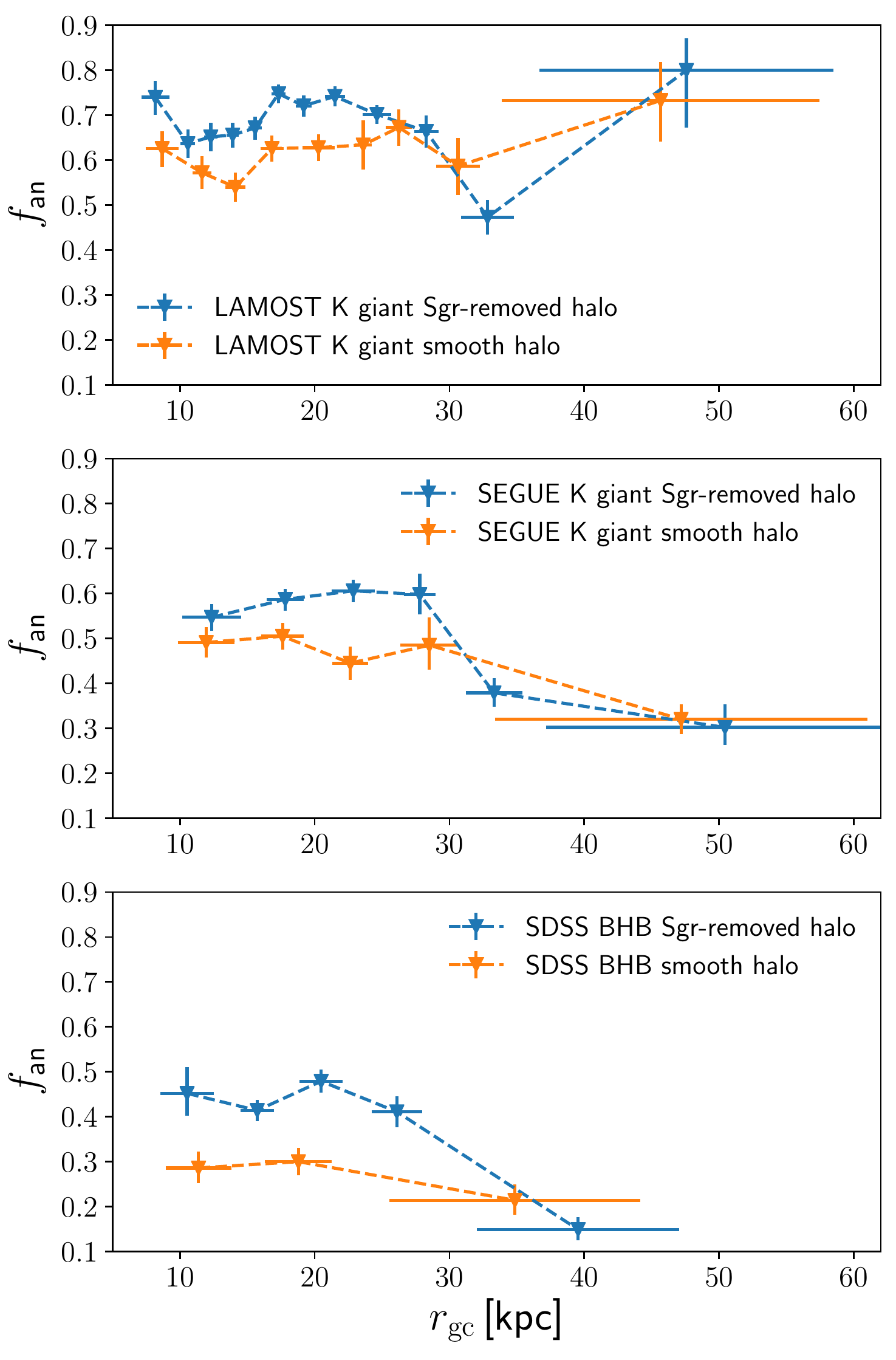}
 	\caption{Contribution of the GMM/Sausage component in Galactocentric radial bins for the Sgr-removed and the smooth stellar halo data of the LAMOST K giants (top), SEGUE K giants (middle), and SDSS BHB stars (bottom). The lower error and upper error of $f_\mathrm{an}$ are obtained from the 16th,
 		50th, 84th percentiles of the marginalized distribution.} 
 	\label{fig:gmm_ratio}
 \end{figure}

 Figure~\ref{fig:gmm_ratio} displays the proportion of the GMM/Sausage component as a function of Galactocentric radius. For the LAMOST K giant sample, the GMM/Sausage component occupies about $64\%-74\%$ of the inner stellar halo. This fraction slowly increases and remains as high as 0.70 at $r_\mathrm{gc}$ = 24.55 kpc. For the SEGUE K giant sample, the fraction ranges from 55\% to 61\% in the inner halo, reaching its peak at $r_\mathrm{gc}$ = 22.78 kpc. For the SDSS BHB sample, the fraction is about $41\%-48\%$ in the inner halo. We notice the difference in $f_\mathrm{an}$ among the three Sgr-removed halo samples. As we mentioned in Section~\ref{sec:data}, the difference of  metallicity distribution of the three halo star catalogs is mainly caused by the selection criteria. Considering that the \textit{Gaia}-Sausage is a relatively more metal-rich component, a more metal-rich halo star sample is very likely to be more heavily influenced by the \textit{Gaia}-Sausage. As the most metal-rich halo star sample, the LAMOST K giant sample has the largest value of $f_\mathrm{an}$. Our result is consistent with $N$-body simulation results of \citet{2021ApJ...923...92N} that the \textit{Gaia}-Sausage merger could deliver about 50\% of the Milky Way's stellar halo. From the GMM fits, we find that stars of the \textit{Gaia}-Sausage is a dominate part of the inner stellar halo. After reaching its peak value, the contribution of the GMM/Sausage component starts to drop sharply beyond $r_\mathrm{gc}$ of $25-30$ kpc, which is also the outermost apocenter of the \textit{Gaia}-Sausage debris as revealed by \citet{2018ApJ...862L...1D}.

 The fraction of the GMM/Sausage component in the outer halo proves to be more complex when compared to the inner halo. The contributions of the GMM/Sausage component to the outer halo are only 30\% for SEGUE K giants and 15\% for SDSS BHB stars. However, for the LAMOST K giant Sgr-removed halo, the contribution of the \textit{Gaia}-Sausage is still significant and suddenly increases from 47\% to 80\%. The sudden increase and high fraction are quite different to the results of the other two halo samples. Anisotropy of the GMM/Sausage component of the outermost distance bin ${r_\mathrm{gc} \in [36.95, 102.93]}$ kpc of the LAMOST K giant Sgr-removed halo is 0.71, which is much smaller than the typical value of anisotropy (${\beta > 0.9}$) of the other distance bins. It is hard to determine whether ${f_\mathbf{an}}$ really corresponds to the the fraction of the stars of the \textit{Gaia}-Sausage or not in that distance bin.

\begin{figure*}
	\centering
	\includegraphics[width=1\textwidth]{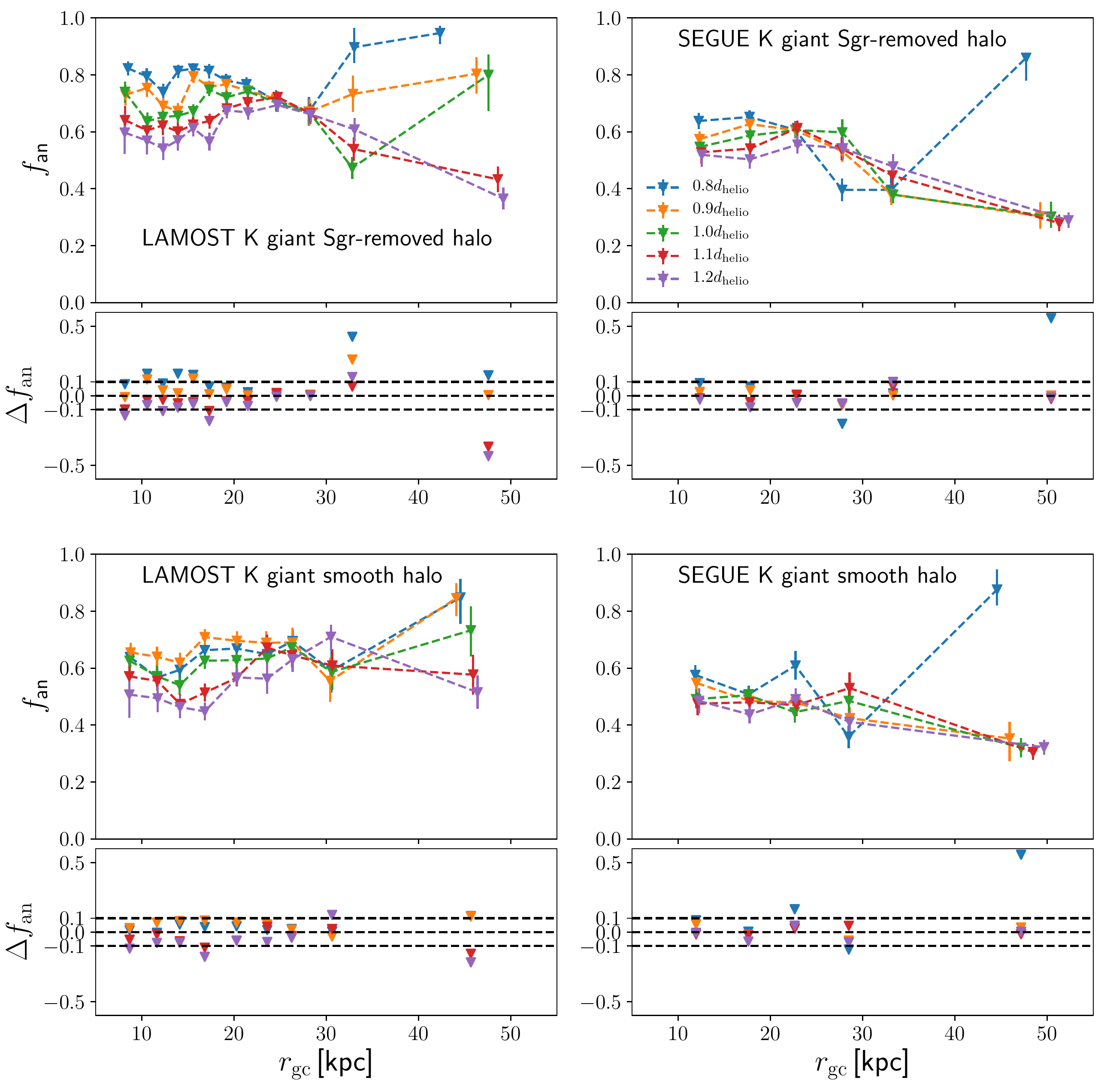}
	\caption{Change of ${f_\mathrm{an}}$ after testing systematic distance corrections made to the distance for the Sgr-removed and smooth halo of LAMOST and SEGUE K giants. Considering the possible bias of K giants distance revealed in \citet{2019ApJ...886..154Y} and the mean distance precision (13\% for LAMOST K giants and 16\% for SEGUE K giants), we systematically change the heliocentric distance within 0.2${d_\mathrm{helio}}$ assuming that our distance estimation is underestimated or overestimated. We find that the influence of the distance correction on ${f_\mathrm{an}}$ is generally within 0.1 in the inner halo, but more obvious in the outer halo. The systematic bias of K giant distance estimation decreases with the increasing magnitudes \citep{2019ApJ...886..154Y}. Since most of our K giants are at the fainter magnitudes and ${f_\mathrm{an}}$ changes within 0.1 at most of radii, we will continue use the results of the \citet{2014ApJ...784..170X} distance estimation in this study.}
	\label{fig:correct}
\end{figure*}

To figure out the reliability of ${f_\mathrm{an}}$ of the outermost distance bin, we apply a test related to the distance accuracy. \citet{2019ApJ...886..154Y} calibrated the heliocentric distance (${d_\mathrm{helio}}$) of K giants and BHB stars with the inverted parallax using halo stars from their identified streams. They found that the \citet{2014ApJ...784..170X} distances of K giants may be underestimated by 15\% on average, but BHB stars show no bias in the distance estimation. In Figure~\ref{fig:correct}, we systematically change the distances of our K giants (${0.8d_\mathrm{helio}, 0.9d_\mathrm{delio}, 1.1d_\mathrm{helio}, 1.2d_\mathrm{helio}}$) to check the influence of the possible systematic bias of distance estimation on ${f_\mathrm{an}}$. The spherical velocities and covariances are recalculated based on the bias-corrected distance. Please note that the bias of the estimated distance of K giants decreases with the increasing magnitudes \citep{2019ApJ...886..154Y}. Since most of our K giants are at the fainter magnitude, it is very likely that this bias is enlarged in the checking. We find that the change of ${f_\mathrm{an}}$ due to the distance correction is generally within 0.1 in the inner halo, while in the outer halo this change could be as high as 0.5. For the outermost distance bin of the LAMOST Sgr-removed halo, values of ${f_\mathrm{an}}$ are 0.95 for 0.8${d_\mathrm{helio}}$, 0.81 for 0.9${d_\mathrm{helio}}$, 0.80 for 1${d_\mathrm{helio}}$, 0.43 for 1.1${d_\mathrm{helio}}$, and 0.37 for 1.2${d_\mathrm{helio}}$. The variation of ${f_\mathrm{an}}$ in the outermost distance bin is very large even when we only correct the distance by ${0.1d_\mathrm{helio}}$. We think that a very accurate stellar distance estimation like RR Lyrae stars is needed to obtain a reliable ${f_\mathrm{an}}$ in the outer halo. Considering the sensitivity of ${f_\mathrm{an}}$ to the accuracy of distance estimation, we choose to ignore our ${f_\mathrm{an}}$ result in the outermost distance bin and abstain from using it in the further analysis since we are not sure about its reliability. For the SEGUE K giant Sgr-removed halo, values of ${f_\mathrm{an}}$ show little variation between the inner and outer halo except for the ${f_\mathrm{an}}$ obtained from 0.8${d_\mathrm{helio}}$. As the estimated distance of K giants tends to be underestimated and this bias is smaller for stars with fainter magnitudes, the distance correction of ${0.8d_\mathrm{helio}}$ is less reasonable compared the other three. We think that ${f_\mathrm{an}}$ of the SEGUE K giant Sgr-removed, outer halo is a more reliable result because it is much less influenced by the possible bias of distance estimation than the LAMOST K giant outer halo.
 
 \citet{2019MNRAS.486..378L} supported a small fraction ($< 0.10$) of stars kinematically associated to the \textit{Gaia}-Sausage beyond $r_\mathrm{gc} \sim$ 30 kpc. \citet{2020MNRAS.495...29E} found differently in their analysis of $\sim$ 150 MW analogs from Illustris. They found that the one ancient radial merger dominating the inner halo, which is thought to be the best \textit{Gaia}-Sausage analog, deposits a significant number of stars in the outer halo. Although the outer halo is significantly less influenced by the \textit{Gaia}-Sausage than the inner halo, the contribution of the GMM/Sausage component beyond $r_\mathrm{gc}$ of 30 kpc is largely uncertain from our results. The new samples of stars observed by the upcoming LAMOST DR8 will add supplementary of K giants to our existing sample, of which some giants may be located in the outer halo. We also expect that a more detailed analysis of elemental abundances will give aid to clarifying the origins of the outer halo stars.

\begin{deluxetable}{cccc}
	\label{tab:Sgr_bic}
	\centering
	\tablecaption{Anisotropy of the anisotropic and isotropic components of the GMM and ${\Delta}$BIC values for the Sgr-removed halo samples.}
	\tablewidth{0pt}
	\tablehead{
		\colhead{$r_\mathrm{gc}^\mathrm{a}\,\mathrm{(kpc)}$}&\colhead{$\beta_\mathrm{an}^\mathrm{b}$}&\colhead{$\beta_\mathrm{iso}^\mathrm{c}$}&\colhead{$\Delta \text{BIC}$}}
	\decimals
	\startdata
	\multicolumn{4}{c}{LAMOST K giant Sgr-removed stellar halo} \\
	\hline
	$~\,5.40-~\,9.60$&$0.90_{-0.01}^{+0.01}$&$0.38_{-0.12}^{+0.10}$&124\\
	$~\,9.60-11.46$&$0.94_{-0.01}^{+0.01}$&$0.43_{-0.08}^{+0.07}$&226\\
	$11.46-13.12$&$0.95_{-0.01}^{+0.01}$&$0.44_{-0.09}^{+0.08}$&218\\
	$13.12-14.72$&$0.95_{-0.01}^{+0.01}$&$0.50_{-0.08}^{+0.07}$&258\\
	$14.72-16.46$&$0.95_{-0.01}^{+0.01}$&$0.48_{-0.08}^{+0.07}$&255\\
	$16.46-18.22$&$0.94_{-0.01}^{+0.01}$&$0.36_{-0.11}^{+0.10}$&350\\
	$18.22-20.24$&$0.94_{-0.01}^{+0.01}$&$0.43_{-0.08}^{+0.08}$&321\\
	$20.24-22.87$&$0.94_{-0.01}^{+0.01}$&$0.32_{-0.09}^{+0.08}$&332\\
	$22.87-26.79$&$0.94_{-0.01}^{+0.01}$&$0.33_{-0.09}^{+0.08}$&499\\
	$26.79-30.00$&$0.93_{-0.01}^{+0.01}$&$0.44_{-0.12}^{+0.10}$&133\\
	$30.00-36.95$&$0.94_{-0.01}^{+0.01}$&$0.52_{-0.07}^{+0.07}$&89\\
	$~\,36.95-102.93$&$0.71_{-0.12}^{+0.07}$&$0.23_{-0.12}^{+0.16}$&29\\
	\hline
	\multicolumn{4}{c}{SEGUE K giant Sgr-removed stellar halo}\\
	\hline
	$~\,6.09-15.43$&$0.95_{-0.01}^{+0.01}$&$0.47_{-0.07}^{+0.06}$&250\\
	$15.43-20.18$&$0.94_{-0.01}^{+0.01}$&$0.41_{-0.08}^{+0.07}$&322\\
	$20.18-25.85$&$0.94_{-0.01}^{+0.01}$&$0.45_{-0.07}^{+0.07}$&329\\
	$25.85-30.00$&$0.94_{-0.02}^{+0.01}$&$0.58_{-0.08}^{+0.07}$&138\\
	$30.00-37.31$&$0.93_{-0.02}^{+0.02}$&$0.51_{-0.07}^{+0.06}$&102\\
	$~\,37.31-125.02$&$0.93_{-0.04}^{+0.03}$&$0.32_{-0.11}^{+0.08}$&60\\
	\hline
	\multicolumn{4}{c}{SDSS BHB Sgr-removed stellar halo}\\
	\hline
	$~\,5.54-13.45$&$0.93_{-0.02}^{+0.02}$&$0.29_{-0.10}^{+0.09}$&103\\
	$13.45-17.91$&$0.96_{-0.01}^{+0.01}$&$0.22_{-0.08}^{+0.08}$&242\\
	$17.91-23.34$&$0.94_{-0.01}^{+0.01}$&$0.27_{-0.09}^{+0.08}$&250\\
	$23.34-30.00$&$0.93_{-0.02}^{+0.02}$&$0.45_{-0.08}^{+0.07}$&129\\
	$30.00-60.00$&$-2.0_{-0.74}^{+0.53}$&$0.43_{-0.07}^{+0.06}$&74\\
	\enddata
	\tablenote{Selected width of the distance in $r_\mathrm{gc}$.}
	\tablenote{$\beta_\mathrm{an}$ is the anisotropy of the GMM/Sausage component.}
	\tablenote{$\beta_\mathrm{iso}$ is the anisotropy of the GMM/isotropic component.}
\end{deluxetable}

\begin{deluxetable}{cccc}
	\label{tab:smooth_bic}
	\centering
	\tablecaption{Anisotropy of the anisotropic and isotropic components of the GMM and ${\Delta}$BIC values for the smooth halo samples.}
	\tablewidth{0pt}
	\tablehead{
		\colhead{$r_\mathrm{gc}^\mathrm{a}\,\mathrm{(kpc)}$}&\colhead{$\beta_\mathrm{an}^\mathrm{b}$}&\colhead{$\beta_\mathrm{iso}^\mathrm{c}$}&\colhead{$\Delta \text{BIC}$}}
	\decimals
	\startdata
	\multicolumn{4}{c}{LAMOST K giant smooth stellar halo} \\
	\hline
	$~\,5.41-10.42$&$0.92_{-0.02}^{+0.01}$&$0.42_{-0.09}^{+0.08}$&145\\
	$10.42-12.90$&$0.95_{-0.01}^{+0.01}$&$0.40_{-0.11}^{+0.09}$&207\\
	$12.90-15.40$&$0.95_{-0.01}^{+0.01}$&$0.49_{-0.07}^{+0.06}$&189\\
	$15.40-18.37$&$0.94_{-0.01}^{+0.01}$&$0.35_{-0.09}^{+0.08}$&260\\
	$18.37-22.53$&$0.94_{-0.01}^{+0.01}$&$0.40_{-0.08}^{+0.08}$&234\\
	$22.53-24.75$&$0.91_{-0.02}^{+0.01}$&$0.45_{-0.12}^{+0.10}$&94\\
	$24.75-28.11$&$0.93_{-0.02}^{+0.01}$&$0.40_{-0.12}^{+0.11}$&132\\
	$28.11-33.78$&$0.93_{-0.02}^{+0.02}$&$0.57_{-0.11}^{+0.09}$&75\\
	$~\,33.78-100.43$&$0.86_{-0.04}^{+0.03}$&$0.24_{-0.23}^{+0.16}$&34\\
	\hline
	\multicolumn{4}{c}{SEGUE K giant smooth stellar halo}\\
	\hline
	$~\,6.18-15.07$&$0.95_{-0.01}^{+0.01}$&$0.46_{-0.07}^{+0.06}$&250\\
	$15.07-20.60$&$0.94_{-0.01}^{+0.01}$&$0.37_{-0.08}^{+0.08}$&323\\
	$20.60-25.25$&$0.91_{-0.02}^{+0.02}$&$0.39_{-0.09}^{+0.08}$&329\\
	$25.25-32.66$&$0.94_{-0.02}^{+0.02}$&$0.51_{-0.09}^{+0.07}$&102\\
	$~\,32.66-111.07$&$0.91_{-0.03}^{+0.02}$&$0.40_{-0.07}^{+0.07}$&60\\
	\hline
	\multicolumn{4}{c}{SDSS BHB smooth stellar halo}\\
	\hline
	$~\,5.54-15.05$&$0.95_{-0.02}^{+0.02}$&$0.33_{-0.08}^{+0.07}$&78\\
	$15.05-23.73$&$0.94_{-0.01}^{+0.01}$&$0.33_{-0.07}^{+0.07}$&109\\
	$23.73-59.00$&$-1.56_{-0.57}^{+0.45}$&$0.70_{-0.05}^{+0.04}$&140\\
	\enddata
	\tablenote{Selected width of the distance in $r_\mathrm{gc}$.}
	\tablenote{$\beta_\mathrm{an}$ is the anisotropy of the GMM/Sausage component.}
	\tablenote{$\beta_\mathrm{iso}$ is the anisotropy of the GMM/isotropic component.}
\end{deluxetable}

\begin{deluxetable*}{ccccccccccccc}
\label{tab:Sgr-removed}
\centering
\tablecaption{Best estimated parameters of the GMM for the Sgr-removed stellar halo samples.}
\tablewidth{0pt}
\tablehead{
	\colhead{$r_\mathrm{gc}^\mathrm{a}$}&\colhead{$N^\mathrm{b}$}&\colhead{$\langle v_{r}^\mathrm{an} \rangle$}&\colhead{$\langle v_\phi^\mathrm{an} \rangle$}&\colhead{$\sigma_{v_r;\mathrm{an}}$}&\colhead{$\sigma_{t;\mathrm{an}}$}&\colhead{$\sigma_{v_r;\mathrm{iso}}$}&\colhead{$\sigma_{t;\mathrm{iso}}$}&\colhead{$\mu_\text{[Fe/H]}^\mathrm{an}$}&\colhead{$\sigma_\text{[Fe/H]}^\mathrm{an}$}&\colhead{$\mu_\text{[Fe/H]}^\mathrm{iso}$}&\colhead{$\sigma_\text{[Fe/H]}^\mathrm{iso}$}&\colhead{$f_\mathrm{an}$}\\
\colhead{kpc}&\colhead{}&\colhead{km $\text{s}^{-1}$}&\colhead{km $\text{s}^{-1}$}&\colhead{km $\text{s}^{-1}$}&\colhead{km $\text{s}^{-1}$}&\colhead{km $\text{s}^{-1}$}&\colhead{km $\text{s}^{-1}$}&\colhead{dex}&\colhead{dex}&\colhead{dex}&\colhead{dex}}
\decimals
\startdata
\multicolumn{13}{c}{LAMOST K giant Sgr-removed stellar halo} \\
\hline
$~\,5.40-~\,9.60$&1000&$127^{+6}_{-6}$&$-17^{+3}_{-3}$&$113^{+5}_{-5}$&$55^{+2}_{-2}$&$168^{+10}_{-9}$&$133^{+7}_{-7}$&$-1.31^{+0.02}_{-0.02}$&$0.42^{+0.01}_{-0.01}$&$-1.67^{+0.04}_{-0.04}$&$0.45^{+0.03}_{-0.03}$&$0.74^{+0.04}_{-0.04}$\\
$~\,9.60-11.46$&1000&$138^{+5}_{-5}$&$-12^{+2}_{-2}$&$93^{+4}_{-4}$&$40^{+2}_{-2}$&$155^{+7}_{-7}$&$117^{+5}_{-4}$&$-1.30^{+0.02}_{-0.02}$&$0.40^{+0.02}_{-0.02}$&$-1.73^{+0.04}_{-0.04}$&$0.46^{+0.02}_{-0.02}$&$0.64^{+0.03}_{-0.03}$\\
$11.46-13.12$&1000&$138^{+5}_{-5}$&$-7^{+2}_{-2}$&$91^{+4}_{-3}$&$37^{+2}_{-2}$&$149^{+7}_{-7}$&$112^{+5}_{-4}$&$-1.39^{+0.02}_{-0.02}$&$0.36^{+0.01}_{-0.01}$&$-1.61^{+0.03}_{-0.04}$&$0.49^{+0.02}_{-0.02}$&$0.65^{+0.03}_{-0.03}$\\
$13.12-14.72$&1000&$119^{+5}_{-5}$&$-11^{+2}_{-2}$&$90^{+4}_{-4}$&$33^{+1}_{-1}$&$148^{+6}_{-6}$&$106^{+4}_{-4}$&$-1.37^{+0.02}_{-0.02}$&$0.33^{+0.01}_{-0.01}$&$-1.70^{+0.03}_{-0.03}$&$0.48^{+0.02}_{-0.02}$&$0.66^{+0.03}_{-0.03}$\\
$14.72-16.46$&1000&$110^{+4}_{-4}$&$-11^{+2}_{-2}$&$84^{+4}_{-3}$&$32^{+1}_{-1}$&$148^{+6}_{-6}$&$106^{+4}_{-4}$&$-1.36^{+0.02}_{-0.02}$&$0.35^{+0.01}_{-0.01}$&$-1.70^{+0.03}_{-0.03}$&$0.48^{+0.02}_{-0.02}$&$0.67^{+0.03}_{-0.03}$\\
$16.46-18.22$&1000&$94^{+4}_{-4}$&$-9^{+1}_{-2}$&$80^{+3}_{-3}$&$29^{+1}_{-1}$&$137^{+7}_{-6}$&$109^{+5}_{-5}$&$-1.39^{+0.02}_{-0.02}$&$0.34^{+0.01}_{-0.01}$&$-1.73^{+0.03}_{-0.03}$&$0.44^{+0.03}_{-0.02}$&$0.74^{+0.02}_{-0.02}$\\
$18.22-20.24$&1000&$86^{+4}_{-4}$&$-7^{+1}_{-1}$&$74^{+4}_{-3}$&$26^{+1}_{-1}$&$127^{+6}_{-6}$&$97^{+4}_{-4}$&$-1.38^{+0.02}_{-0.02}$&$0.34^{+0.01}_{-0.01}$&$-1.73^{+0.03}_{-0.03}$&$0.43^{+0.02}_{-0.02}$&$0.72^{+0.02}_{-0.02}$\\
$20.24-22.87$&1000&$75^{+5}_{-6}$&$-7^{+1}_{-1}$&$79^{+5}_{-5}$&$27^{+1}_{-1}$&$131^{+7}_{-6}$&$100^{+5}_{-2}$&$-1.41^{+0.02}_{-0.02}$&$0.31^{+0.01}_{-0.01}$&$-1.72^{+0.03}_{-0.03}$&$0.43^{+0.03}_{-0.03}$&$0.74^{+0.02}_{-0.02}$\\
$22.87-26.79$&1000&$66^{+4}_{-6}$&$-6^{+1}_{-1}$&$70^{+6}_{-4}$&$25^{+1}_{-1}$&$142^{+7}_{-6}$&$117^{+5}_{-4}$&$-1.45^{+0.02}_{-0.02}$&$0.29^{+0.01}_{-0.01}$&$-1.69^{+0.03}_{-0.03}$&$0.46^{+0.02}_{-0.02}$&$0.70^{+0.02}_{-0.02}$\\
$26.79-30.00$&455&$44^{+20}_{-29}$&$-5^{+2}_{-2}$&$93^{+9}_{-11}$&$27^{+2}_{-2}$&$128^{+9}_{-8}$&$96^{+6}_{-5}$&$-1.48^{+0.03}_{-0.03}$&$0.31^{+0.02}_{-0.02}$&$-1.79^{+0.05}_{-0.05}$&$0.42^{+0.04}_{-0.03}$&$0.66^{+0.04}_{-0.05}$\\
$30.00-36.95$&500&$51^{+16}_{-32}$&$-7^{+3}_{-3}$&$82^{+12}_{-12}$&$23^{+2}_{-2}$&$125^{+6}_{-5}$&$87^{+4}_{-3}$&$-1.44^{+0.04}_{-0.03}$&$0.28^{+0.02}_{-0.02}$&$-1.79^{+0.03}_{-0.03}$&$0.39^{+0.03}_{-0.02}$&$0.47^{+0.04}_{-0.05}$\\
$~\,36.95-102.93$&464&$75^{+5}_{-5}$&$-16^{+4}_{-4}$&$60^{+4}_{-3}$&$44^{+5}_{-8}$&$115^{+15}_{-11}$&$102^{+15}_{-11}$&$-1.58^{+0.05}_{-0.03}$&$0.37^{+0.02}_{-0.03}$&$-1.80^{+0.10}_{-0.23}$&$0.39^{+0.07}_{-0.21}$&$0.80^{+0.07}_{-0.13}$\\
\hline
\multicolumn{13}{c}{SEGUE K giant Sgr-removed stellar halo}\\
\hline
$~\,6.09-15.43$&1000&$130^{+5}_{-5}$&$-14^{+2}_{-2}$&$91^{+4}_{-4}$&$37^{+2}_{-2}$&$146^{+6}_{-5}$&$104^{+3}_{-3}$&$-1.44^{+0.02}_{-0.02}$&$0.26^{+0.02}_{-0.02}$&$-1.94^{+0.03}_{-0.03}$&$0.52^{+0.02}_{-0.02}$&$0.55^{+0.03}_{-0.03}$\\
$15.43-20.18$&1000&$104^{+4}_{-5}$&$-7^{+2}_{-2}$&$80^{+4}_{-4}$&$29^{+1}_{-1}$&$126^{+5}_{-5}$&$96^{+3}_{-3}$&$-1.44^{+0.01}_{-0.01}$&$0.22^{+0.01}_{-0.01}$&$-1.90^{+0.03}_{-0.03}$&$0.51^{+0.02}_{-0.02}$&$0.59^{+0.02}_{-0.02}$\\
$20.18-25.85$&1000&$68^{+6}_{-11}$&$-2^{+2}_{-2}$&$79^{+9}_{-6}$&$27^{+1}_{-1}$&$132^{+5}_{-5}$&$98^{+4}_{-3}$&$-1.46^{+0.02}_{-0.02}$&$0.28^{+0.01}_{-0.01}$&$-1.91^{+0.03}_{-0.03}$&$0.54^{+0.02}_{-0.02}$&$0.61^{+0.02}_{-0.02}$\\
$25.85-30.00$&480&$47^{+18}_{-30}$&$1^{+2}_{-2}$&$89^{+10}_{-12}$&$25^{+2}_{-2}$&$123^{+7}_{-7}$&$80^{+5}_{-4}$&$-1.42^{+0.02}_{-0.03}$&$0.22^{+0.03}_{-0.02}$&$-1.92^{+0.05}_{-0.05}$&$0.48^{+0.03}_{-0.03}$&$0.60^{+0.05}_{-0.05}$\\
$30.00-37.31$&500&$49^{+17}_{-31}$&$4^{+3}_{-3}$&$84^{+12}_{-13}$&$26^{+2}_{-2}$&$113^{+5}_{-5}$&$78^{+3}_{-3}$&$-1.35^{+0.02}_{-0.02}$&$0.13^{+0.02}_{-0.02}$&$-1.98^{+0.04}_{-0.04}$&$0.51^{+0.02}_{-0.02}$&$0.38^{+0.03}_{-0.03}$\\
$~\,37.31-125.02$&591&$84^{+7}_{-9}$&$-1^{+4}_{-5}$&$63^{+6}_{-5}$&$28^{+6}_{-4}$&$94^{+5}_{-4}$&$77^{+4}_{-3}$&$-1.44^{+0.03}_{-0.03}$&$0.20^{+0.04}_{-0.03}$&$-2.10^{+0.04}_{-0.04}$&$0.48^{+0.03}_{-0.02}$&$0.30^{+0.05}_{-0.04}$\\
\hline
\multicolumn{13}{c}{SDSS BHB Sgr-removed stellar halo}\\
\hline
$~\,5.54-13.45$&800&$117^{+9}_{-10}$&$-2^{+3}_{-3}$&$101^{+8}_{-7}$&$42^{+4}_{-3}$&$128^{+6}_{-5}$&$107^{+5}_{-4}$&$-1.74^{+0.03}_{-0.03}$&$0.30^{+0.03}_{-0.03}$&$-1.99^{+0.03}_{-0.03}$&$0.37^{+0.02}_{-0.02}$&$0.45^{+0.06}_{-0.05}$\\
$13.45-17.91$&800&$98^{+5}_{-6}$&$0^{+2}_{-2}$&$75^{+5}_{-4}$&$25^{+1}_{-1}$&$116^{+4}_{-4}$&$103^{+3}_{-3}$&$-1.75^{+0.02}_{-0.02}$&$0.23^{+0.02}_{-0.02}$&$-2.10^{+0.02}_{-0.02}$&$0.33^{+0.02}_{-0.02}$&$0.41^{+0.02}_{-0.03}$\\
$17.91-23.34$&800&$72^{+5}_{-5}$&$-3^{+2}_{-2}$&$63^{+5}_{-4}$&$24^{+1}_{-1}$&$113^{+4}_{-4}$&$96^{+3}_{-3}$&$-1.75^{+0.02}_{-0.02}$&$0.25^{+0.02}_{-0.02}$&$-2.10^{+0.02}_{-0.02}$&$0.34^{+0.02}_{-0.02}$&$0.48^{+0.03}_{-0.03}$\\
$23.34-30.00$&505&$34^{+17}_{-23}$&$2^{+2}_{-2}$&$72^{+9}_{-13}$&$21^{+2}_{-2}$&$120^{+5}_{-5}$&$89^{+3}_{-3}$&$-1.80^{+0.03}_{-0.03}$&$0.26^{+0.03}_{-0.03}$&$-2.10^{+0.03}_{-0.03}$&$0.41^{+0.02}_{-0.02}$&$0.41^{+0.03}_{-0.03}$\\
$30.00-60.00$&612&$35^{+24}_{-24}$&$24^{+23}_{-22}$&$90^{+11}_{-12}$&$172^{+13}_{-12}$&$91^{+3}_{-3}$&$69^{+3}_{-3}$&$-1.33^{+0.10}_{-0.11}$&$0.50^{+0.07}_{-0.07}$&$-2.04^{+0.02}_{-0.02}$&$0.31^{+0.02}_{-0.02}$&$0.15^{+0.03}_{-0.03}$\\
\enddata
\tablenote{Selected width of the distance in $r_\mathrm{gc}$.}
\tablenote{$N$ is the number of stars of the observational data in the corresponding distance bin for modeling.}
\end{deluxetable*}

 \subsection{Contribution of the \textit{Gaia}-Sausage to the Smooth Halo}\label{sub:smooth}

 Stars stripped from a satellite during the merging process can form substructures such as stellar streams, shells, and clouds in density space \citep{2003ApJ...599.1082M,2006ApJ...642L.137B}, velocity phase space \citep{2009ApJ...698..567S,2016ApJ...816...80J}, and stellar age distribution \citep{2015ApJ...813L..16S,2016NatPh..12.1170C}. Previous studies have found several halo substructures which are closely associated with the \textit{Gaia}-Sausage \citep{2019A&A...631L...9K,2020ApJ...898L..37Y,2020ApJ...891...39Y,2021ApJ...908..191C}. In this study, halo substructures are selected in the IoM space. Although IoM are a power tool to identify halo substructures, not all star particles associated to the accretion events can be recovered by this method \citep{2000MNRAS.319..657H}. Therefore, not all stars of the \textit{Gaia}-Sausage may be located in substructures, and some fraction may be found in the smooth halo. In this subsection, we will explore the proportion of the \textit{Gaia}-Sausage stars in the stellar halo with all substructures found in the IoM space excluded.

The steps to obtain the contribution of the \textit{Gaia}-Sausage are consistent with the previously detailed, which are selecting distance bins, applying the GMM to the data, and making the mock data based on the best estimated parameters. For the LAMOST K giant smooth halo, 9 bins are defined, and the edges of these bins are $r_\mathrm{gc}$ = 5.41, 10.42, 12.90, 15.40, 18.37, 22.53, 24.75, 28.11, 33.78, and 100.43 kpc. Each of the first 5 bins contain 1000 stars, while the last four bins have 400, 400, 400, and 444 stars. Five bins are defined for the SEGUE K giant smooth halo, and their edges are $r_\mathrm{gc}$ = 6.18, 15.07, 20.60, 25.25, 32.66, and 111.07 kpc. The five bins contain 800, 800, 500, 500, and 675 stars, respectively. The SDSS BHB smooth halo has 3 bins with edges $r_\mathrm{gc}$ = 5.54, 15.05, 23.73, and 59 kpc. The three bins have 800, 800, and 681 stars, respectively. The best estimated parameters of the GMM/Sausage component for the smooth halo samples are shown in Table~\ref{tab:smooth}.

\begin{figure*}
 	\centering
 		\includegraphics[width = 1.0\textwidth]{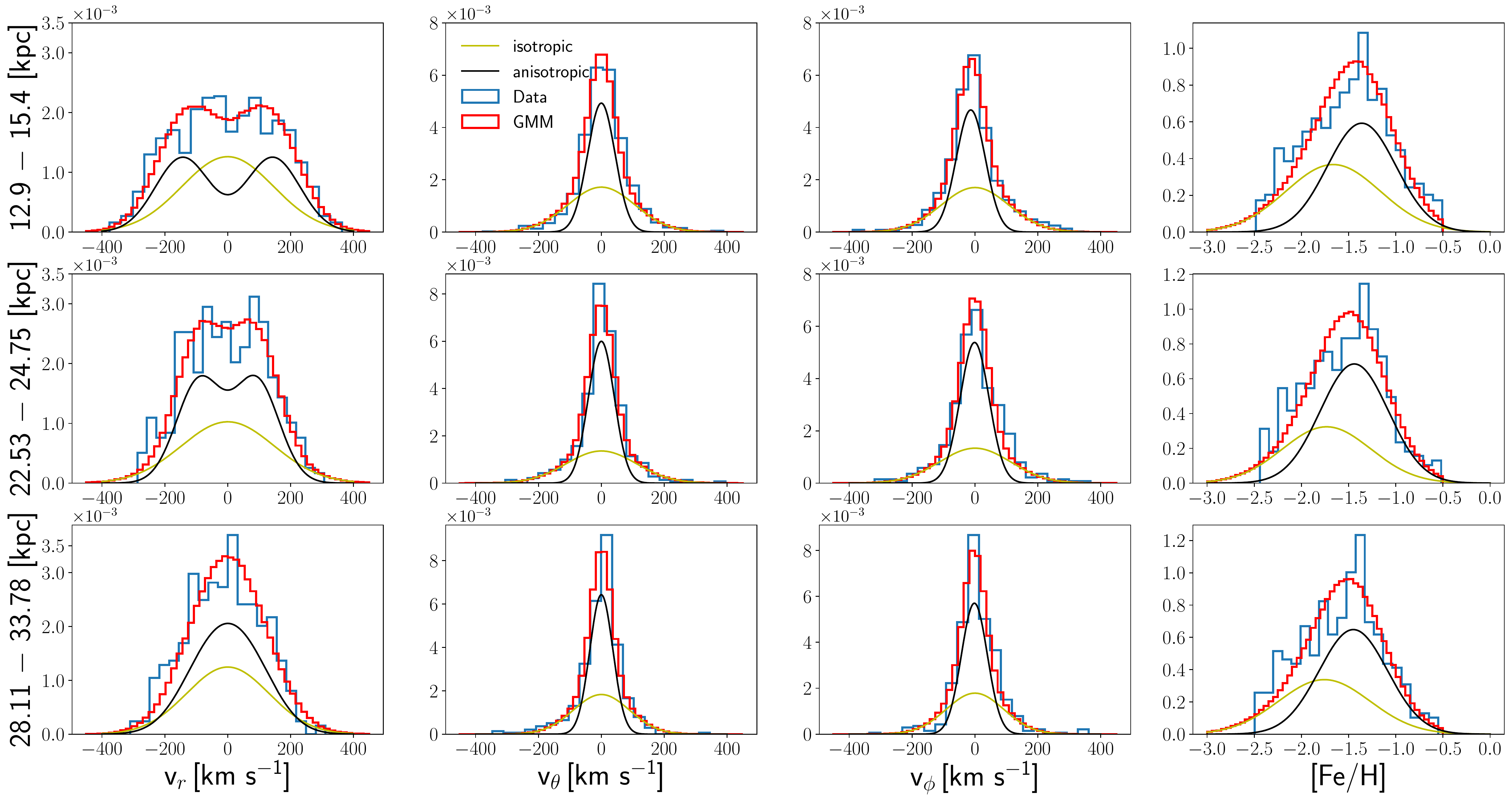}
 		\caption{GMM fitting result of the LAMOST K giant smooth stellar halo.}
 		\label{fig:gmm_sm_lm}
 	\includegraphics[width = 1.0\textwidth]{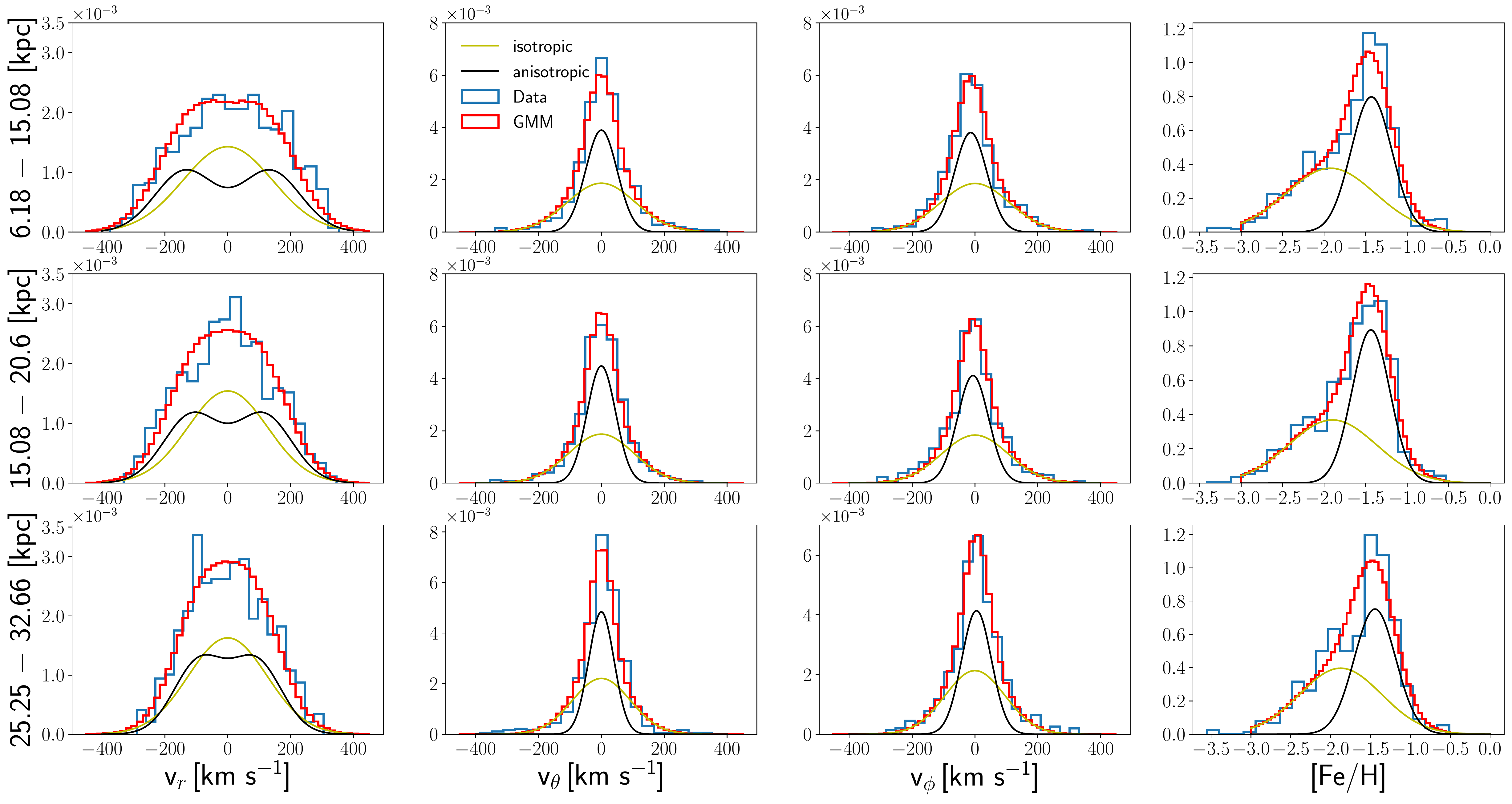}
 	\caption{GMM fitting result of the SEGUE K giant smooth stellar halo.}
 	\label{fig:gmm_sm_se}
 \end{figure*}

\begin{figure*}
	\centering
	\includegraphics[width = 1.0\textwidth]{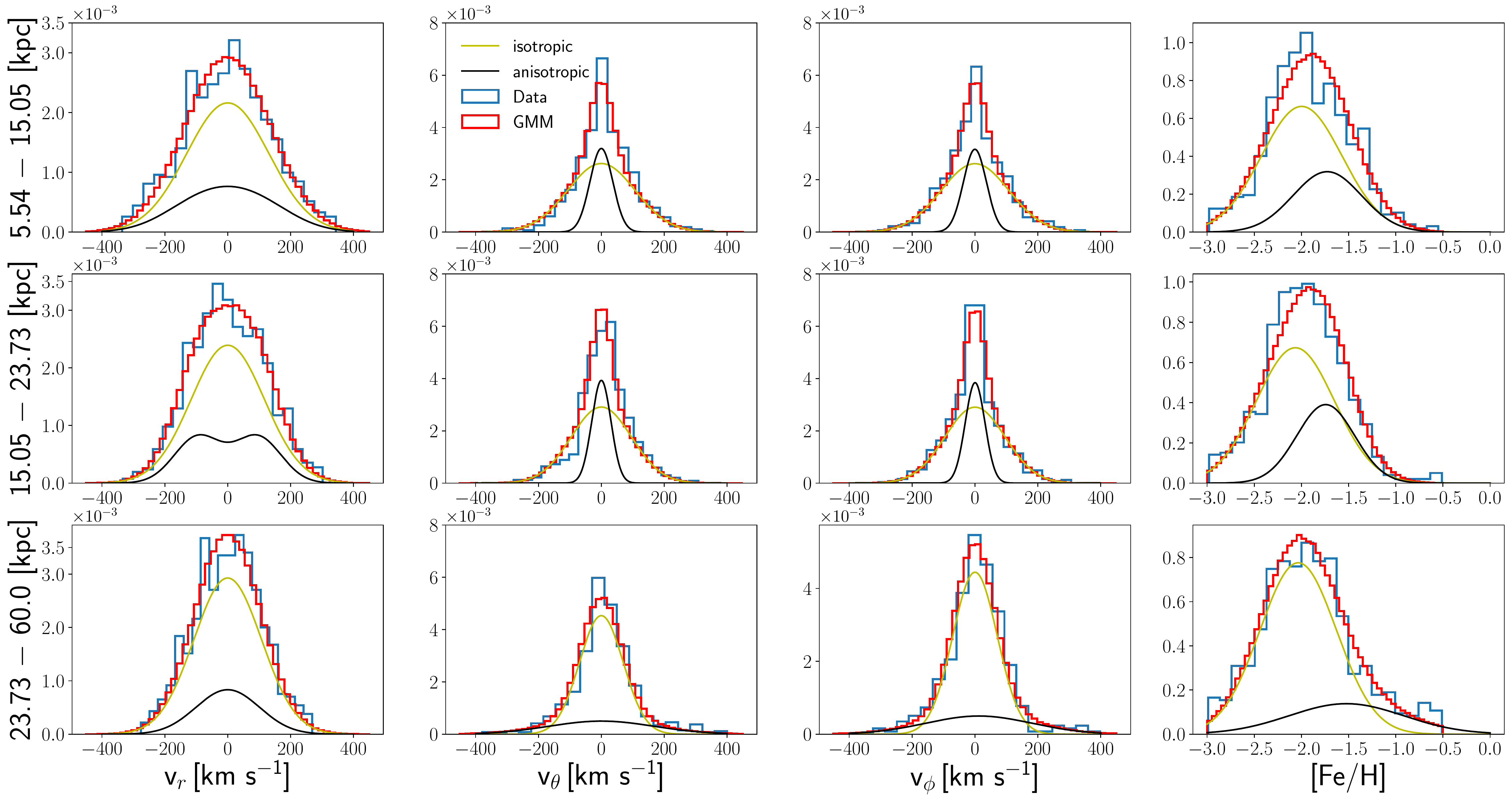}
	\caption{GMM fitting result of the SDSS BHB smooth stellar halo.}
	\label{fig:gmm_sm_sd}
\end{figure*}

 Comparisons between the observational and the mock data are made in Figures~\ref{fig:gmm_sm_lm}, ~\ref{fig:gmm_sm_se}, and ~\ref{fig:gmm_sm_sd}. From Figure~\ref{fig:gmm_ratio}, we can see that the fraction of the GMM/Sausage component declines in the inner halo after the removal of halo substructures. \citet{2021ApJ...919...66B} found that removing substructures causes $\beta$ to be sightly less radial within 20 kpc from the Galactic Center. In their view, it is likely due to the removal of obvious substructures related to the \textit{Gaia}-Sausage. The decrease of $f_\mathrm{an}$ shown in Figure~\ref{fig:gmm_ratio} is supportive of their view. However, we find that the smooth, inner halo is still largely influenced by the \textit{Gaia}-Sausage,  where the fraction of the \textit{Gaia}-Sausage stars is $57\%-67\%$ for LAMOST K giants, $45\%-51\%$ for SEGUE K giants, and $29\%-30\%$ for SDSS BHB stars. 

Integrals of motion are a powerful tool to identify halo substructure for the case that the stars share roughly similar orbital information. However, stars stripped from a massive satellite over time may loose their coherence in the energy/angular momentum space because of dynamical friction and tidal effects between the satellite and the host galaxy \citep{2005MNRAS.359...93M,2017A&A...604A.106J}. The removal of halo substructures using IoM is inadequate to exclude all stars of the \textit{Gaia}-Sausage. Therefore, $\beta$ remains high in the smooth, inner halo. The slight decrease of the fraction of the GMM/Sausage component is consistent with the slight decrease of $\beta$ shown in \citet{2021ApJ...919...66B}. 

The situation in the outer halo is more complex than the inner halo. For the SEGUE K giant and SDSS BHB star smooth, outer halo samples, the fraction of the GMM/Sausage component has a slight increase after the removal of halo substructures, but the GMM/isotropic component is still the dominant part. As we mentioned in Sec~\ref{sub:diffuse}, the outer halo is significantly less influenced by the \textit{Gaia}-Sausage when compared to the inner halo, and the outermost apocenter of stars of the \textit{Gaia}-Sausage is about 25 kpc \citep{2018ApJ...862L...1D}. Therefore, we suspect that most of the halo substructures in the outer halo are unrelated to the \textit{Gaia}-Sausage. The removal of substructures may reduce the number of stars belonging to the GMM/isotropic component in the outer halo, which could lead to the increase of the fraction of the GMM/Sausage component. In the LAMOST K giant smooth, outer halo, the fractions of the GMM/Sausage are 59\% and 73\%, which are much larger than the results of SEGUE K giant and SDSS BHB star samples. Figure~\ref{fig:correct} shows that values of ${f_\mathrm{an}}$ in the LAMOST K giant smooth, outer halo change with the distance correction, especially for the outermost distance bin of ${r_\mathrm{gc} \in [33.78, 100.43]}$ kpc. The sensitivity of $f_\mathrm{an}$ to the accuracy of distance estimation should be noticed for the outermost distance bin of LAMOST K giant smooth halo. 

\textit{N}-body hydrodynamical simulations suggest that the ${\beta}$ profile of MW-like disk galaxies is steady and rises slowly with the increasing Galactocentric radius in the absence of halo substructures \citep{2005MNRAS.364..367D,2006MNRAS.365..747A,2012ApJ...761...98K}. Anisotropy of the simulated outer stellar halo can be as high as ${0.6-0.8}$ without the participation of major merger events \citep{2013ApJ...773L..32R,2018ApJ...853..196L}. The slight increase of ${f_\mathrm{an}}$ in SEGUE K giant and SDSS BHB smooth, outer halo and large value of ${f_\mathrm{an}}$ in LAMOST K giant smooth, outer halo may correspond to the highly radial biased stellar halo at very large ${r_\mathrm{gc}}$ in the absence of substructures. Future work which applies the GMM to simulated MW-like galaxies with all substructures excluded is needed to figure out this question.

\begin{deluxetable*}{ccccccccccccc}
	\label{tab:smooth}
	\centering
	\tablecaption{Best estimated parameters of the GMM for the smooth stellar halo samples.}
	\tablewidth{0pt}
	\tablehead{
		\colhead{$r_\mathrm{gc}^\mathrm{a}$}&\colhead{$N^\mathrm{b}$}&\colhead{$\langle v_{r}^\mathrm{an} \rangle$}&\colhead{$\langle v_\phi^\mathrm{an} \rangle$}&\colhead{$\sigma_{v_r;\mathrm{an}}$}&\colhead{$\sigma_{t;\mathrm{an}}$}&\colhead{$\sigma_{v_r;\mathrm{iso}}$}&\colhead{$\sigma_{t;\mathrm{iso}}$}&\colhead{$\mu_\text{[Fe/H]}^\mathrm{an}$}&\colhead{$\sigma_\text{[Fe/H]}^\mathrm{an}$}&\colhead{$\mu_\text{[Fe/H]}^\mathrm{iso}$}&\colhead{$\sigma_\text{[Fe/H]}^\mathrm{iso}$}&\colhead{$f_\mathrm{an}$}\\
		\colhead{kpc}&\colhead{}&\colhead{km $\text{s}^{-1}$}&\colhead{km $\text{s}^{-1}$}&\colhead{km $\text{s}^{-1}$}&\colhead{km $\text{s}^{-1}$}&\colhead{km $\text{s}^{-1}$}&\colhead{km $\text{s}^{-1}$}&\colhead{dex}&\colhead{dex}&\colhead{dex}&\colhead{dex}}
	\decimals
	\startdata
\multicolumn{13}{c}{LAMOST K giant smooth stellar halo} \\
\hline
$~\,5.41-10.42$&1000&$141^{+6}_{-6}$&$-21^{+3}_{-3}$&$108^{+5}_{-5}$&$52^{+2}_{-3}$&$168^{+8}_{-7}$&$129^{+6}_{-5}$&$-1.28^{+0.03}_{-0.03}$&$0.42^{+0.02}_{-0.02}$&$-1.67^{+0.04}_{-0.04}$&$0.45^{+0.02}_{-0.02}$&$0.63^{+0.04}_{-0.04}$\\
$10.42-12.90$&1000&$156^{+5}_{-5}$&$-10^{+3}_{-3}$&$90^{+4}_{-4}$&$42^{+3}_{-2}$&$152^{+6}_{-6}$&$118^{+5}_{-4}$&$-1.33^{+0.03}_{-0.02}$&$0.37^{+0.02}_{-0.02}$&$-1.66^{+0.03}_{-0.03}$&$0.49^{+0.02}_{-0.02}$&$0.57^{+0.04}_{-0.04}$\\
$12.90-15.40$&1000&$144^{+5}_{-5}$&$-13^{+2}_{-2}$&$86^{+4}_{-3}$&$36^{+2}_{-2}$&$145^{+6}_{-6}$&$104^{+4}_{-4}$&$-1.36^{+0.02}_{-0.02}$&$0.33^{+0.02}_{-0.02}$&$-1.66^{+0.03}_{-0.03}$&$0.48^{+0.02}_{-0.02}$&$0.54^{+0.03}_{-0.03}$\\
$15.40-18.37$&1000&$119^{+4}_{-4}$&$-12^{+2}_{-2}$&$84^{+4}_{-3}$&$35^{+2}_{-2}$&$136^{+6}_{-6}$&$110^{+4}_{-4}$&$-1.37^{+0.02}_{-0.02}$&$0.34^{+0.01}_{-0.01}$&$-1.68^{+0.03}_{-0.03}$&$0.45^{+0.02}_{-0.02}$&$0.63^{+0.03}_{-0.03}$\\
$18.37-22.53$&1000&$107^{+4}_{-4}$&$-6^{+2}_{-2}$&$79^{+4}_{-3}$&$32^{+1}_{-1}$&$130^{+6}_{-5}$&$100^{+4}_{-4}$&$-1.39^{+0.02}_{-0.02}$&$0.30^{+0.01}_{-0.01}$&$-1.75^{+0.03}_{-0.03}$&$0.42^{+0.02}_{-0.02}$&$0.63^{+0.03}_{-0.03}$\\
$22.53-24.75$&400&$91^{+6}_{-7}$&$-1^{+3}_{-3}$&$72^{+7}_{-6}$&$35^{+3}_{-3}$&$147^{+10}_{-9}$&$109^{+8}_{-7}$&$-1.44^{+0.03}_{-0.03}$&$0.31^{+0.02}_{-0.02}$&$-1.73^{+0.05}_{-0.05}$&$0.43^{+0.04}_{-0.04}$&$0.63^{+0.05}_{-0.06}$\\
$24.75-28.11$&400&$74^{+9}_{-17}$&$-3^{+3}_{-3}$&$79^{+14}_{-8}$&$29^{+2}_{-2}$&$142^{+10}_{-9}$&$110^{+7}_{-7}$&$-1.48^{+0.03}_{-0.03}$&$0.30^{+0.02}_{-0.02}$&$-1.69^{+0.05}_{-0.05}$&$0.50^{+0.04}_{-0.04}$&$0.67^{+0.04}_{-0.05}$\\
$28.11-33.78$&400&$58^{+18}_{-36}$&$-2^{+3}_{-3}$&$94^{+15}_{-14}$&$29^{+3}_{-4}$&$131^{+9}_{-8}$&$87^{+7}_{-6}$&$-1.45^{+0.04}_{-0.04}$&$0.30^{+0.03}_{-0.03}$&$-1.76^{+0.05}_{-0.05}$&$0.44^{+0.04}_{-0.03}$&$0.59^{+0.06}_{-0.06}$\\
$~\,33.78-100.43$&444&$79^{+5}_{-5}$&$-4^{+3}_{-3}$&$63^{+5}_{-4}$&$39^{+4}_{-3}$&$107^{+10}_{-8}$&$94^{+13}_{-9}$&$-1.60^{+0.03}_{-0.03}$&$0.35^{+0.02}_{-0.02}$&$-1.65^{+0.08}_{-0.07}$&$0.47^{+0.05}_{-0.04}$&$0.73^{+0.08}_{-0.09}$\\
\hline
\multicolumn{13}{c}{SEGUE K giant smooth stellar halo}\\
\hline
$~\,6.18-15.07$&800&$136^{+6}_{-6}$&$-14^{+3}_{-3}$&$94^{+5}_{-4}$&$38^{+2}_{-2}$&$142^{+6}_{-6}$&$104^{+4}_{-3}$&$-1.43^{+0.02}_{-0.02}$&$0.22^{+0.02}_{-0.02}$&$-1.92^{+0.04}_{-0.04}$&$0.53^{+0.02}_{-0.02}$&$0.49^{+0.03}_{-0.03}$\\
$15.07-20.60$&800&$115^{+6}_{-6}$&$-6^{+3}_{-3}$&$90^{+5}_{-5}$&$36^{+2}_{-2}$&$127^{+5}_{-5}$&$101^{+4}_{-3}$&$-1.44^{+0.02}_{-0.02}$&$0.20^{+0.02}_{-0.02}$&$-1.90^{+0.03}_{-0.03}$&$0.52^{+0.02}_{-0.02}$&$0.51^{+0.03}_{-0.03}$\\
$20.60-25.25$&500&$93^{+9}_{-12}$&$0^{+4}_{-4}$&$86^{+10}_{-8}$&$38^{+3}_{-2}$&$129^{+6}_{-6}$&$101^{+4}_{-4}$&$-1.41^{+0.02}_{-0.02}$&$0.20^{+0.02}_{-0.02}$&$-1.88^{+0.04}_{-0.04}$&$0.53^{+0.03}_{-0.03}$&$0.45^{+0.04}_{-0.04}$\\
$25.25-32.66$&500&$91^{+8}_{-10}$&$5^{+3}_{-3}$&$80^{+8}_{-6}$&$30^{+3}_{-2}$&$125^{+6}_{-6}$&$88^{+6}_{-4}$&$-1.44^{+0.04}_{-0.04}$&$0.23^{+0.04}_{-0.04}$&$-1.88^{+0.04}_{-0.04}$&$0.50^{+0.03}_{-0.03}$&$0.49^{+0.06}_{-0.05}$\\
$~\,32.66-111.07$&675&$90^{+6}_{-6}$&$2^{+4}_{-4}$&$65^{+5}_{-5}$&$33^{+4}_{-4}$&$102^{+4}_{-4}$&$79^{+3}_{-2}$&$-1.44^{+0.02}_{-0.02}$&$0.17^{+0.03}_{-0.03}$&$-2.06^{+0.03}_{-0.04}$&$0.51^{+0.02}_{-0.02}$&$0.33^{+0.03}_{-0.03}$\\
\hline
\multicolumn{13}{c}{SDSS BHB smooth stellar halo}\\
\hline
$~\,5.54-15.05$&800&$82^{+20}_{-49}$&$0^{+4}_{-4}$&$119^{+21}_{-15}$&$34^{+5}_{-4}$&$131^{+5}_{-4}$&$107^{+3}_{-3}$&$-1.73^{+0.04}_{-0.04}$&$0.29^{+0.03}_{-0.03}$&$-2.00^{+0.02}_{-0.02}$&$0.37^{+0.02}_{-0.02}$&$0.29^{+0.04}_{-0.03}$\\
$15.05-23.73$&800&$95^{+7}_{-8}$&$0^{+3}_{-3}$&$74^{+7}_{-6}$&$29^{+2}_{-2}$&$117^{+4}_{-4}$&$95^{+3}_{-2}$&$-1.74^{+0.03}_{-0.03}$&$0.22^{+0.03}_{-0.03}$&$-2.06^{+0.02}_{-0.02}$&$0.35^{+0.02}_{-0.02}$&$0.30^{+0.03}_{-0.03}$\\
$23.73-59.00$&681&$35^{+23}_{-24}$&$12^{+15}_{-14}$&$93^{+10}_{-11}$&$163^{+11}_{-10}$&$107^{+4}_{-3}$&$58^{+3}_{-3}$&$-1.52^{+0.10}_{-0.09}$&$0.57^{+0.05}_{-0.05}$&$-2.03^{+0.02}_{-0.02}$&$0.34^{+0.02}_{-0.02}$&$0.21^{+0.03}_{-0.03}$\\
\enddata
\tablenote{Selected width of the distance in $r_\mathrm{gc}$.}
\tablenote{$N$ is the number of stars in the corresponding distance bin of the observational data for modeling.}
\end{deluxetable*}

\subsection{Comparison between Sausage-Related Substructures and the GMM/Sausage Component} \label{sub:substructures}

The decline of $f_\mathrm{an}$ in the smooth, inner stellar halo indicates that the obvious substructures originating from the ancient radial merger have been removed. Stars of the \textit{Gaia}-Sausage stripped during the merger event are characterized by highly eccentric orbits. We therefore characterize substructures as likely belonging to the ancient merger based on eccentricity.
 
\begin{figure}
 	\centering
 	\includegraphics[width = 0.5\textwidth]{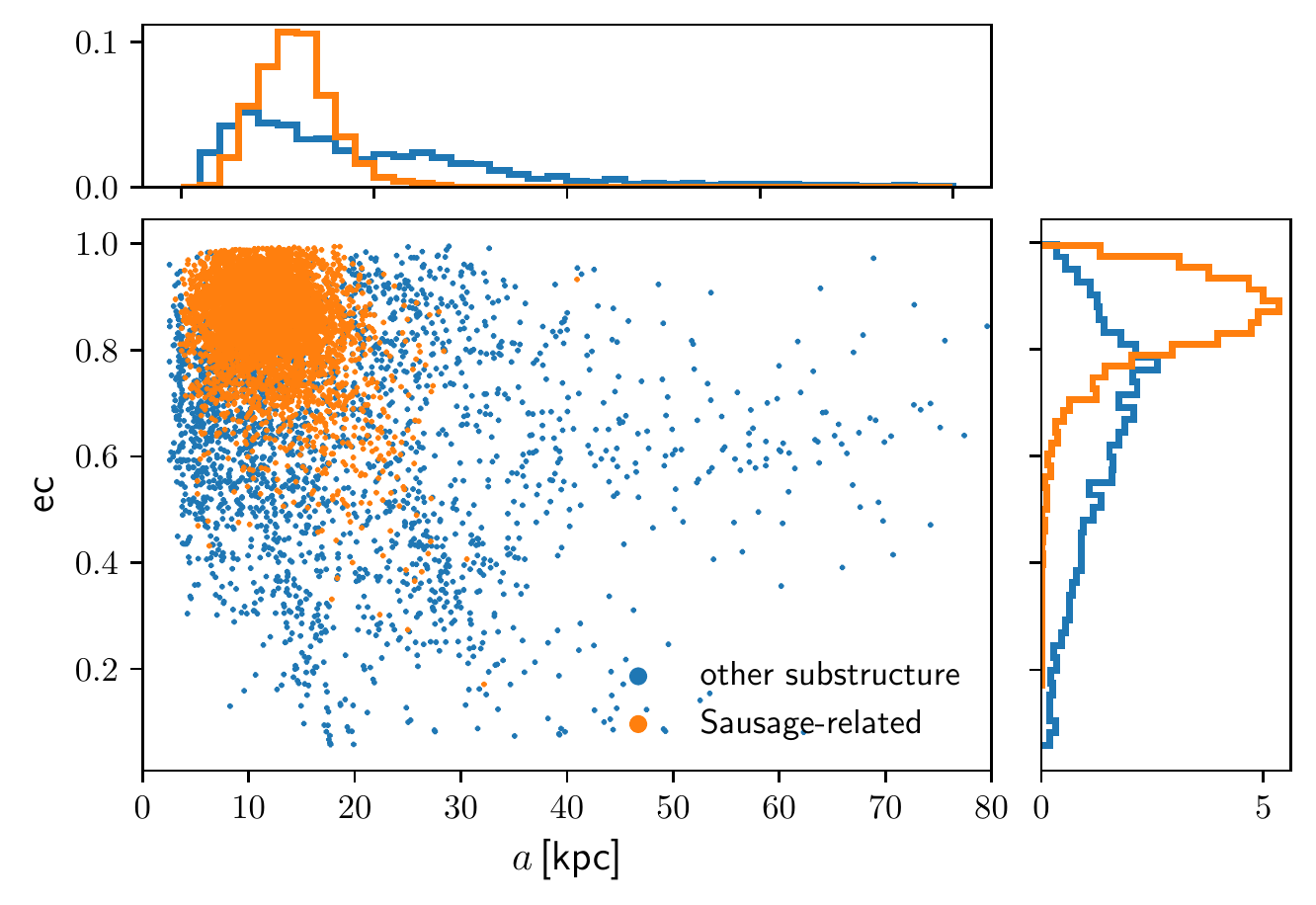}
 	\caption{Distribution of stars of the likely Sausage-related and other substructures in the semi major axis versus eccentricity diagram, and the side panels show the normalized histograms of semi-major axis and eccentricity. The Sausage-related substructure stars are mainly located at the upper left region with semi-major axis ${a}$ $<$ 30 kpc and ec $<$ 0.7.}
 	\label{fig:a-ec}
 \end{figure}

\begin{figure}
 	\centering
     \includegraphics[width = 0.5\textwidth]{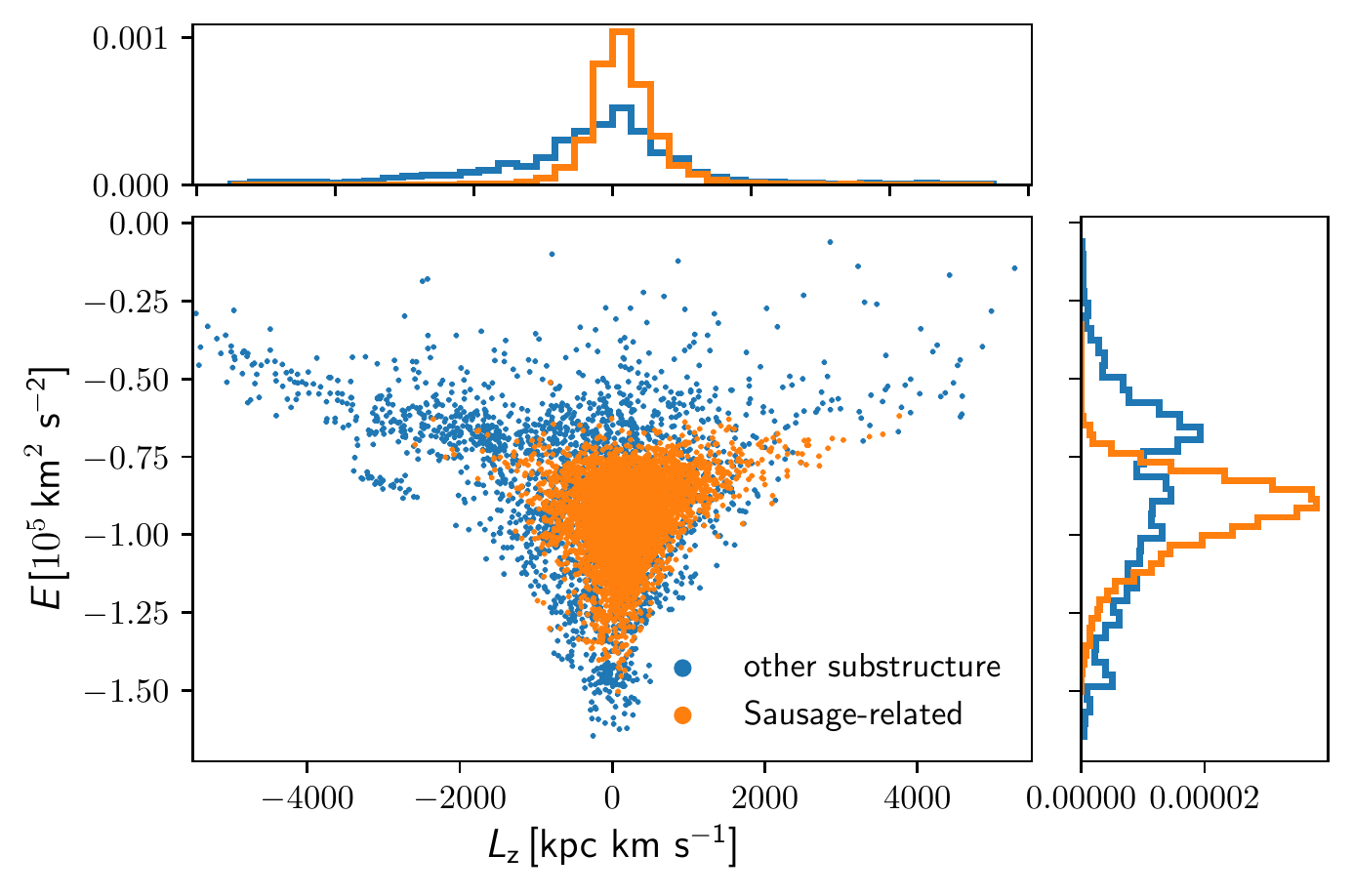}
     \caption{Distribution of stars of the likely Sausage-related and other substructures in the \textit{E} versus $\textit{L}_\mathrm{z}$ diagram, and the side panels show the normalized histograms of \textit{E} and $\textit{L}_\mathrm{z}$. The Sausage-related substructure stars have a mean energy ${\langle \textit{E} \rangle = -9.3\,\times\,10^4\,\text{km}^2\,\text{s}^{-2}}$, dispersion of energy ${\sigma_\textit{E} = 1.3\,\times\,10^4\,\text{km}^2\,\text{s}^{-2}}$, mean angular momentum ${\langle \textit{L}_\mathrm{z} \rangle = 125\,\text{kpc}\,\text{km}\,\text{s}^{-1}}$, and dispersion in ${\textit{L}_\mathrm{z}}$ ${\sigma_{\textit{L}_\mathrm{z}} = 498\,\text{kpc}\,\text{km}\,\text{s}^{-1}}$.}
     \label{fig:E-Lz}
 \end{figure}

 We select several possible Sausage-related substructures based on the work of Xue et al. (2021, in preparation). The specific selection criteria are semi-major axis $a$ $<$ 30 kpc and ec $>$ 0.8. Stars satisfying these criteria are included as possible members of the \textit{Gaia}-Sausage. Substructures containing such stars are taken into consideration. We only keep substructures with median ec $> 0.8$ and median $a < 20$ kpc. From the total number of substructure stars (not including Sgr stream members), the likely Sausage-related substructures have a total of 4611 stars (including LAMOST and SEGUE K giants, and SDSS BHB stars), and 2778 stars remain which belong to the other substructures. Figure~\ref{fig:a-ec} shows the distributions of stars of the Sausage-related and other substructures in the ${a}-\text{ec}$ diagram, where 93\% of the Sausage-related substructure stars are located at the space with ${a} <$ 30 kpc and ec $>$ 0.7. From Figure~\ref{fig:E-Lz}, we can see that stars of the Sausage-related substructures have a much narrower distribution in the $\textit{E}-\textit{L}_\mathrm{z}$ diagram when compared the other substructures. These Sausage-related substructure stars peak at ${\textit{E} \approx -1\,\times\,10^5\,\text{km}^2\,\text{s}^{-2}}$ and ${\textit{L}_\mathrm{z} \approx 125\,\text{kpc}\,\text{km}\,\text{s}^{-1}}$.  
 
\begin{figure*}
     \centering
     \includegraphics[width = 1.0\textwidth]{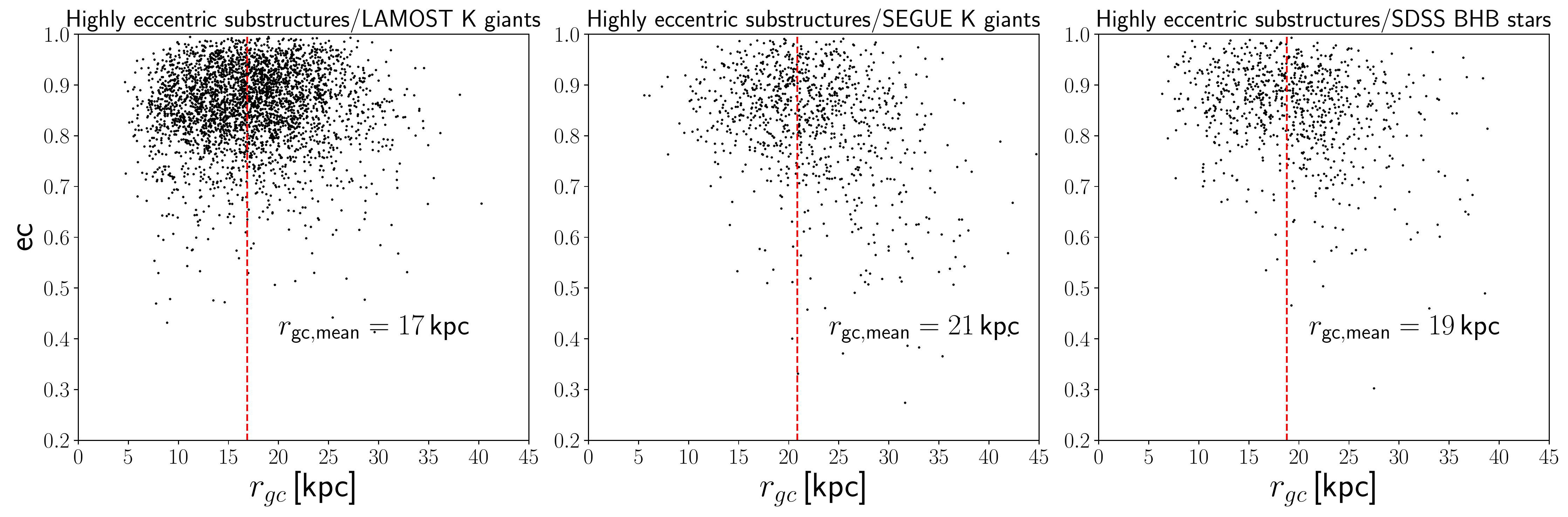}
     \caption{Galactocentric radius versus eccentricity of the highly eccentric substructure stars found in the LAMOST K giants (left), SEGUE K giants (middle), SDSS BHB stars (right). The red dashed line represents the mean Galactocentric radius ($r_\mathrm{gc,mean}$) of these stars. The majority of stars are located in the inner halo and are on highly eccentric orbits.}
     \label{fig:ec_rgc}
 \end{figure*}

\begin{figure}
     \centering
     \includegraphics[width = 0.5\textwidth]{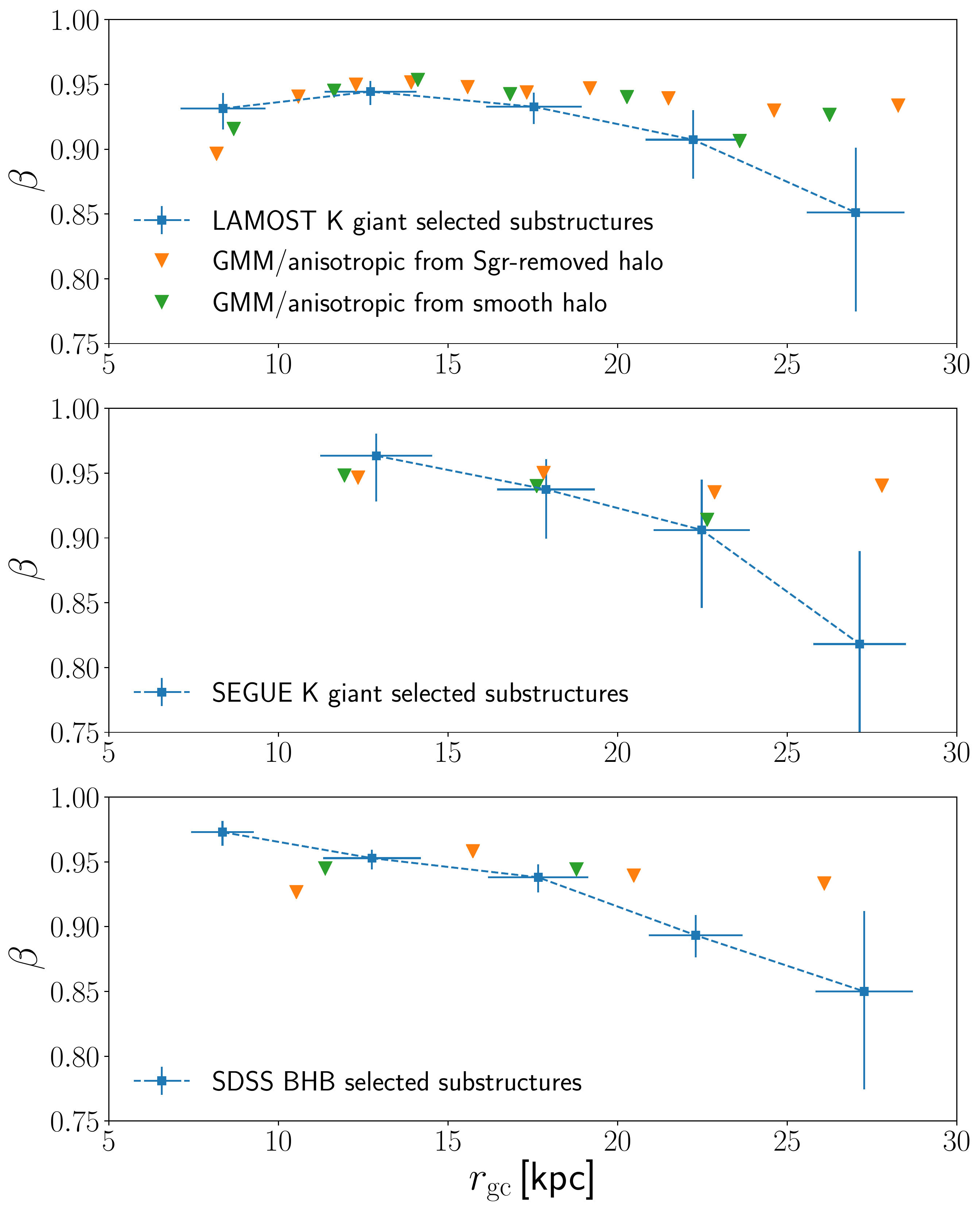}
     \caption{${\beta}$ of stars in the Sausage-related substructures versus $\beta$ of the GMM/anisotropic component of the Sgr-removed and the smooth halo samples for LAMOST K giants (top), SEGUE K giants (middle), and SDSS BHB stars (bottom).}
     \label{fig:beta_group}
     \end{figure}

 Figure~\ref{fig:ec_rgc} shows the Galactocentric radius and the eccentricity of stars belonging to these halo substructures. Only 6\% of the Sausage-related substructure stars are located in the outer halo, and they will be excluded in the following analysis. Figure~\ref{fig:beta_group} shows that the selected Sausage-related substructures are highly consistent with the GMM/Sausage component, both characterized by highly radially biased anisotropy.  

\begin{figure*}
    \centering
    \includegraphics[width=0.8\textwidth]{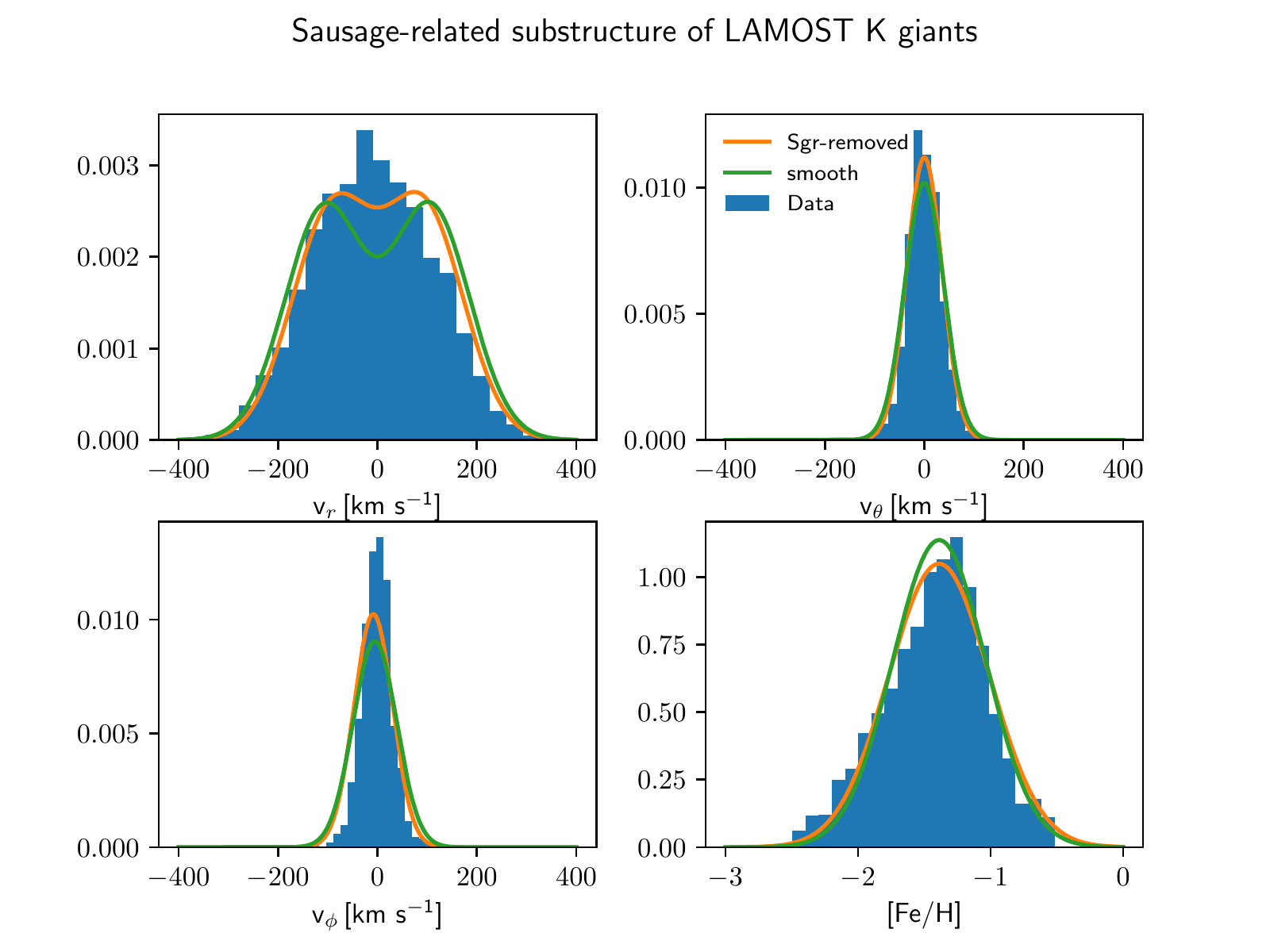}
    \includegraphics[width=0.8\textwidth]{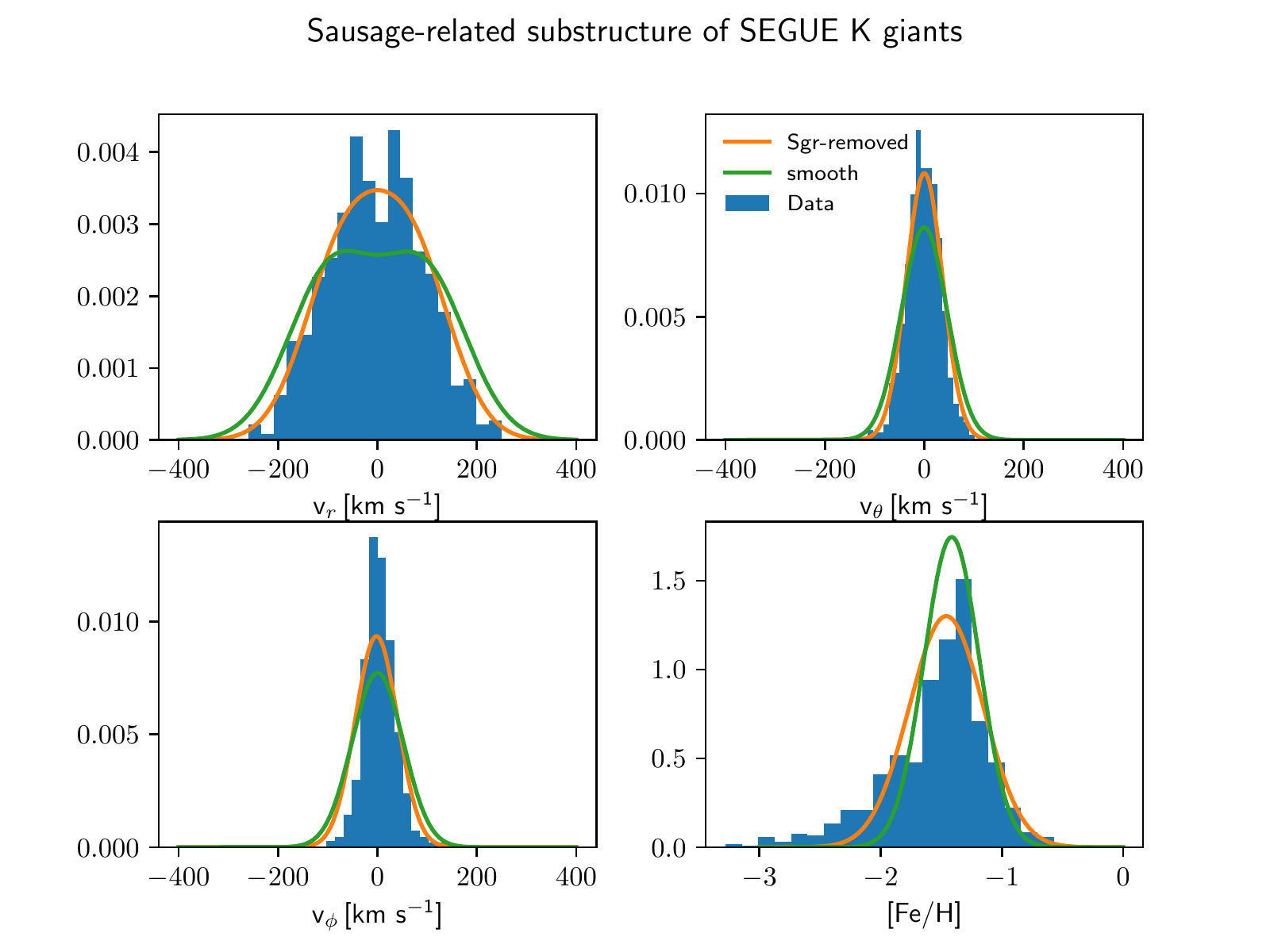}
    \caption{Velocity and metallicity distributions for the selected Sausage-related substructures (blue histogram) found in the LAMOST/SEGUE K giant sample. The GMM/anisotropic components (orange and green curves) are added as a comparison.}
    \label{fig:substructures_kg}
\end{figure*}

\begin{figure*}
    \centering
    \includegraphics[width=0.8\textwidth]{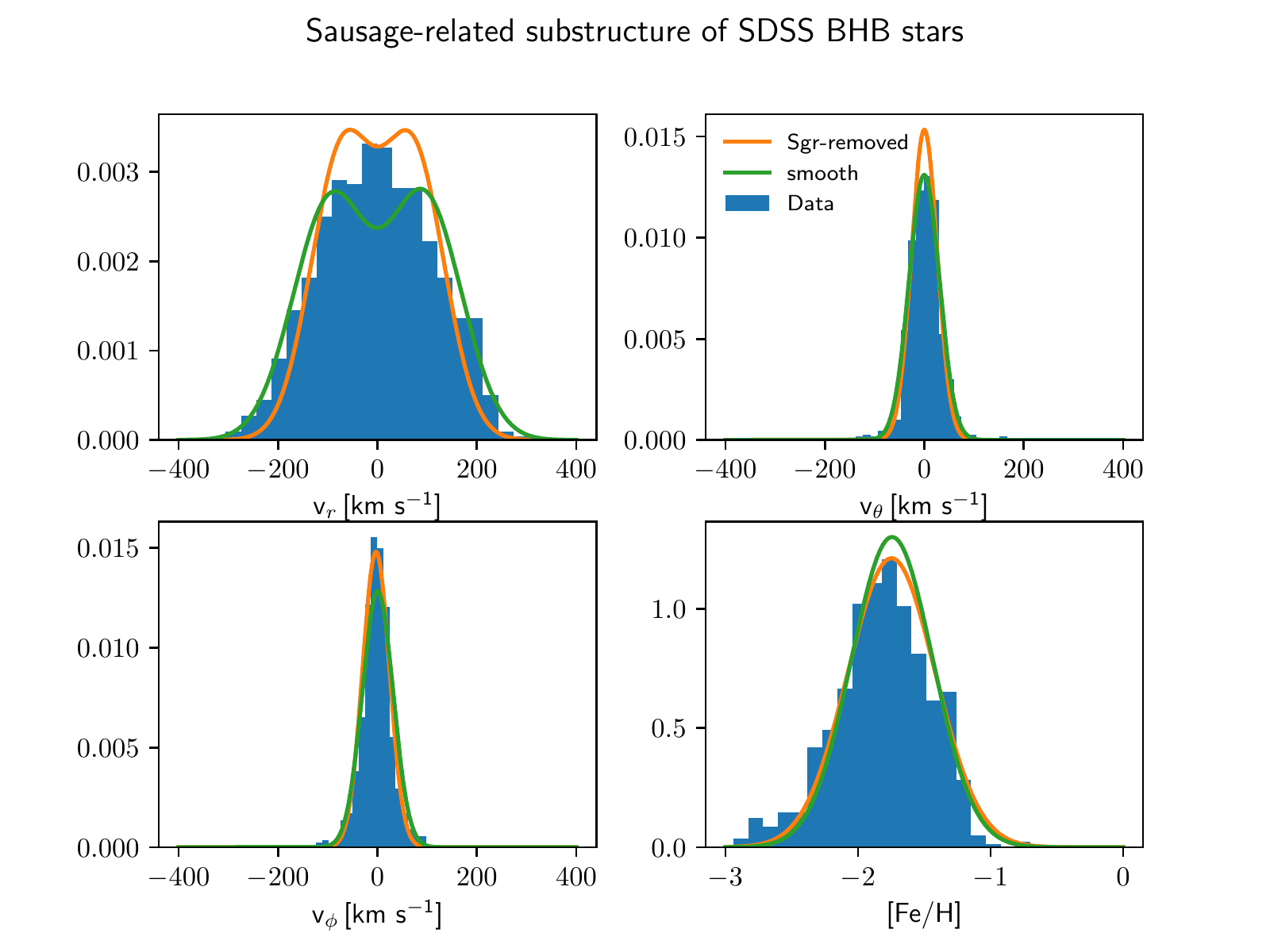}
    \caption{Velocity and metallicity distributions for the selected Sausage-related substructures found in the SDSS BHB sample.}
    \label{fig:substructures_bhb}
\end{figure*}

Substructures composed of stars on highly eccentric orbits can be used to test our GMM fits. The shape of the velocity ellipsoid and the metallicity distribution of Sausage-related substructure stars are shown in Figure~\ref{fig:substructures_kg} and ~\ref{fig:substructures_bhb}. The GMM/Sausage components of the Sgr-removed and the smooth halo samples are obtained in Section~\ref{sub:diffuse} and ~\ref{sub:smooth}.  For the comparison, we select the GMM/Sausage component of the distance bin in which the mean $r_\mathrm{gc}$ of the Sausage-related substructure stars is located. For example, ${r}_\mathrm{gc,mean}$ of the Sausage-related substructure stars found in the LAMOST K giant sample is 17 kpc. These stars are compared to the GMM/Sausage component of the LAMOST K giant Sgr-removed halo ${r}_\mathrm{gc} \in [16.46, 18.22]\,\text{kpc}$ and smooth halo ${r}_\mathrm{gc} \in [15.40, 18.37]\,\text{kpc}$. We also tried to compare the selected substructure stars to the GMM/Sausage component in the region $r_\mathrm{gc} \in [5,30]$ kpc. However, for the LAMOST K giant Sgr-removed halo sample, $r_\mathrm{gc}$ of 65\% of the stars in the inner halo is smaller than ${r}_{\mathrm{gc},\mathrm{mean}}$ of 17 kpc. Therefore, the GMM/Sausage component in the region $r_\mathrm{gc} \in [5,30]$ kpc tends to represent the Sausage-related substructure stars inside $r_\mathrm{gc}$ of 17 kpc, but seems to inadequately represent stars with $r_\mathrm{gc} >$  17 kpc.

The selection criteria of Sausage-related substructures in this study include a robust cut in the eccentricity. The GMM in Section~\ref{sub:GMM} is also a simplified model of the stellar halo, where the metallicity distribution of the \textit{Gaia}-Sausage is simply described by a Gaussian function. Therefore, slight differences between the GMM/Sausage component and the Sausage-related substructure stars are expected. In general, the GMM/Sausage components are consistent with the Sausage-related substructure stars both in the shape of the velocity ellipsoid and the stellar metallicity distribution, especially for the GMM/Sausage component of Sgr-removed halo. The inclusion of the Sausage-related substructure stars in the fitting process may cause the better performance of the GMM/Sausage component of Sgr-removed halo in the comparison.

 \citet{2017A&A...604A.106J} modeled the accretion of a series of satellites by means of high-resolution $N$-body simulations, and studied the kinematics of the accreted streams including the energy and angular momentum. They found that satellites of different masses can produce similar substructures and in-situ stars also form substructures in response to the accretion events. Therefore, stars originating from other smaller satellites and the in-situ halo may hide undetected in our selected Sausage-related substructures. However, we conclude that an in-situ halo component does not add a significant contribution since stars within the possible Sausage-related substructures are well matched by the GMM/Sausage component in the stellar metallicity distribution. A more detailed study of the elemental abundance of these stars will allow us to identify the substructures associated to the \textit{Gaia}-Sausage more accurately.

\begin{figure}
     \centering
     \includegraphics[width = 0.5\textwidth]{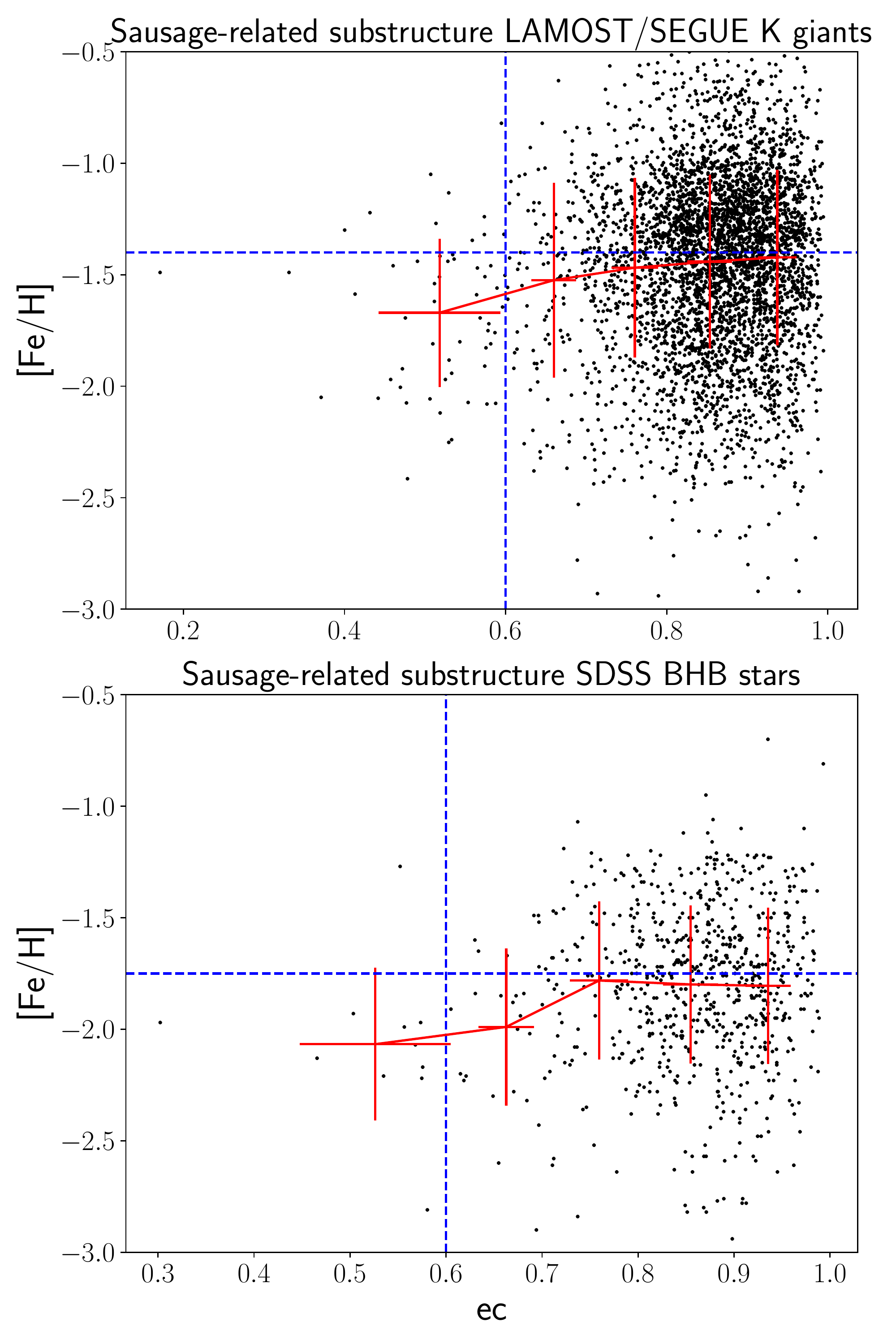}
     \caption{Eccentricity versus metallicity of stars belonging to the Sausage-related substructures for LAMOST/SEGUE K giants (top) and SDSS BHB stars (bottom). The vertical dash line separates stars of the Sausage-related substructures into two parts: low eccentricity (ec $\leq 0.6$) and high eccentricity (ec $\geq 0.6$). The horizontal blue dashed line represents the mean metallicity of the GMM/Sausage component. The red crosses show the mean stellar metallicity as a function of eccentricity. Stars with lower eccentricity tend to be more metal poor.} 
     \label{fig:ec_feh}
 \end{figure}

\citet{2020A&A...642L..18K} analyzed the kinematic properties of the debris of the \textit{Gaia}-Sausage through a detailed analysis of simulations from \citet{2008MNRAS.391.1806V}. In their preferred simulations, 75\% of the debris have eccentricity larger than 0.8, roughly 9\% have eccentricity smaller than 0.6. Our selection criteria tend to include stars with large eccentricity. However, stars with low eccentricity ($\leq 0.6$) can still be found in our selected substructures, which account for 1\% of the Sausage-related substructure stars. Figure~\ref{fig:ec_feh} displays the eccentricity and the stellar metallicity [Fe/H] of stars belonging to the Sausage-related substructures. We notice that stars with lower eccentricity are inclined to be more metal poor. For the combination of LAMOST and SEGUE K giants, the difference of the mean metallicity between the stars of low eccentricity (ec $\leq 0.6$) and high eccentricity ($0.9 \leq$ ec $\leq 1.0$) is 0.28, and for SDSS BHB stars this difference is 0.20. In the work of \citet{2020A&A...642L..18K}, star particles lost early from the \textit{Gaia}-Sausage analogue have large retrograde motions, and some of these as well have low eccentricity. Such stars are expected to be relatively metal poor since they originate from the outskirts of the satellite. This may explain the metal-poor stars with low eccentricity found in Figure~\ref{fig:ec_feh}.

\section{Discussion: Spectroscopic Survey Selection Function} \label{Sec:selection}

Halo stars used in this study are obtained from different spectroscopic surveys. Compared to photometric surveys, spectroscopic surveys are less complete and can only cover a smaller fraction of the stars in the Milky Way because the stars must be selected prior to collecting their spectra. The selection effect of spectroscopic surveys may affect the velocity and metallicity distributions of our halo star samples, which could further affect our result of the contribution of the \textit{Gaia}-Sausage. 

To correct the bias generated by selecting subsamples, many studies have developed selection functions in which the probability of a star being included in a spectroscopic survey is given by its Galactic location (\textit{l}, \textit{b}), color (\textit{c}), and apparent magnitude (\textit{m}). \citet{2017RAA....17...96L} and \citet{2015ApJ...809..144X} provided the selection functions for LAMOST DR5 and SEGUE-\uppercase\expandafter{\romannumeral2} K giants, respectively. However, the selection functions are not completely known for SEGUE-\uppercase\expandafter{\romannumeral1} and SDSS Legacy survey. Therefore, we can only use our LAMOST K giant sample to explore the influence of selection effects.  

The selection function \textit{S}(\textit{c}, \textit{m}, \textit{l}, \textit{b}) is defined as
\begin{equation}
    \textit{S}(\textit{c}, \textit{m}, \textit{l}, \textit{b}) = \frac{\textit{n}_\mathrm{sp}(\textit{c}, \textit{m}, \textit{l}, \textit{b})}{\textit{n}_\mathrm{ph}(\textit{c}, \textit{m}, \textit{l}, \textit{b})},
    \label{eq:selec_function}
\end{equation}
where $\textit{n}_\mathrm{sp}(\textit{c}, \textit{m}, \textit{l}, \textit{b})$ and $\textit{n}_\mathrm{ph}(\textit{c}, \textit{m}, \textit{l}, \textit{b})$ are the star counts of the spectroscopic and photometric data in the color-magnitude diagram. 

The likelihood of the GMM for a halo sample after the correction of selection effects is defined as
 \begin{equation}
 \begin{aligned}
 	 \mathcal{L}(D|\theta) = \prod_{i=1}^{N}(f_\mathrm{iso}\mathcal{L}_\mathrm{iso}(D_i|\theta) + f_\mathrm{an}\mathcal{L}_\mathrm{an}(D_i|\theta))^{\frac{1}{\textit{S}_i}},
 	 \end{aligned}
 	 \label{eq:GMM_selec}
 \end{equation}
where $\textit{S}_i$ is the selection function of a star $i$. 

\begin{figure}
    \centering
    \includegraphics[width=0.5\textwidth]{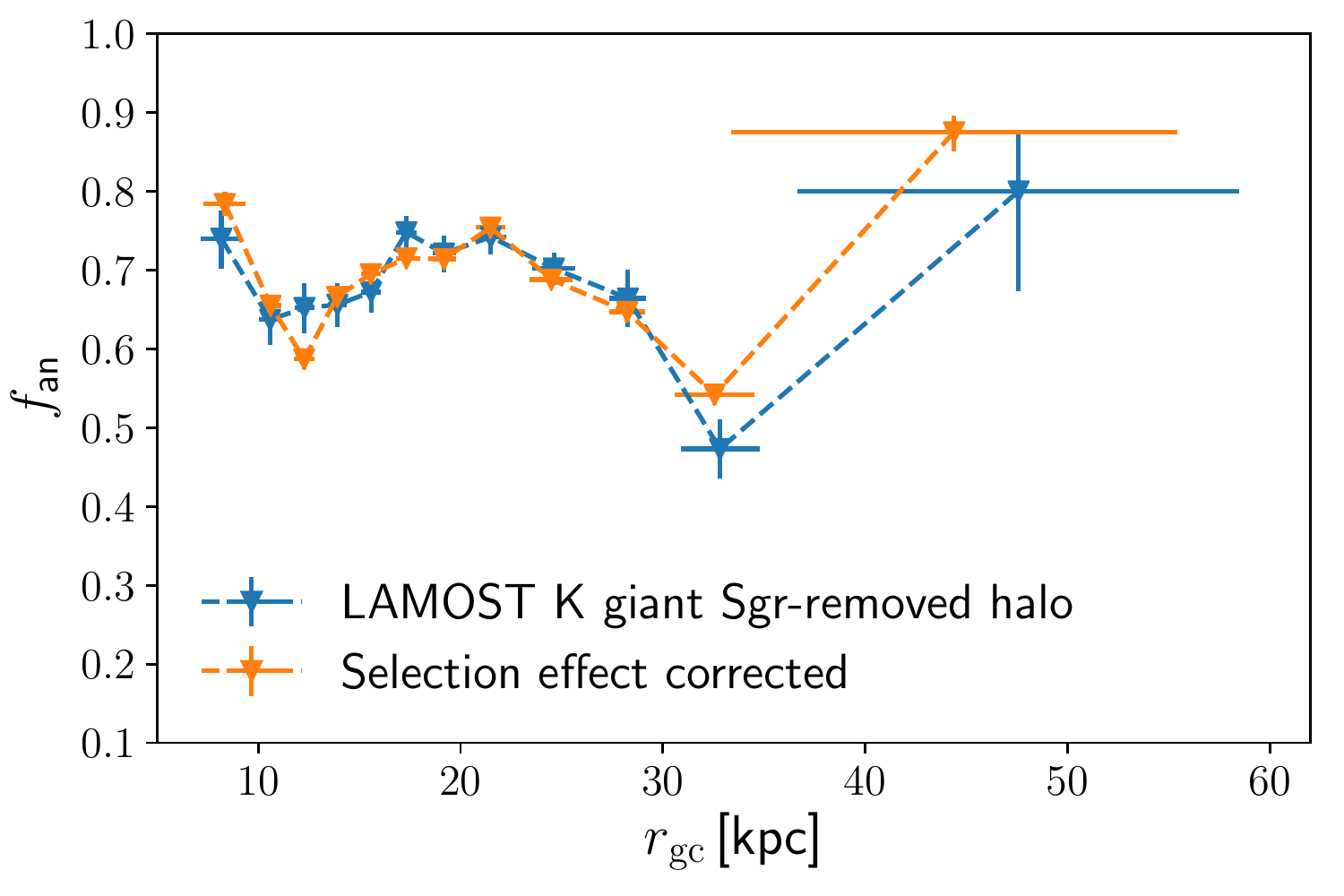}
    \caption{Comparison of the contribution of the GMM/Sausage component to the Sgr-removed stellar halo of LAMOST K giants before and after the correction of selection effects.}
    \label{fig:selection_corrected}
\end{figure}

For the LAMOST K giant Sgr-removed halo, 9756 stars satisfying $1/\textit{S} > 1$ and $1/\textit{S} < 40$ are selected and used in the GMM fit. From Figure~\ref{fig:selection_corrected}, we find that the influence of the selection effects on the contribution of the GMM/Sausage component is negligible for the LAMOST K giant sample, especially in the inner halo.     

\section{Conclusions} \label{dis_con}

In this study, we explore the contribution of the \textit{Gaia}-Sausage to the Galactic stellar halo by analyzing the chemodynamical  properties of LAMOST K giants, SEGUE K giants, and SDSS BHB stars. Xue et al. (2021, in preparation) selected halo substructures in IoM space by the Friend-of-Friend algorithm. In this study, we define the Sgr-removed halo sample as stars with the Sgr stream members excluded and the smooth halo sample as stars with all obvious IoM substructures removed. The Gaussian Mixture model, which divides the stellar halo into isotropic and anisotropic components, is used to fit the shape of the velocity ellipsoid and the metallicity distribution of the halo stars. We define substructures consisting of stars with high eccentricity (ec $>$ 0.8) as likely related to the \textit{Gaia}-Sausage. We compare these substructure stars with the GMM/anisotropic component. We summarize our main results as follows:

1. The GMM is confirmed as a better match of the stellar halo than the Single Gaussian model. The GMM/Sausage component is shown to be more metal-rich and radially biased than the GMM/isotropic component. The anisotropy of the GMM/Sausage component remains large ($\geq 0.8$) and changes little with $r_\mathrm{gc}$, while the anisotropy of the GMM/isotropic component changes more sporadically along $r_\mathrm{gc}$. Two distinct high $v_r$ lobes found in \citet{2018MNRAS.478..611B} are thought to be related to the debris of the \textit{Gaia}-Sausage. In our GMM fits, the absolute mean radial velocity $\langle v_r^\mathrm{an}\rangle$ of the two lobes decreases with increasing $r_\mathrm{gc}$. This is characteristic of ``shell" structures which result from the merger of high-mass-ratio satellite galaxies on highly radial orbits.   

2. The contribution of the \textit{Gaia}-Sausage to the Sgr-removed halo within 30 kpc of the Galactic Center is $64\%-74\%$ for LAMOST K giants, $55\%-61\%$ for SEGUE K giants, and $41\%-48\%$ for SDSS BHB stars. As the most metal-rich halo star sample in this study, the LAMOST K giant sample is proven to be the most heavily influenced by the \textit{Gaia}-Sausage. Our results support that the inner stellar halo is dominated by the last significant merger event, the \textit{Gaia}-Sausage. After reaching its peak value, the fraction of the GMM/Sausage component starts to decline beyond $r_\mathrm{gc} \sim$ $25-30$ kpc. In contrast to the inner halo, we find that the outer halo is significantly less influenced by the \textit{Gaia}-Sausage, and is best fit with \textit{Gaia}-Sausage contributing about $15\%-47\%$. 


3. The contribution of the \textit{Gaia}-Sausage to the smooth, inner stellar halo declines after the removal of the substructures, but is still as high as $54\%-63\%$ for LAMOST K giants, $45\%-51\%$ for SEGUE K giants, and $29\%-30\%$ for SDSS BHB stars. The chemodynamical properties of the GMM/Sausage component are not strongly influenced by the substructure removal process, especially the metallicity distribution shows negligible change.

4. We select substructures likely related to the \textit{Gaia}-Sausage and test our GMM fits. We find that the GMM/Sausage component is a good fit of the possible Sausage-related substructure stars in both the shape of the velocity ellipsoid and the metallicity distribution.

The contribution of the \textit{Gaia}-Sausage to the outer halo is largely uncertain in this study. Results obtained from the study of the SEGUE K giants and SDSS BHB stars support a small fraction, while results of the LAMOST K giants suggest that the \textit{Gaia}-Sausage still accounts for over half of the outer stellar halo. Larger samples of halo stars with detailed information of elemental abundances and ages, especially for the outer halo stars, are needed to further investigate the contribution of the \textit{Gaia}-Sausage to the stellar halo. Large-scale stellar spectroscopic surveys, for example further LAMOST data release, SDSS-\uppercase\expandafter{\romannumeral5}, and 4MOST Consortium Survey of the Milky Way, are expected to bring new insights into the assembly history of the Galactic halo \citep{2017arXiv171103234K,2019Msngr.175...23H}. \citet{2021ApJ...919...66B}
\acknowledgements
We thank Chris Flynn, Lachlan Lancaster, Lan Zhang for useful help and discussions. This study is supported by the National Natural Science Foundation of China under grant Nos. 11988101, 11890694, and 11873052 and National Key R$\&$D Program of China No. 2019YFA0405500. S.A.B. acknowledges support from the Aliyun Fellowship and Chinese Academy of Sciences President’s International Fellowship Initiative Grant (no. 2021PM0055). Telescope (the Large Sky Area Multi-Object Fiber Spectroscopic Telescope LAMOST) is a National Major Scientific Project built by the Chinese Academy of Sciences. Funding for the project has been provided by the National Development and Reform Commission. LAMOST is operated and managed by the National Astronomical Observatories, Chinese Academy of Sciences. This work presents results from the European Space Agency (ESA) space mission Gaia. Gaia data are being processed by the Gaia Data Processing and Analysis Consortium (DPAC). Funding for the DPAC is provided by national institutions, in particular the institutions participating in the Gaia MultiLateral Agreement (MLA). Funding for SDSS-III has been provided by the Alfred P. Sloan Foundation, the Participating Institutions, the National Science Foundation, and the U.S. Department of Energy Office of Science. The SDSS-III web site is \url{http://www.sdss3.org/.} SDSS-III is managed by the Astrophysical Research Consortium for the Participating Institutions of the SDSS-III Collaboration including the University of Arizona, the Brazilian Participation Group, Brookhaven National Laboratory, Carnegie Mellon University, University of Florida, the French Participation Group, the German Participation Group, Harvard University, the Instituto de Astrofisica de Canarias, the Michigan State/Notre Dame/JINA Participation Group, Johns Hopkins University, Lawrence Berkeley National Laboratory, Max Planck Institute for Astrophysics, Max Planck Institute for Extraterrestrial Physics, New Mexico State University, New York University, Ohio State University, Pennsylvania State University, University of Portsmouth, Princeton University, the Spanish Participation Group, University of Tokyo, University of Utah, Vanderbilt University, University of Virginia, University of Washington, and Yale University.


\bibliography{sample63}
\bibliographystyle{aasjournal}



\end{document}